\documentclass[sigconf, nonacm]{acmart}

\newcommand\vldbdoi{XX.XX/XXX.XX}

\newcommand\vldbavailabilityurl{https://github.com/RocmFang/FeLoG}
\usepackage[T1]{fontenc}
\usepackage{aecompl}
\usepackage{epsfig}
\usepackage{epstopdf}
\usepackage{cleveref}
\usepackage{mathtools}
\usepackage{algorithm}
\usepackage{multirow}
\usepackage{dsfont}
\usepackage{varwidth}
\usepackage[noend]{algpseudocode}
\algrenewcommand\algorithmicrequire{\textbf{Input:}}
\algrenewcommand\algorithmicensure{\textbf{Output:}}

\DeclareMathSymbol{\mlq}{\mathord}{operators}{``}
\DeclareMathSymbol{\mrq}{\mathord}{operators}{`'}
\usepackage{marginnote}

\usepackage{graphicx}
\usepackage{booktabs}
\usepackage{amsmath}
\usepackage{array}
\usepackage{amsfonts}
\usepackage{epsfig}
\usepackage{graphicx}
\usepackage{url}
\usepackage[bottom]{footmisc}
\usepackage{multirow}
\usepackage{xspace}
\usepackage{tabularx}
\usepackage{pifont}% http://ctan.org/pkg/pifont
\newcommand{\cmark}{\ding{51}}%
\newcommand{\xmark}{\ding{55}}%
\usepackage{url}
\usepackage{booktabs,adjustbox}

\newcommand{\spara}[1]{\smallskip\noindent{\bf #1}}

\renewcommand\arraystretch{1.3}

\newcommand{\squishlist}{
 \begin{list}{$\bullet$}
  {  \setlength{\itemsep}{0pt}
     \setlength{\parsep}{3pt}
     \setlength{\topsep}{3pt}
     \setlength{\partopsep}{0pt}
     \setlength{\leftmargin}{2em}
     \setlength{\labelwidth}{1.5em}
     \setlength{\labelsep}{0.5em}
} }
\newcommand{\squishlisttight}{
 \begin{list}{$\bullet$}
  { \setlength{\itemsep}{0pt}
    \setlength{\parsep}{0pt}
    \setlength{\topsep}{0pt}
    \setlength{\partopsep}{0pt}
    \setlength{\leftmargin}{2em}
    \setlength{\labelwidth}{1.5em}
    \setlength{\labelsep}{0.5em}
} }

\newcommand{\squishdesc}{
 \begin{list}{}
  {  \setlength{\itemsep}{0pt}
     \setlength{\parsep}{3pt}
     \setlength{\topsep}{3pt}
     \setlength{\partopsep}{0pt}
     \setlength{\leftmargin}{1em}
     \setlength{\labelwidth}{1.5em}
     \setlength{\labelsep}{0.5em}
} }

\newcommand{\squishend}{
  \end{list}
}

\newcommand{\eat}[1]{}

\newcounter{ccc}

\newcommand{\bigO}{\mathcal{O}}

\usepackage{pifont} 
\usepackage{subcaption}
\usepackage{xspace}
\newcommand{\felog}{{\sf FeLoG}\xspace}
\newcommand{\feest}{{\sf FeeST}\xspace}
\newcommand{\acm}{{\sf ACM}\xspace}
\newcommand{\fabc}{{\sf FaBC}\xspace}

\usepackage{enumitem}
\usepackage{balance}

\usepackage{wrapfig}
\usepackage{listings}
\usepackage[most]{tcolorbox}

\usepackage{multicol}
\newtcolorbox[auto counter]{mybox}[1][]{%
breakable,
enhanced,
sharp corners,
colback=white,
fonttitle=\bfseries,
enlarge bottom at break by=5mm,
enlarge top at break by=5mm,
overlay first={%
    \draw[black, line width=0.5mm](frame.south west)--(frame.south east);
    \node[anchor=north east] at (frame.south east) {continued on next page};
    },
overlay middle={%
    \draw[black, line width=0.5mm](frame.south west)--(frame.south east);
    \draw[black, line width=0.5mm](frame.north west)--(frame.north east);
    \node[anchor=north east] at (frame.south east) {continued on next page};
    \node[anchor=south west] at (frame.north west) {continued from next page};
    },
overlay last={%
    \draw[black, line width=0.5mm](frame.north west)--(frame.north east);
    \node[anchor=south west] at (frame.north west) {continued from next page};},
#1
}

  {\list{}{\leftmargin=0.15in\rightmargin=0.15in}\item[]}%
  {\endlist}

\usepackage{multicol}

\definecolor{metacolor}{HTML}{0072B2} % blue
\definecolor{R1color}{HTML}{800080} % purple
\definecolor{R2color}{HTML}{008866} % green
\definecolor{R3color}{HTML}{D55E00} % orange

\newtcolorbox{myquoteMeta}[1][]{
    colback=black!5,
    colframe=black!5,
    notitle,
    sharp corners,
    borderline west={1.5pt}{0pt}{blue},
    enhanced,
    breakable,
    left=2pt,
    right=2pt,
    top=0pt,
    bottom=0pt,
    ignore nobreak,
}

\newtcolorbox{myquoteR1}[1][]{
    colback=black!5,
    colframe=black!5,
    notitle,
    sharp corners,
    borderline west={1.5pt}{0pt}{R1color},
    enhanced,
    breakable,
    left=2pt,
    right=2pt,
    top=0pt,
    bottom=0pt,
    ignore nobreak,
}

\newtcolorbox{myquoteR2}[1][]{
    colback=black!5,
    colframe=black!5,
    notitle,
    sharp corners,
    borderline west={1.5pt}{0pt}{R2color},
    enhanced,
    breakable,
    left=2pt,
    right=2pt,
    top=0pt,
    bottom=0pt,
    ignore nobreak,
    #1
}

\newtcolorbox{myquoteR3}[1][]{
    colback=black!3,
    colframe=black!3,
    notitle,
    sharp corners,
    borderline west={1.5pt}{0pt}{R3color},
    enhanced,
    breakable,
    left=2pt,
    right=2pt,
    top=0pt,
    bottom=0pt,
    ignore nobreak,
}

\begin{document}

\title{FeLoG: Scalable and Efficient Distributed Graph Embedding with Feedback Loop Mechanism}
% \author{Peng Fang$^{\dagger}$, Arijit Khan$^{\ddagger}$, Ziqiang Wu$^{\dagger}$, Zhenli Li$^{\dagger}$, Yibo Zhou$^{\dagger}$, Fang Wang$^{\dagger}$, Dan Feng$^{\dagger}$}
% \affiliation{$^{\dagger}$Huazhong University of Science and Technology, China $^{\ddagger}$Bowling Green State University, USA}

\settopmatter{authorsperrow=4}
\author{Peng Fang}
\affiliation{Huazhong University of Science and Technology}
\email{fangpeng@hust.edu.cn}

\author{Arijit Khan}
\affiliation{Bowling Green State University}
\email{arijitk@bgsu.edu}

\author{Ziqiang Wu}
\affiliation{Huazhong University of Science and Technology}
\email{ziqiang_wu@hust.edu.cn}

\author{Zhenli Li}
\affiliation{Huazhong University of Science and Technology}
\email{ztruth@qq.com}

\author{Yibo Zhou}
\affiliation{Huazhong University of Science and Technology}
\email{zhouyibo@hust.edu.cn}

\author{Fang Wang}
\authornote{Corresponding Author.}
\affiliation{Huazhong University of Science and Technology}
\email{wangfang@hust.edu.cn}

\author{Dan Feng}
\affiliation{Huazhong University of Science and Technology}
\email{dfeng@hust.edu.cn}

\begin{abstract}

Graph embedding maps graph nodes into low-dimensional vectors to support applications such as recommendation, fraud detection, and retrieval-augmented generation.
As graphs scale to billions of edges, scalable and efficient graph embedding has become increasingly important.
Existing frameworks commonly adopt a sampling-training paradigm, in which mini-batches are constructed by sampling nodes and their neighbors.
However, sampling is typically decoupled from evolving embedding quality, causing redundant exploration of well-trained regions while under-sampling undertrained nodes.
At the system level, such decoupling further leads to excessive communication, serialized execution, and low resource utilization in distributed settings.
We present {\sf FeLoG}, a feedback loop-driven system for distributed graph embedding.
{\bf (1)} {\sf FeLoG} introduces feedback-coupled sampling and training, dynamically prioritizing undertrained nodes according to real-time embedding-quality feedback, reducing redundant computation and accelerating convergence.
{\bf (2)} It employs activity-aware communication that compresses frequently occurring node sequences to reduce intra-machine PCIe traffic and selectively synchronizes frequently updated embeddings to reduce inter-machine communication.
{\bf (3)} It adopts a round-interleaved pipeline that overlaps next-round sampling with current-round training to improve CPU-GPU utilization.
Experiments against state-of-the-art baselines on large-scale graphs show that {\sf FeLoG} achieves an average speedup of 27.9$\times$, reduces communication cost by more than 53.1\%, and sustains over 80\% CPU-GPU utilization.

% In this paper, we propose {\sf FeLoG}, a feedback loop-driven system for scalable and efficient distributed graph embedding, with three key technical innovations.
% {\bf (1)} {\sf FeLoG} introduces a feedback-coupled sampling-training model that dynamically adapts sampling using real-time embedding quality metrics. 
% Consequently, the undertrained nodes receive higher sampling priority and thereby
% reduces wasted computation and accelerates convergence.
% {\bf (2)} {\sf FeLoG} 
% implements
% an activity-aware communication mechanism to reduce
% communication overhead by analyzing data flow characteristics.
% Specifically, it compresses high-frequency node sequences to 
% %alleviate
% mitigate intra-machine PCIe traffic and selectively synchronize frequently updated embeddings to minimize inter-machine communication.
% {\bf (3)} {\sf FeLoG} adopts a round-interleaved pipeline that parallelizes
% sampling in the next round with training in the current round, thereby improving CPU-GPU utilization.
% We conduct extensive experiments on {\sf FeLoG} and six state-of-the-art baselines
% using large-scale real-world and synthetic graph datasets.
% The results
% show that {\sf FeLoG} achieves an average acceleration of 27.9$\times$ over the baselines, reduces communication cost by over 53.1\%, and sustains over 80\% CPU-GPU utilization, demonstrating its efficiency and scalability.

\end{abstract}


\maketitle

% %%% do not modify the following VLDB block %%
% %%% VLDB block start %%%
% \pagestyle{\vldbpagestyle}
% \begingroup\small\noindent\raggedright\textbf{PVLDB Reference Format:}\\
% Peng Fang, Arijit Khan, Ziqiang Wu, Zhenli Li, Yibo Zhou, Fang Wang, Dan Feng.
% \vldbtitle. PVLDB, \vldbvolume(\vldbissue): \vldbpages, \vldbyear. %\\ \href{https://doi.org/\vldbdoi}{doi:\vldbdoi}
% \endgroup
% \begingroup
% \renewcommand\thefootnote{}\footnote{\noindent
% This work is licensed under the Creative Commons BY-NC-ND 4.0 International License. Visit \url{https://creativecommons.org/licenses/by-nc-nd/4.0/} to view a copy of this license. For any use beyond those covered by this license, obtain permission by emailing \href{mailto:info@vldb.org}{info@vldb.org}. Copyright is held by the owner/author(s). Publication rights licensed to the VLDB Endowment. \\
% \raggedright Proceedings of the VLDB Endowment, Vol. \vldbvolume, No. \vldbissue\ %
% ISSN 2150-8097. \\
% %\href{https://doi.org/\vldbdoi}{doi:\vldbdoi} \\
% }\addtocounter{footnote}{-1}\endgroup
% %%% VLDB block end %%%

% %%% do not modify the following VLDB block %%
% %%% VLDB block start %%%
% \ifdefempty{\vldbavailabilityurl}{}{
% \vspace{.3cm}
% \begingroup\small\noindent\raggedright\textbf{PVLDB Artifact Availability:}\\
% The source code and data have been made available at \url{\vldbavailabilityurl}.
% \endgroup
% }
% %%% VLDB block end %%%

\section{Introduction}
\label{sec:intro}

Graph embedding \cite{cai2018comprehensive},
a powerful tool for graph analytics, has become a cornerstone for learning node representations from graph data. Specifically, it transforms high-dimensional, sparse graph structures into low-dimensional, dense vector spaces, preserving both structural and attribute information of the original graph.
Beyond traditional data-driven tasks, e.g., recommendation \cite{zhang2021graph}, link prediction \cite{rossi2021knowledge}, and node classification \cite{wu2022nodeformer}, their versatility has recently extended to
retrieval-augmented generation (RAG) \cite{edge2024local}.
Embeddings act as a bridge between graph-structured data and large language models (LLMs), enabling efficient retrieval and grounding over large-scale graphs.
In such applications, graphs can be huge with millions of nodes and billions of edges. 
For example, the Stanford {\em OGB-papers100M} citation network contains over 111 million nodes and 1.6 billion edges, posing significant challenges for tasks such as link prediction and classification \cite{hu2020open}.
Similarly, GraphRAG frameworks leverage knowledge graphs with comparable or greater scale, often containing millions of entities and relationships \cite{luo2025gfm}.

\begin{figure}[t!]
   \centering
   \includegraphics[width= 2.9 in]{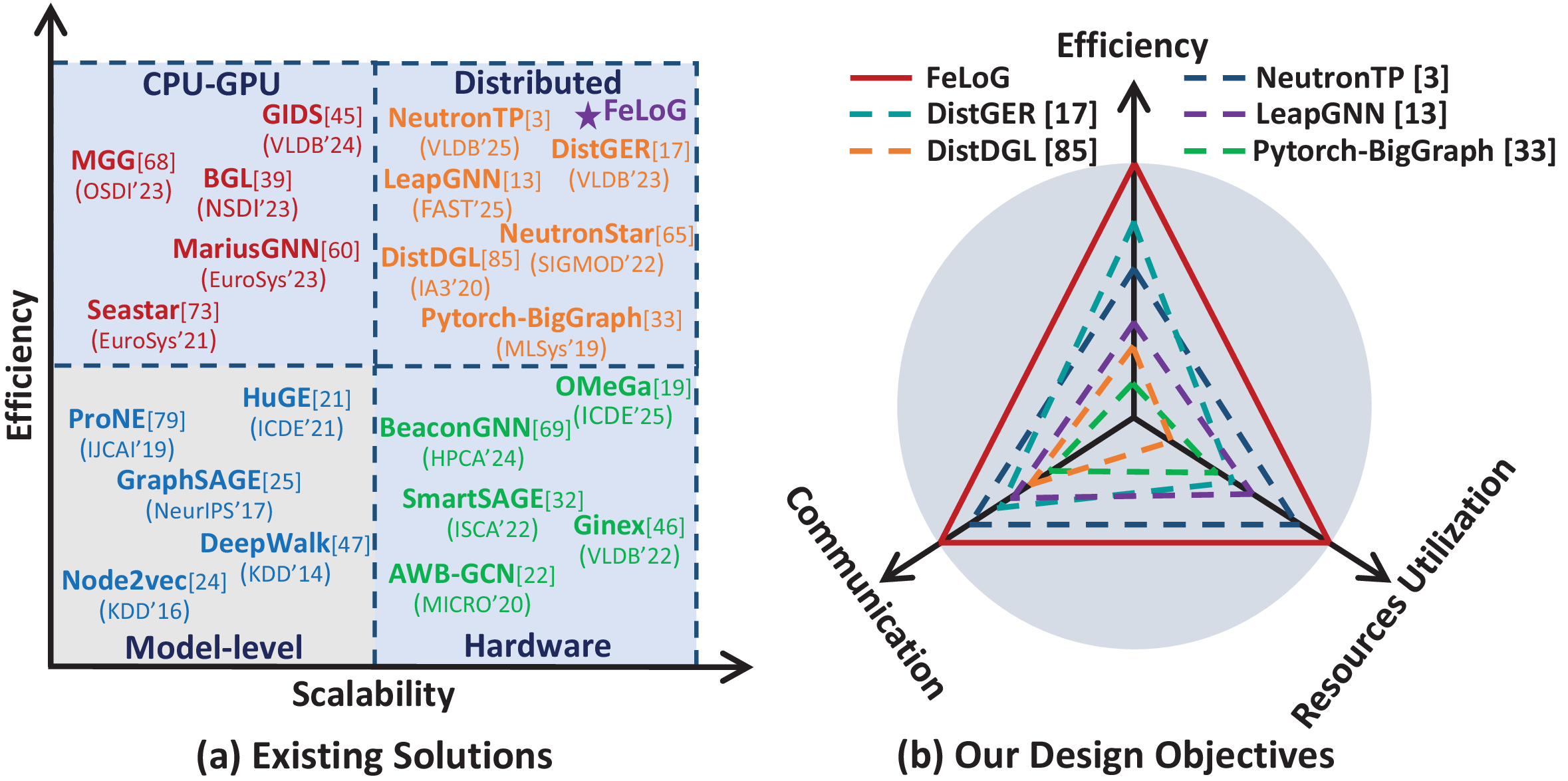}
   \vspace{-2mm}
   \caption{(a) Existing approaches vs. {\sf FeLoG} (ours) w.r.t. model- and system-level design choices and efficiency-scalability performance; (b) our design objectives for the proposed system {\sf FeLoG}.}
   \label{Sys_approaches}
\vspace{-4mm}
 \end{figure}

The commonly-used graph embedding paradigms follow a two-stage process: sampling and training. 
In random walk-based methods \cite{HuGE_2021,node2vec_2016,DeepWalk_2014}, node sequences or subgraphs are sampled from the original graph, and are used to train embedding models. 
In graph neural network (GNN)-based approaches \cite{GraphSAGE_2017,Graph_attention_2018,liao2022scara}, 
the sampling step selects a subset of neighbors for each node to construct computation subgraphs, and the training stage then performs message passing and aggregation on these sampled subgraphs to learn the final embeddings.
When the scale of graph data increases, the fundamental challenges in graph embedding primarily lie in efficiency and scalability, as depicted in Figure~\ref{Sys_approaches}(a).
At the core of these challenges is the decoupled sampling-training paradigm, which serves as the root cause of inefficiencies across both the model and system levels.

From the model perspective, the separation of sampling from training generates large amounts of redundant or uninformative training data without guidance from the evolving embedding states, thereby driving both computational and memory inefficiencies.
Classical methods such as {\sf DeepWalk} \cite{DeepWalk_2014} and {\sf node2vec} \cite{node2vec_2016} require months to process graphs with hundreds of millions of nodes and edges \cite{ProNE_2019}, and their scalability is further constrained by excessive memory consumption \cite{HuGE_2021}. 
Even neighbor-sampling methods such as {\sf GraphSAGE} \cite{GraphSAGE_2017} spend over 80\% of runtime on sampling due to the lack of feedback from training.
More recent methods like {\sf HuGE} \cite{HuGE_2021, HuGE+_2022} guide sampling adaptively but remain limited by the decoupled paradigm, hindering redundancy control and efficiency.

From the system perspective, existing efforts exploit heterogeneous architectures or distributed clusters by leveraging stronger GPUs \cite{liu2023bgl, mariusgnn, wang2023mgg}, larger SSDs \cite{Ginex_2022, park2024accelerating, SMARSAGE_2022}, or scale-out designs \cite{DistDGL_2020, fang2023distributed, wang2022neutronstar}. 
However, the decoupled sampling-training paradigm is also the key reason behind system-level inefficiencies, which persist despite these advances. One critical challenge is high communication overhead, as the bandwidth gap between computation and interconnect severely limits scalability. Since sampling is decoupled from training, the system must transfer all sampled data, including redundant neighborhoods, across CPU-GPU or machine boundaries, which significantly inflates communication costs. For example, the NVIDIA A100 provides 2 TB/s of GPU memory bandwidth, whereas PCIe bandwidth remains around 20 GB/s, making CPU-side sampling the bottleneck. In distributed settings, synchronizing embeddings for billion-edge graphs such as {\em Twitter} \cite{twitter_2010} requires hundreds of GBs of data exchange, incurring multi-second latencies and reducing GPU utilization to as low as 10\% in {\sf DistDGL} \cite{DistDGL_2020}. Another challenge is low resource utilization, since the sequential dependency enforced by decoupled sampling and training prevents overlap between the two stages and results in long idle periods. In {\sf DistGER} \cite{fang2023distributed}, CPU and GPU utilization remain only 15.8\% and 25.5\%, respectively, even under heavy workloads. Pipeline and caching strategies adopted in {\sf MariusGNN} \cite{mariusgnn} and {\sf Ginex} \cite{Ginex_2022} partially mitigate this issue, but their decoupled designs lack runtime feedback to dynamically adapt sampling to training needs.

To address the above challenges, we present {\sf FeLoG}, a distributed system built around a sampling-training \underline{\bf Fe}edback \underline{\bf Lo}op for \underline{\bf G}raph embedding. 
This feedback loop directly targets the root cause of inefficiencies and serves as the foundation that enables our three technical contributions.
Figure~\ref{Sys_approaches}(b) summarizes the design objectives of {\sf FeLoG}, which correspond to improving execution efficiency, reducing communication overhead, and enhancing resource utilization.
{\bf First}, {\sf FeLoG} introduces a \underline{Fee}dback-coupled \underline{S}ampling-\underline{T}raining model ({\sf FeeST}) that tightly integrates sampling with training. Unlike existing methods that sample blindly or heuristically, {\sf FeeST} leverages real-time feedback from embedding quality to guide subsequent sampling. By prioritizing nodes with undertrained embeddings, it captures more informative context, reduces redundancy, and accelerates convergence.
{\bf Second}, to mitigate communication bottlenecks, {\sf FeLoG} incorporates an \underline{A}ctivity-aware \underline{C}ommunication \underline{M}echanism ({\sf ACM}). Here, activity refers to dynamic access and update patterns of nodes during embedding training, e.g., how frequently a node or its context is sampled and how often its embedding is modified. By exploiting these patterns, {\sf ACM} analyzes data flow characteristics and applies targeted optimizations. Specifically, it uses a \underline{F}requency-\underline{a}ware \underline{B}itmap \underline{C}ompression strategy ({\sf FaBC}) to reduce PCIe traffic between CPU and GPU, and a \underline{H}otspot-\underline{a}ware \underline{Syn}chronization strategy ({\sf HaSyn}) to minimize inter-machine communication by focusing only on embeddings with high update activity.
{\bf Last}, to overcome resource underutilization, {\sf FeLoG} employs \underline{R}ound-\underline{i}nterleaved \underline{P}ipeline \underline{P}arallelism ({\sf RiPP}), which overlaps sampling in the next round with training in the current one. This design decouples strict sequential execution, allowing CPU-based samplers and GPU-based trainers to operate concurrently across rounds,  improving system throughput.

{\bf Contributions and roadmap.}
We present {\sf FeLoG}, a scalable, efficient, end-to-end distributed graph-embedding system, which, to our best knowledge, is the first general-purpose system that integrates feedback loop coordination to couple sampling and training.

$\bullet$ We propose the feedback-coupled sampling-training model ({\sf FeeST}) for adaptive sampling based on embedding quality (\S\ref{sec:FeedST}).

$\bullet$ We develop an activity-aware communication mechanism ({\sf ACM}), including frequency-aware bitmap compression ({\sf FaBC}) and hotspot-aware synchronization ({\sf HaSyn}), to reduce intra- and inter-machine communication overheads (\S\ref{sec:acm}).

$\bullet$ We design the round-interleaved pipeline parallelism ({\sf RiPP}) to overlap sampling and training, improving resource utilization (\S\ref{sec:rip}).

$\bullet$ We demonstrate the generalization of {\sf FeLoG}, supporting diverse techniques and graph types, including purely structural graphs and graphs with node features 
(\S \ref{sec:integration} \& \S \ref{sec:experiments}).

$\bullet$ We conduct extensive experiments on seven large, real-world 
and synthetic graphs to demonstrate that {\sf FeLoG} achieves superior efficiency, scalability, and effectiveness over state-of-the-art, popular distributed graph embedding frameworks.
% namely, {\sf DistDGL} \cite{DistDGL_2020}, {\sf Pytorch-BigGraph} \cite{PBG_2019}, {\sf DistGER} \cite{fang2023distributed}, {\sf DistGER-Pipe} \cite{fang2025information} {\sf LeapGNN} \cite{chen2025leapgnn}, and {\sf NeutronTP} \cite{ai2024neutrontp}. 
Moreover, we are the first to exhibit a Graph-based RAG case study using distributed graph embeddings, showcasing {\sf FeLoG}’s integration into a retrieval-augmented generation pipeline 
(\S  \ref{sec:experiments}).

% $\bullet$ \textcolor{red}{[AK: mention about GraphRAG  case study - highlight that we are the first to demonstrate such case study among distributed graph embedding works.]} \textcolor{blue}{[Fang: updated in experiment.]}

% The rest of the paper is organized as follows. We discuss preliminaries and a baseline approach in \S \ref{sec:preliminaries}, related work in \S \ref{sec:related}, motivations in \S \ref{sec:motivations}, and we conclude in \S \ref{sec:conclusions}.
% \vspace{-2mm}
\section{Preliminaries}
\label{sec:preliminaries}

\begin{figure}[tb!]
 \centering
 \includegraphics[width= 3 in]{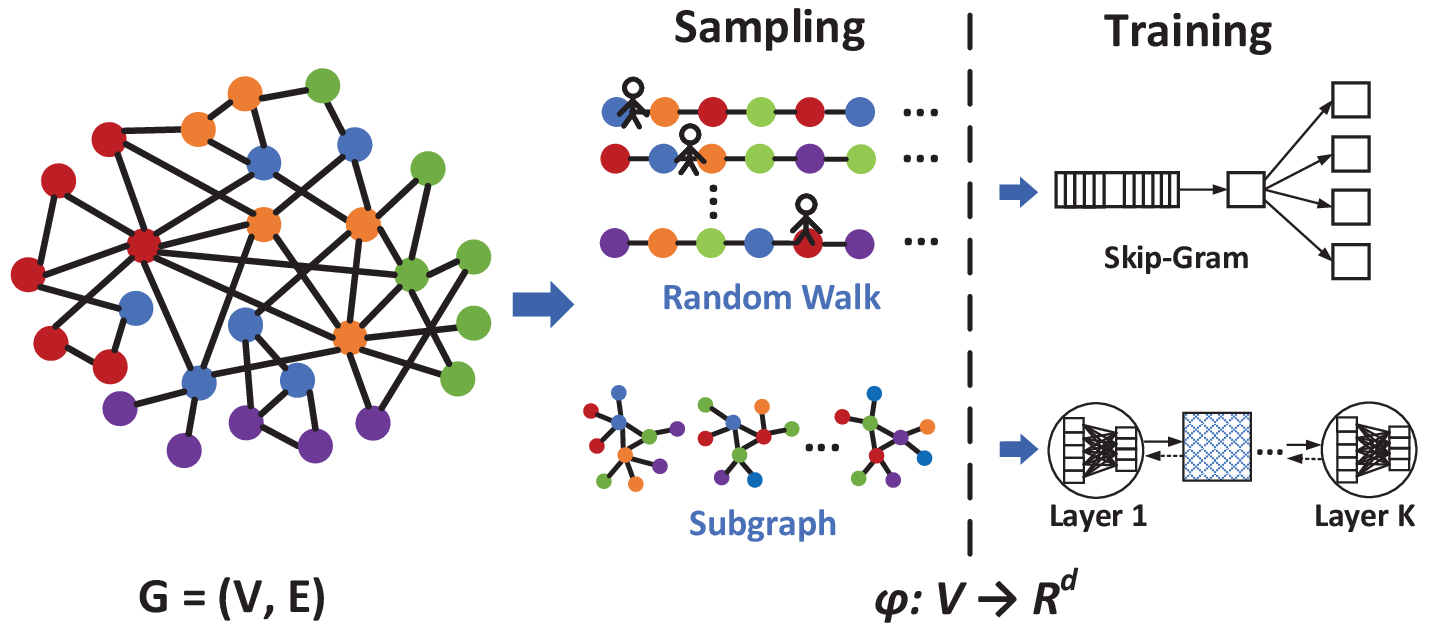}
 \caption{Sampling and training in graph embedding.}
 \vspace{-2mm}
 \label{sampling_training_execution}
 \vspace{-2mm}
\end{figure}

Most graph embedding methods use a two-stage pipeline: a sampling stage that extracts node sequences or subgraphs from the input graph, followed by a training stage that learns node embeddings from those samples (Figure ~\ref{sampling_training_execution}). The stages run independently, with sampling driven only by the graph structure or previous samples and receiving no feedback from the training process.
 
\underline{Sampling process} is commonly implemented via truncated walks. 
Starting from each node, multiple fixed-length walks are performed to explore the local graph structure. Walk directions are determined based on structural heuristics, such as node degree or common neighbors \cite{HuGE_2021}. To ensure coverage, sampling is repeated across the entire graph. While effective, this process often generates redundant or low-quality information, especially when applied uniformly across all nodes regardless of their structural roles \cite{DeepWalk_2014}.
For GNN-based approaches, the sampling step corresponds to selecting a subset of neighbors for each node at every layer, thereby constructing a smaller computation subgraph for subsequent training. This reduces the input size compared to using the full neighborhood, but redundancy still exists when large neighborhoods are uniformly sampled.

\underline {Training process} typically uses models inspired by natural language processing, such as word2vec~\cite{Word2Vec_2013}. Here, sampled node sequences form a corpus, with nodes as words and their co-occurrences defining context. The Skip-gram model predicts surrounding nodes (context) given a target node (center), learning embedding vectors that capture structural and semantic proximity. Formally, it maximizes the co-occurrence probability within a window \(w\):
\begin{small}
% \vspace{-2mm}
\begin{equation}
\operatorname*{argmax}_{\vec{V}} \frac{1}{|V|}\sum\limits_{j=1}^{|V|}\sum\limits_{-w\leq i\leq w}\log{p(u_{j+i}| u_j)}
\label{skim_gram_eq}
\end{equation}
\end{small}
The generated walks are treated as a corpus with vocabulary $V$, 
where $u_{j+i}$ denotes a context node in a window $w$.
Existing methods speed up training via negative sampling \cite{negative_sampling_2013}.
It adjusts the embeddings such that the target node is close to its true context nodes in the vector space, while the negative samples are pushed farther away.
\begin{small}
\begin{equation}
\label{negative_sampling_Eq}
\begin{aligned}
\log p(u_j| u_{j+i}) &\approx \log\sigma(\varphi_{in}(u_{j+i})\cdot\varphi_{out}(u_{j})) \\
&+\sum\limits_{k=1}^{K}\mathbb{E}_{u_k \sim Pn(u)}[\log\sigma(-\varphi_{in}(u_{j+i})\cdot\varphi_{out}(u_{k}))]
\end{aligned}
\end{equation}
\end{small}
\begin{table}[tb!]
\centering
\caption{Frequently used notations.}
\label{tab:Notations_paper}
\vspace{-2mm}
\footnotesize
\setlength{\tabcolsep}{2pt}
\renewcommand{\arraystretch}{1.0}
\begin{tabular}{ll}
\toprule
\textbf{Notation} & \textbf{Meaning} \\
\midrule
$G=(V,E)$        & Undirected, unweighted graph: $V$ nodes and $E$ edges \\
$\varphi(u)$     & Embedding of node $u$ with dimension $d$ \\
$\mathcal{N}(v)$ & Sampled nodes accessed for message passing of node $v$ \\
$V_a^r$          & Active nodes in round $r$ \\
$\Psi(v)$        & Embedding quality of node $v$ \\
$L$              & Random walk length \\
$l$              & $l$-th GNN layer \\
$r$              & Number of random walks per node \\
\bottomrule
\end{tabular}
\vspace{-4mm}
\end{table}

Here, $\sigma(x)=\frac{1}{1+exp(-x)}$ is the sigmoid function, and the expectations
are computed by drawing random nodes from a sampling distribution $Pn(u)$, $\forall u \in V$.
Typically, the number of negative samples $K$ is much smaller than $|V|$ (e.g., $K\in[5,20]$).
However, since the training process does not influence how samples are collected, it cannot correct or avoid redundant or ineffective sampling.

For GNN-based methods, the training stage performs iterative message passing and aggregation on the sampled subgraphs. Each node updates its embedding by aggregating the representations of its sampled neighbors through parameterized functions (e.g., mean pooling, attention), followed by non-linear transformations.
Formally, the $l$-th layer of a GNN is expressed as:  
\begin{equation}
\small
h_v^{(l)} = \sigma \Big( W^{(l)} \cdot \text{AGG}\big( \{ h_u^{(l-1)} : u \in \mathcal{N}(v) \cup \{v\} \} \big) \Big)
\label{GNN_train}
\end{equation}
where $h_v^{(l)}$ is the embedding of node $v$ at layer $l$, $\mathcal{N}(v)$ are the sampled neighbors, $\text{AGG}(\cdot)$ is a permutation-invariant function, $W^{(l)}$ is a learnable weight matrix, and $\sigma(\cdot)$ is a non-linear activation.

\underline{Complexity analysis.} For random walk-based approaches, assume that the number of walks per node is $r$, walk length $L$, embedding dimensions $d$, window size $w$,
and the number of negative samples $K$. The time complexity of sampling procedure
is $\bigO(r\cdot L\cdot |V|)$.
For training procedure, the corpus size $C = r\cdot L$. Let us denote the complexity of the unit operation of predicting and updating one node's
embedding as $o$. The {\sf Skip-Gram} with the negative sampling only needs $K+1$ words to obtain a probability distribution
(Eq.~\ref{negative_sampling_Eq}), thus the time complexity of
training procedure is $\bigO(C \cdot w \cdot (K+1) \cdot o)$.
For GNN-based methods, let $|\mathcal{N}(v)|$ denote the total number of nodes accessed for message passing for node $v$ across all $l$ layers (including multi-hop neighbors selected by sampling).
The sampling complexity is $\bigO(|V|\cdot|\mathcal{N}(v)|)$, 
and the training complexity from Eq.~\ref{GNN_train} (i.e., one forward pass in GNN) is $\bigO(|V| \cdot |\mathcal{N}(v)| \cdot d)$.

\section{Related Work}
\label{sec:related}

\spara{Graph embedding algorithms.}
Graph embeddings~\cite{cai2018comprehensive} have been applied to uncertain graphs~\cite{uncertain_graph_embedding_2017}, dynamic graphs~\cite{dynamic_embedding_2021}, heterogeneous networks~\cite{wang2022survey}, retrieval-augmented generation~\cite{zhang2025survey}, and knowledge graphs~\cite{cao2024knowledge}.
Broadly, recent algorithms fall into two families.
\emph{GNN-based} approaches~\cite{TuCWY018,GraphSAGE_2017,Graph_attention_2018,Graphgan_2018} learn embeddings via iterative message passing and aggregation over sampled neighborhoods, enabling semi-supervised or supervised learning; representative methods include {\sf GraphSAGE}~\cite{GraphSAGE_2017}, {\sf GAT}~\cite{Graph_attention_2018}, and {\sf GraphGAN}~\cite{Graphgan_2018}.
\emph{Random-walk-based} methods~\cite{DeepWalk_2014,node2vec_2016,HuGE_2021,HuGE+_2022} transform graphs into node sequences via random walks and train with {\sf Skip-Gram}~\cite{Word2Vec_2013}; e.g., {\sf DeepWalk}~\cite{DeepWalk_2014} pioneered the idea and {\sf node2vec}~\cite{node2vec_2016} introduced biased walks, while {\sf HuGE}~\cite{HuGE_2021,HuGE+_2022} adapts sampling configurations on the fly.
Despite their differences, both paradigms adopt a sequential, decoupled execution of sampling and training, which can induce redundant sampling and high training overhead, ultimately limiting efficiency and scalability~\cite{fang2023distributed}.

\spara{Single-Machine Graph Embedding Systems.}
To overcome efficiency bottlenecks in large-scale graph embedding, prior work ranges from full-GPU systems (e.g., GENTI \cite{yu2024genti}, GAMMA \cite{qiu2024gpu}, and XGNN \cite{tang2024xgnn}) that maximize on-device throughput to heterogeneous architectures that address end-to-end bottlenecks.
Early systems such as {\sf BGL} \cite{liu2023bgl} %and Tencent's graph embedding platform \cite{Tencent_GE_2020} 
adopt a CPU-GPU hybrid design, with CPUs performing random-walk sampling and GPUs handling embedding training. 
{\sf MariusGNN} \cite{mariusgnn} further optimizes this paradigm with disk-efficient pipelines, while {\sf Seastar} \cite{seastar_2021} provides a GPU-centric execution framework with a vertex-centric programming model \cite{luo2023multi}.
% More recent efforts improve single-node efficiency by coordinating heterogeneous resources and the I/O path.
Recently, NeutronOrch \cite{ai2024neutronorch} orchestrates computation and communication across devices via runtime scheduling, DAHA \cite{li2024daha} adapts hardware assignment based on workload characteristics, and GIDS \cite{park2024accelerating} enables GPU-direct storage access to accelerate data loading.
{\sf LPS-GNN} \cite{cheng2025lps} further scales single-node GNN training via improved partitioning.
Despite these advances, single-machine systems still face fundamental scalability limits, primarily due to CPU-GPU imbalance and the constrained GPU memory capacity.
% the performance gap between CPUs and GPUs and the limited memory capacity of GPUs remain fundamental scalability bottlenecks.  
Beyond CPU-GPU pipelines, several systems leverage specialized hardware to further enhance scalability and efficiency. 
{\sf Ginex} \cite{Ginex_2022} exploits SSDs to extend memory for large-scale embeddings; {\sf SmartSAGE} \cite{SMARSAGE_2022} and {\sf AWB-GCN} \cite{geng2020awb} design hardware-conscious scheduling strategies for GNN workloads; {\sf BeaconGNN} \cite{wang2024beacongnn} incorporates hardware accelerators for higher throughput; and the very recent {\sf OMeGa} \cite{fang2025omega} utilizes heterogeneous memory processing to deliver high-performance embedding at scale.  
While these systems train on larger datasets, they often sacrifice generalizability and face compatibility challenges across diverse heterogeneous environments.

\spara{Distributed Graph Embedding Systems.}  
To scale graph embeddings, distributed systems have been developed to leverage multi-machine and multi-GPU clusters \cite{NuPS_2022,AliGraph_2019}. 
{\sf MSPipe}~\cite{sheng2024mspipe} accelerates memory-based temporal GNN training on multi-GPUs via staleness-aware pipeline scheduling.
% , focusing on temporal memory dependencies that are orthogonal to {\sf FeLoG}.
{\sf GraNNDis}~\cite{song2024granndis} accelerates distributed GNN training on multi-server clusters by reducing inter-server communication, redundant computation, and sampling-induced inefficiency.
% {\sf NuPS} \cite{NuPS_2022} addresses two major sources of non-uniform parameter accesses during training, improving efficiency of distributed parameter management.  
% {\sf AliGraph} \cite{AliGraph_2019} optimizes distributed GNN training by accelerating sampling operators and reducing communication through node caching on local machines.  
Amazon’s {\sf DistDGL} \cite{DistDGL_2020} extends the {\sf Deep Graph Library} \cite{dgl_2019} with distributed GNN training via mini-batch execution.  
Similarly, {\sf PyTorch-BigGraph} \cite{PBG_2019} scales embedding training to large graphs using graph partitioning and parameter servers.  
However, both face communication bottlenecks that limit GPU utilization and scalability (\S\ref{sec:experiments}). 
Recent works enhance distributed efficiency via improved scheduling and communication.
{\sf ByteGNN} \cite{ZhengCCSWLCYZ22} employs two-level scheduling and tailored graph partitioning to improve parallelism.
{\sf DistGER} \cite{fang2023distributed} and {\sf DistGER-Pipe} \cite{fang2025information} adopt information-oriented random walks to reduce redundant sampling and communication.
{\sf LeapGNN} \cite{chen2025leapgnn} proposes a feature-centric distributed GNN training framework to optimize feature access and communication.
However, these systems largely suffer from decoupled execution and high transfer overheads under heterogeneous clusters.
{\sf NeutronHeter} \cite{cao2025neutronheter} accelerates GNN training through hierarchical workload mapping and adaptive communication, while {\sf NeutronTP} \cite{ai2024neutrontp} optimizes tensor parallelism for GNNs by improving communication and memory usage.
However, {\sf NeutronTP} applies uniform processing across all node features and lacks feedback-driven embedding evaluation, leading to redundant computation and slower convergence compared to {\sf FeLoG} (\S\ref{sec:experiments}).
% Recent efforts focus on improving efficiency in distributed settings through better scheduling and communication optimization.  
% {\sf ByteGNN} \cite{ZhengCCSWLCYZ22} employs two-level scheduling to improve parallelism and resource utilization, along with tailored graph partitioning for GNN workloads.  
% {\sf DistGER} \cite{fang2023distributed} is a distributed system based on information-oriented random walks, with its pipeline extension {\sf DistGER-Pipe} \cite{fang2025information} further optimizes execution. However, their decoupled execution model and heavy data transfers under heterogeneous architecture hinder resource utilization and scalability (\S\ref{sec:DistGPU}).
% {\sf NeutronHeter} \cite{cao2025neutronheter} accelerates GNN training on heterogeneous clusters through hierarchical workload mapping and adaptive communication migration.
% {\sf NeutronTP} \cite{ai2024neutrontp} improves the efficiency of tensor parallelism for GNNs by optimizing communication and memory usage. 
% However, it applies uniform processing across all node features and lacks feedback-driven embedding quality assessment, resulting in redundant computation and slower convergence compared to {\sf FeLoG} (\S\ref{sec:experiments}).  
% To emphasize the distinctive advantages of {\sf FeLoG}, 
Table~\ref{tab:system_comparison} compares representative distributed graph embedding systems across two techniques: random walk-based ({\emph{RW}) and GNN-based embeddings ({\emph{GNN}), along with three system-level dimensions: sampling-training coupling ({\emph{Coupling}), communication optimization ({\emph{Commun.}), and pipeline optimization (\emph{Pipeline}). Existing systems address only a subset, while {\sf FeLoG} unifies all embedding paradigms with comprehensive system-level optimizations. 
% with sampling-training coupling as its unique design foundation.

\begin{table}[t]
\centering
\caption{Comparison of distributed graph embedding systems.}
\label{tab:system_comparison}
\vspace{-1mm}
\footnotesize
\renewcommand{\arraystretch}{1.0}
\begin{adjustbox}{max width=\columnwidth}
\begin{tabular}{lccccc}
\toprule
System & RW & GNN & Coupling & Commun. & Pipeline \\
\midrule
{\sf DistDGL} \cite{DistDGL_2020}             & \xmark & \cmark & \xmark & {\em Partial} & \xmark \\
{\sf PBG} \cite{PBG_2019}                    & \xmark & \xmark & \xmark & \xmark & \xmark \\
{\sf ByteGNN} \cite{ZhengCCSWLCYZ22}         & \xmark & \cmark & \xmark & \cmark & \xmark \\
{\sf DistGER} \cite{fang2023distributed}     & \cmark & \xmark & \xmark & {\em Partial} & {\em Partial} \\
{\sf LeapGNN} \cite{chen2025leapgnn}         & \xmark & \cmark & \xmark & \cmark & \xmark \\
{\sf NeutronHeter} \cite{cao2025neutronheter}& \xmark & \cmark & \xmark & \cmark & {\em Partial} \\
{\sf NeutronTP} \cite{ai2024neutrontp}       & \xmark & \cmark & \xmark & \cmark & \xmark \\
\midrule
{\bf {\sf FeLoG} (ours)}                     & \cmark & \cmark & \cmark & \cmark & \cmark \\
\bottomrule
\end{tabular}
\end{adjustbox}
\vspace{-5mm}
\end{table}

\vspace{-2mm}
\section{Motivations}
\label{sec:motivations}

% \textcolor{red}{[add 1-2 sentence what this section is about and its organization.]}

% We design a scalable and efficient distributed graph-embedding system that couples sampling and training through a feedback loop. To motivate this, we first analyze the limitations of the decoupled sampling-training paradigm and identify system-level bottlenecks that hinder efficiency and scalability in distributed settings.
To motivate our proposed scalable and efficient distributed graph-embedding system, we first analyze the limitations of the decoupled sampling-training paradigm and identify system-level bottlenecks that hinder efficiency and scalability in distributed settings.

\subsection{Limitations of the Sampling-Training Decoupled Paradigm} 
\label{sec:execution_analysis}

Despite efforts to improve sampling quality, existing graph embedding methods still operate under the decoupled sampling-training paradigm, limiting efficiency and scalability on large graphs. For example, {\sf node2vec} \cite{node2vec_2016} uses fixed configurations for biased random walks (e.g., walk length $L=80$, $r=10$ random walks per node), effective for smaller graphs but generating redundant walks for large-scale graphs.
To address this, {\sf HuGE} \cite{HuGE_2021} introduces an information-theoretic random walk mechanism that adaptively determines walk configurations. 
While reducing redundancy, it still adheres to a sequential, decoupled sampling-training design, resulting in slow training times. For instance, processing a billion-edge {\em Twitter} graph \cite{twitter_2010} takes over a week on a modern server.
The most recent system, {\sf DistGER} \cite{fang2023distributed}, extends {\sf HuGE} with incremental information-centric computation in distributed environments, but still evaluates the quality of sampled information solely based on historical statistics, leading to redundancy in sampling. In our observation, {\sf DistGER} spends over 50\% of its total execution time on sampling, with training contributing only 27\%, demonstrating inefficiency in the decoupled execution paradigm.

% The most recently proposed information-oriented distributed graph embedding framework, {\sf DistGER} \cite{fang2023distributed}, extends {\sf HuGE} by introducing incremental information-centric computing in distributed environments. However, it still evaluates the quality of sampled information solely based on historical sampling statistics, without interaction with the training process. As a result, it cannot accurately determine sampling redundancy. Our observation shows that {\sf DistGER} spends over half of its total execution time in the sampling stage, for example, on the large-scale {\em Twitter} graph, sampling accounts for 52\% of the total overhead, while training contributes only 27\%, indicating a clear imbalance due to the decoupled execution paradigm.

\begin{figure}[t!]
 \centering
 \includegraphics[width= 3.2 in]{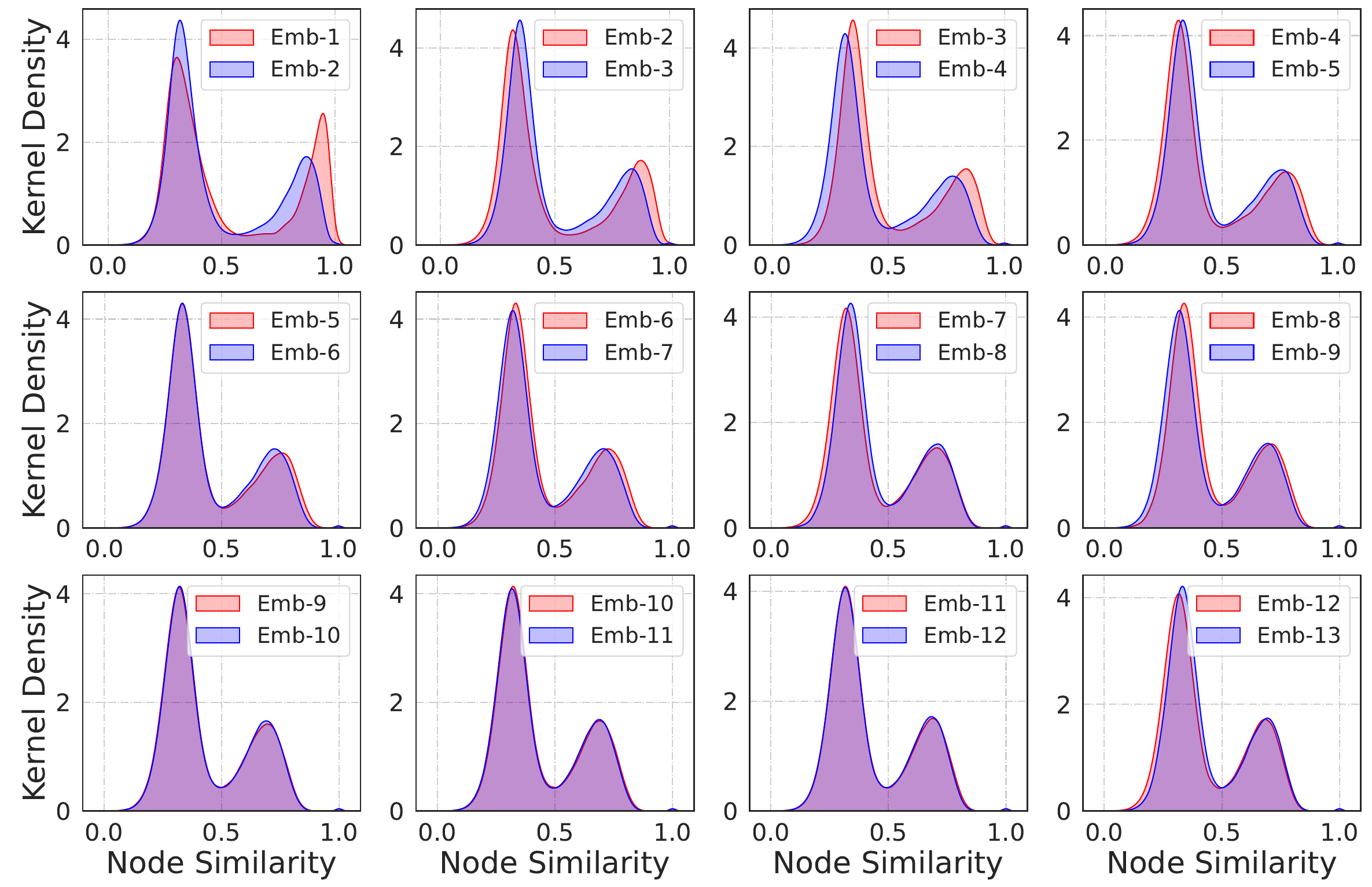}
 \vspace{-3mm}
 \caption{Kernel density distribution of node embedding similarity across different sampling-training rounds on {\em Flickr} graph.
Each subplot compares the similarity distributions of embeddings from round $i$ (red) and round $i+1$ (blue). 
The x-axis represents cosine similarity, the y-axis represents density.
} 
 \label{sampling_training_similarity_analysis}
 \vspace{-8mm}
\end{figure}

We evaluate whether heuristic-driven sampling strategies like {\sf HuGE} and {\sf DistGER} effectively capture informative data without redundancy.
Using the {\em Flickr} graph \cite{Flickr_Youtube_Graph} and following the default configuration of {\sf DistGER}, 
which heuristically determines the $r$ by comparing the distribution of node occurrences in sampled walks with the node degree distribution. In our experiment the sampling procedure terminated after 12 rounds.
After each sampling-training round, we compute the pairwise cosine similarity of node embeddings and use kernel density estimation \cite{worton1989kernel} to obtain the distribution of these similarity values.
Figure~\ref{sampling_training_similarity_analysis} visualizes how these distributions evolve, where each subplot compares consecutive rounds.
We observe that the distributions converge around the 6th round, with minimal changes in subsequent rounds. 
This indicates that the remaining sampling rounds introduce little additional contribution, demonstrating redundancy in both computation and communication.

Current methods make sampling decisions based on structural heuristics or historical statistics, which are indirect proxies for representation quality and do not reveal whether the samples are effectively utilized. A mechanism is needed to identify converged nodes and those that remain undertrained, enabling efficient sampling. Without runtime feedback, samplers continue redundant exploration. This highlights the importance of training feedback for eliminating redundancy and enabling efficient, scalable graph embedding.

\vspace{-2mm}
\subsection{System-Level Bottlenecks in Distributed Graph Embedding}
\label{sec:DistGPU}

To address scalability challenges, existing graph embedding systems have been extended to distributed settings and CPU-GPU heterogeneous architectures. In these systems, sampling is typically executed on CPUs, while training is offloaded to GPUs. However, such designs still suffer from significant system inefficiencies. 
To systematically analyze these bottlenecks, we implemented {\sf DistGER} on an 8-machine cluster with a CPU-GPU heterogeneous architecture.

\begin{figure}
  \centering
  \begin{subfigure}{.48\linewidth}
    \centering
    \includegraphics[width=\linewidth, height=1 in]{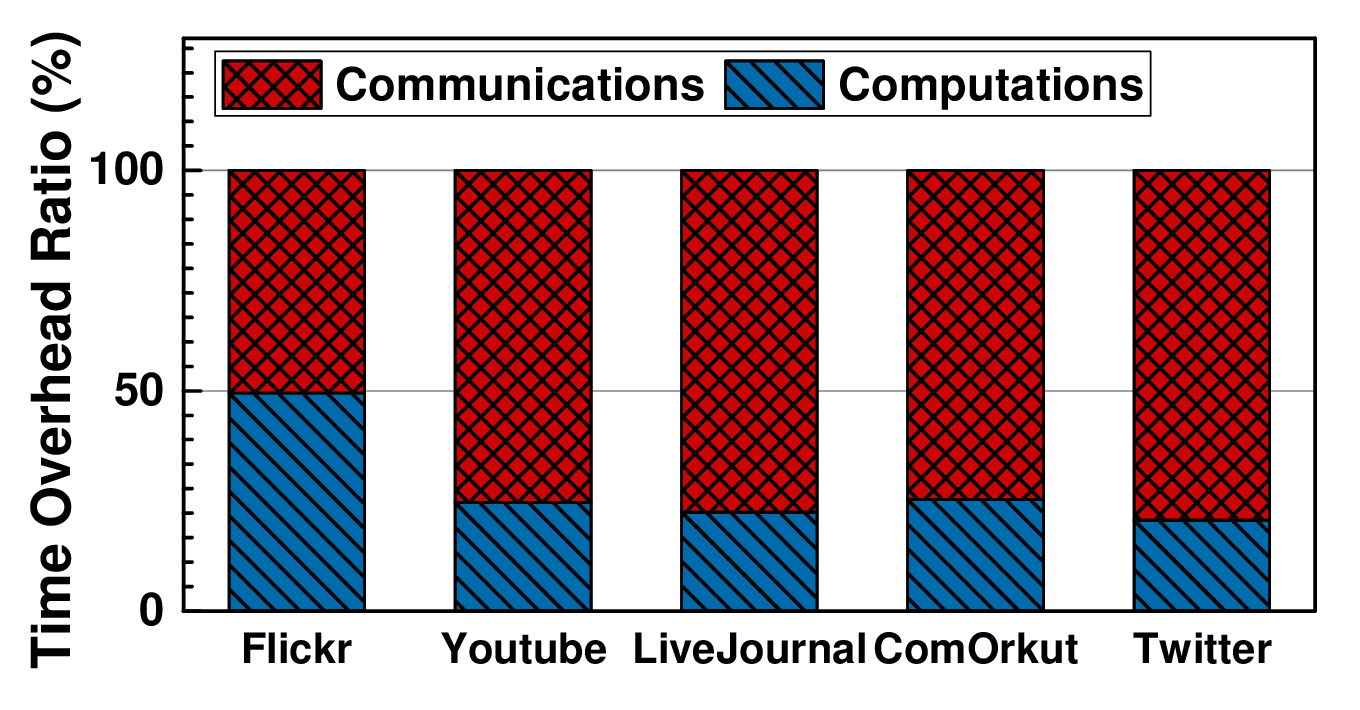}
    \vspace{-5mm}
    \caption{Runtime Breakdown}
    % \label{performance_bottleneck}
  \end{subfigure} 
  \begin{subfigure}{.48\linewidth}
    \centering
    \includegraphics[width=\linewidth, height=1 in]{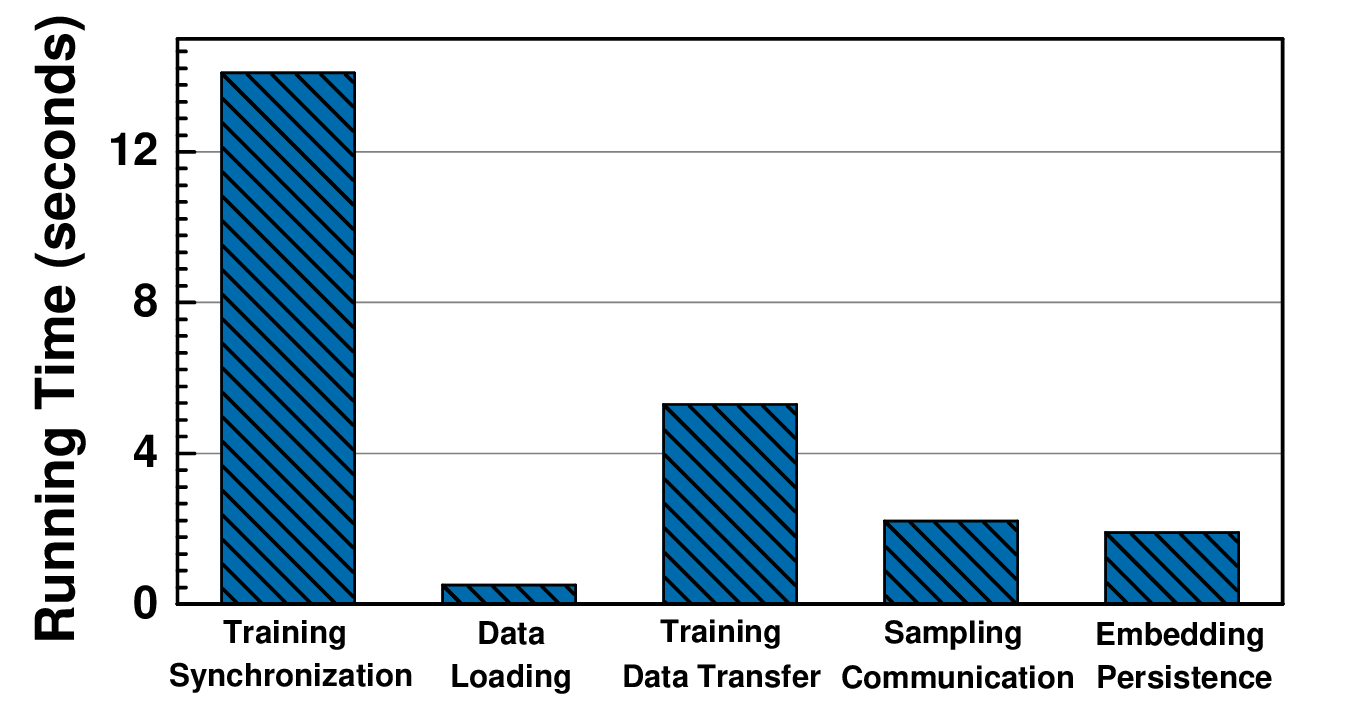}
    \vspace{-5mm}
    \caption{Communication Type}
    % \label{fig:loss-bit-alpha}
  \end{subfigure} 
  \vspace{-3mm}
  \caption{Analysis of communication overheads.}
  \label{performance_bottleneck}
  \vspace{-5mm}
\end{figure}

\spara{Data Communication Overhead.}
We conducted experiments on five real-world datasets and profiled the end-to-end runtime of {\sf DistGER}, decomposing it into computation (sampling and training) and communication (both intra- and inter-machine). As shown in Figure~\ref{performance_bottleneck}(a), communication dominates the total runtime, accounting for an average of 71\%, with its proportion increasing as the dataset size grows.
We further analyzed communication costs using the {\em Flickr} dataset by categorizing the communication into distinct types (Figure~\ref{performance_bottleneck}(b)). Among them, training synchronization and training data transfer emerge as major performance bottlenecks. 

During training synchronization, updated embedding gradients must be synchronized across machines over Ethernet to ensure consistency. The communication volume increases linearly with the number of machines. %Specifically, for $|V|$ nodes, embedding dimension $d$, and $n$ machines, each synchronization round transfers approximately $4nd|V|$ bytes (in single precision). 
For instance, in our 8-machine setup, synchronizing 128-dimensional embeddings on the {\em Twitter} graph using 10 Gbps NICs may introduce up to 4.6 seconds of delay per round,
causing communication to dominate the runtime and accounting for 78\% of the total execution time.
% \textcolor{red}{[AK: how many machines? also clarify why is this a significant delay for the overall embedding learning?]}.
Training data transfer also incurs considerable cost. 
Since sampling is decoupled from training, all sampled data including redundant neighborhoods must be moved from CPU memory to GPU memory and persisted to storage. This overhead grows proportionally with the size of training data and is constrained by PCIe and I/O bandwidth.

% Sampled data must be moved from CPU to GPU memory and persisted to storage, which is constrained by PCIe and I/O bandwidth and grows proportionally with the size of training data.
% (as discussed in \S\ref{sec:execution_analysis} \textcolor{red}{[AK: I thought \S 4.1 is discussing something else, not communication overhead. This is confusing, where is it discussed in \S 4.1?]}) \textcolor{red}{[AK: Is this `communication overhead' concern enhanced due to sampling-training decoupled system -- then it should be clarfied/ demonstrated.]}.

% \begin{figure}[h!]
%  \centering
%  \includegraphics[width=3.2 in]{Figures/DistGER_CPU_GPU_utilization_1.eps}
%  \vspace{-4mm}
%  \caption{Resource utilization on the {\em YouTube} dataset under decoupled sampling and training.}
%  \label{cpu_gpu_util}
% \end{figure}

\spara{Decoupled Execution of Sampling and Training.}
In {\sf DistGER}, sampling and training follow the sampling-training decoupled paradigm and are separated by a round-level barrier: training cannot begin until sampling fully completes and materializes the node sequences for the current round.
Once sampling finishes, the resulting node sequences are passed to the training stage as input.
To ensure fault tolerance, {\sf DistGER} checkpoints the sampled data to storage before launching training, introducing additional I/O overhead.
As a result, sampling and checkpointing phases lie on the critical path; any imbalance or delay in sampling directly stalls GPU execution, creating bubbles in the end-to-end pipeline.
Figure~\ref{cpu_gpu_util_active_nodes}(a) shows that CPU and GPU are significantly underutilized on the {\em YouTube} graph, with average utilizations of only 15.8\% and 25.5\%, respectively.

\begin{figure}
  \centering
  \begin{subfigure}{.32\linewidth}
    \centering
    \includegraphics[width=\linewidth]{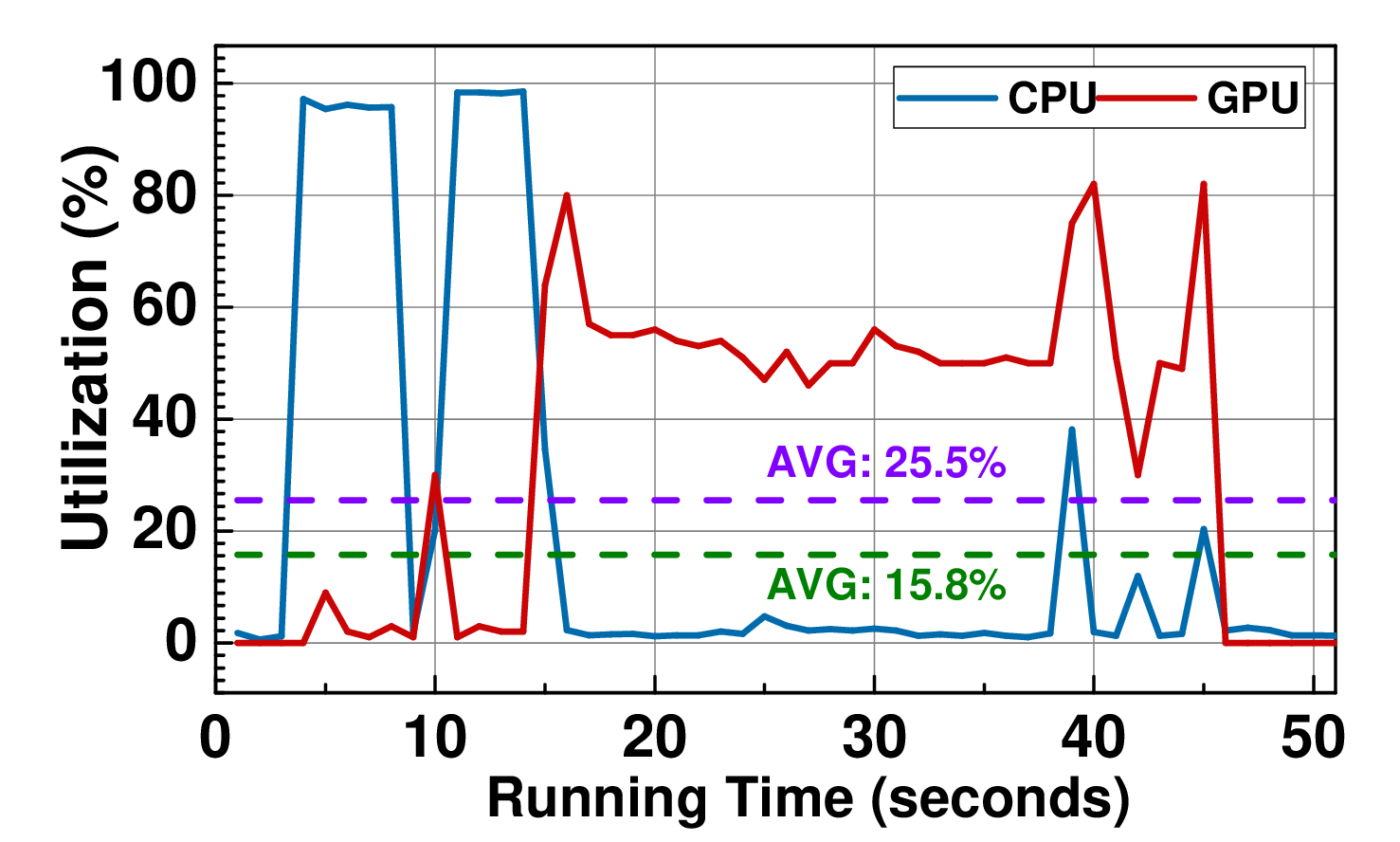}
    \vspace{-4mm}
    \caption{{\footnotesize CPU/GPU Utilization}}
    % \label{performance_bottleneck}
  \end{subfigure} 
  \begin{subfigure}{.32\linewidth}
    \centering
    \includegraphics[width=\linewidth]{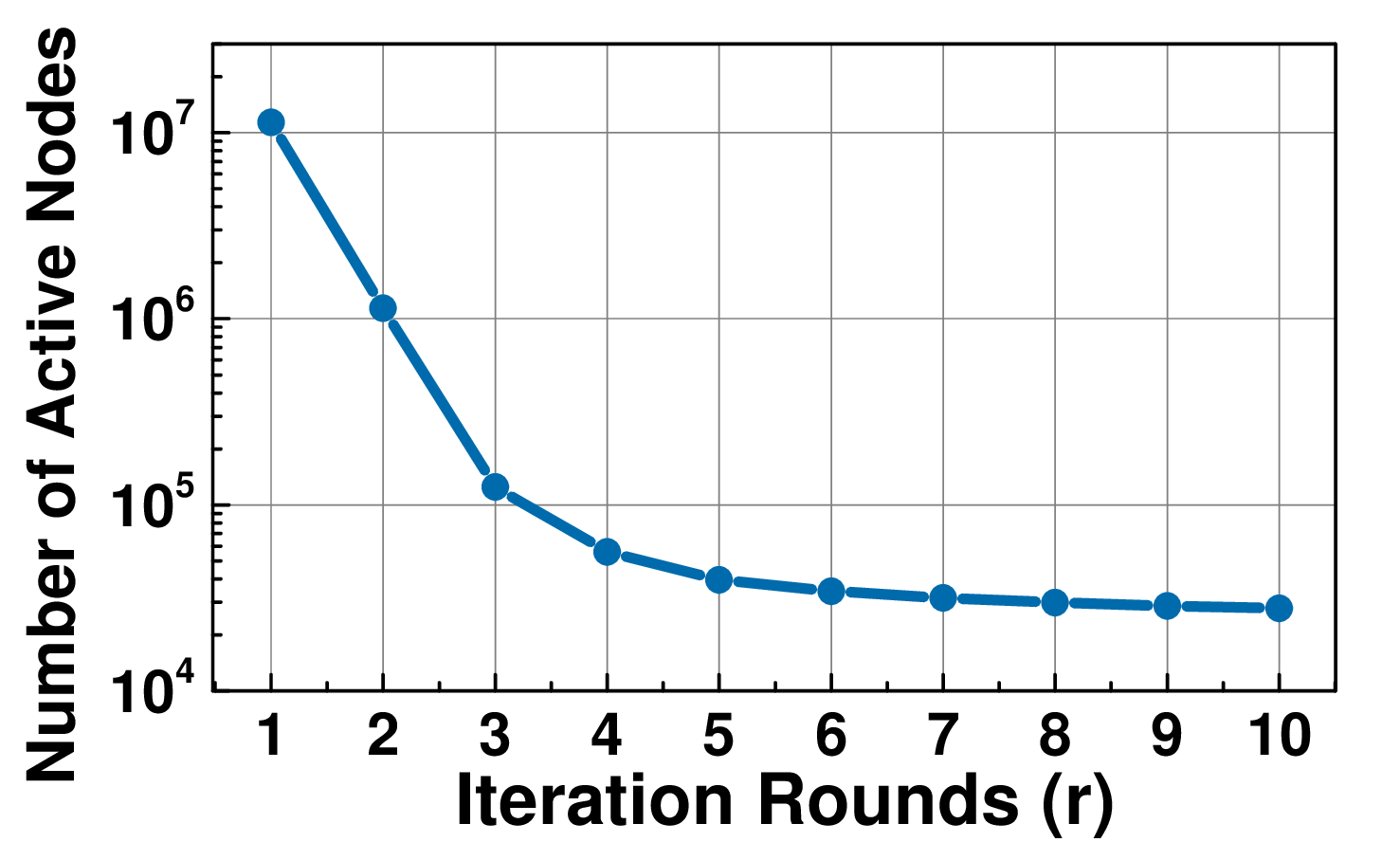}
    % \vspace{-5pt}
    \vspace{-4mm}
    \caption{{\footnotesize Active Nodes}}
    % \label{fig:loss-bit-alpha}
  \end{subfigure}
  \begin{subfigure}{.32\linewidth}
    \centering
    \includegraphics[width=\linewidth]{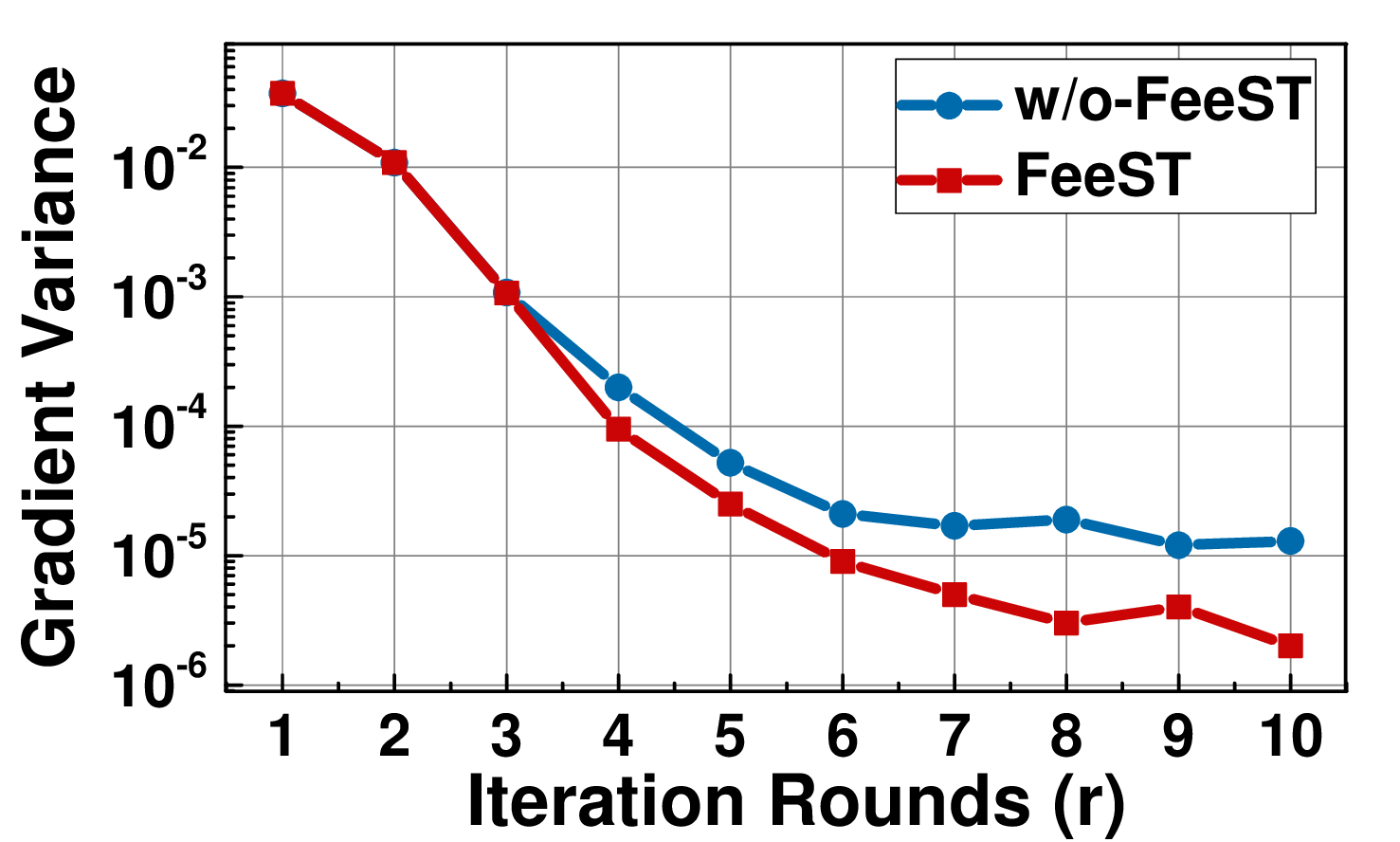}
    \vspace{-4mm}
    \caption{{\footnotesize Gradient Variance}}
    % \label{fig:loss-bit-alpha}
  \end{subfigure}
  % \vskip -5pt
  \vspace{-3mm}
  \caption{{Resource utilization in decoupled sampling-training, and active nodes and gradient variance on {\em YouTube} graph.}}
  \label{cpu_gpu_util_active_nodes}
  \vspace{-2mm}
\end{figure}

\section{ F\lowercase{e}L\lowercase{o}G}
\label{sec:FeLoG}

To address the challenges from sampling-training decoupled paradigm and system-level bottlenecks, 
we design \felog, a scalable and efficient distributed graph embedding system that integrates feedback loop coordination to couple sampling and training.
Figure \ref{FeLoG_framework} summarizes the workflow of {\sf FeLoG}.
We discuss our feedback-coupled sampling-training model ({\sf FeeST}) in \S \ref{sec:FeedST},
and activity-aware communication mechanism ({\sf ACM}) in \S \ref{sec:acm}, including the frequency-aware bitmap compression strategy ({\sf FaBC}) and the hotspot-aware synchronization strategy ({\sf HaSyn}),
while our novel round-interleaved pipeline ({\sf RiPP}) is given in \S \ref{sec:rip}.
Finally, we present how {\sf FeLoG} can be integrated into existing graph embedding frameworks in \S \ref{sec:integration}.

\begin{figure}[h!]
   % \vspace{-4mm}
   \centering
   \includegraphics[width= 3.1 in]{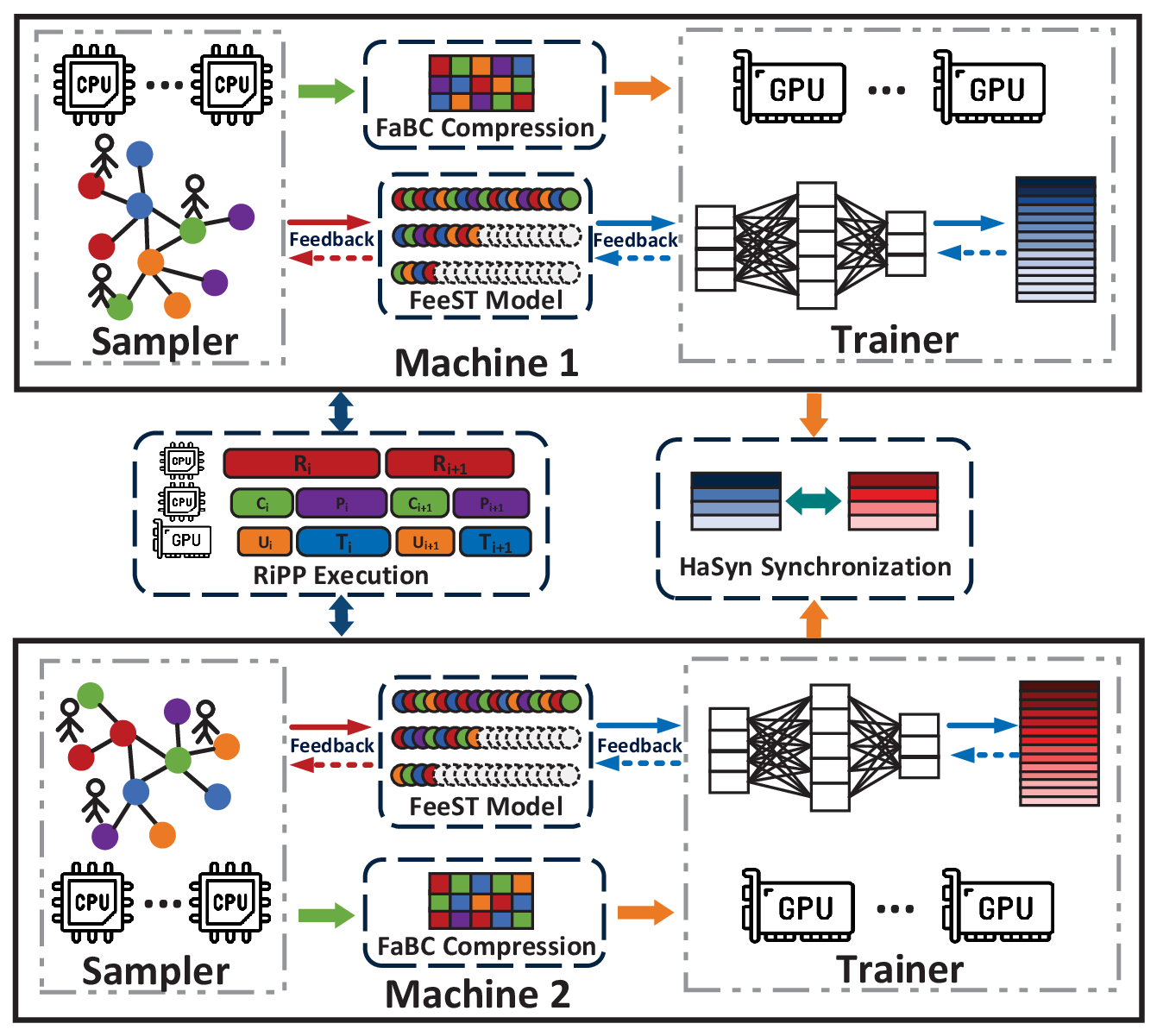}
   \vspace{-2mm}
   \caption{The workflow of {\sf FeLoG}.}
   \vspace{-3mm}
   \label{FeLoG_framework}
% \vspace{-3mm}
 \end{figure}

% \vspace{-2mm}
\subsection {Feedback-coupled Sampling-Training Model}
\label{sec:FeedST}
Traditional graph embedding pipelines treat sampling and training as separate stages, which often leads to redundant sampling and unnecessary computation on large-scale graphs (\S\ref{sec:execution_analysis}). 
Existing fixes \cite{HuGE_2021, fang2023distributed} decide per-node sampling rounds using structural heuristics or historical statistics, yet lack training feedback and may still incur substantial redundancy.
To overcome this limitation, we propose the {\em Feedback-coupled Sampling-Training Model} ({\sf FeeST}), which integrates real-time training feedback into sampling. 
% The key idea is to use embedding convergence as a direct indicator of representation quality, so that only undertrained nodes remain active for subsequent sampling.
The key idea is to derive a neighborhood-consistency metric from training feedback as a direct proxy for representation quality, so that only under-trained nodes remain active for subsequent sampling.

\spara{Overview of {\sf FeeST}.}
As illustrated in Figure~\ref{FeedST_process}, {\sf FeeST} closes the loop between sampling and training via an \emph{active node set}.
At round $r$, {\sf FeeST} samples only from active nodes to generate walks or neighborhood subgraphs as training input.
After training updates embeddings, {\sf FeeST} computes the neighborhood-consistency metric to assess representation quality.
Nodes that meet the consistency criterion are removed from the active set, while under-trained nodes remain active for the next round.
By iteratively shrinking the active set, {\sf FeeST} dynamically allocates sampling effort where it is still beneficial, reducing redundant exploration and speeding up convergence.

\begin{figure}
   \centering
   \includegraphics[width= 3.35 in]{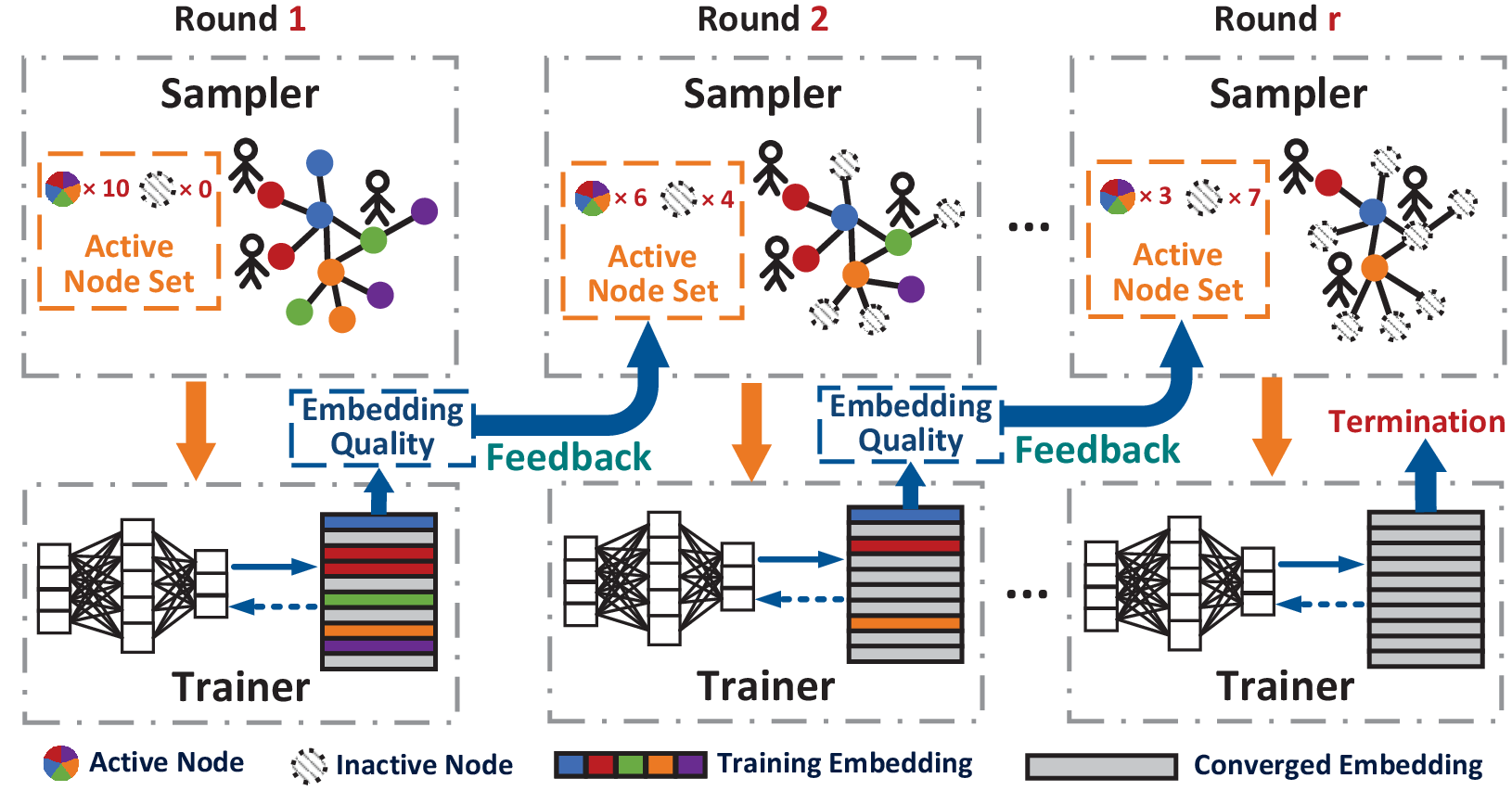}
   \vspace{-4mm}
   \caption{Feedback-driven sampling-training coupling model.}
   \vspace{-2mm}
   \label{FeedST_process}
\vspace{-2mm}
 \end{figure}

\spara{Feedback Mechanism.}
Graph embedding aims to preserve node attributes and graph structure in a low-dimensional space, such that nodes that are proximate or structurally related in the original graph remain similar in the embedding space.
We therefore use embedding similarity with neighbors as a feedback signal to assess representation quality.
Specifically, for a node \(v_i\), if its embedding is sufficiently consistent with those of its neighbors, we treat \(v_i\) as well-trained and exclude it from future sampling rounds.
We quantify this neighborhood-consistency as the average cosine similarity between \(v_i\) and its neighbors.
Let \( N(v_i) \) represent the set of neighbors of node \( v_i \), and \( \mu \) be a predefined similarity parameter. 
To mitigate the computational overhead of computing similarities for pair-wise nodes, {\sf FeeST} adopts a neighbor subsampling strategy. 
Instead of evaluating similarities across the entire neighborhood, we estimate the neighborhood-consistency score \( \Psi(v_i) \) using a sampled subset \( \tilde{N}(v_i) \subseteq N(v_i) \):  
\vspace{-1.2mm}
\begin{equation}
\small
\Psi(v_i) \approx \frac{1}{|\tilde{N}(v_i)|} \sum_{v_j \in \tilde{N}(v_i)} 
\frac{\vec{v_i} \cdot \vec{v_j}}{\|\vec{v_i}\| \|\vec{v_j}\|}
\label{quality_identify}
\end{equation}
% \vspace{-1.2mm}
This approximation is effective because real-world graphs often exhibit redundant local neighborhoods.
In many graphs, nodes with similar attributes or roles tend to connect, making neighboring nodes highly correlated \cite{mcpherson2001birds, kipf2017semi}; meanwhile, power-law degree distributions create hubs with many overlapping neighbors \cite{BA99}.
Empirical studies on GNNs also suggest that aggregating a small subset of neighbors is often sufficient, while using all neighbors yields diminishing returns but incurs extra overhead \cite{GraphSAGE_2017, chen2018fastgcn}.

An alternative is to detect convergence via self-stability, i.e., comparing a node's embedding across consecutive rounds.
However, this requires storing historical embeddings, incurring additional $O(|V|\!\cdot\! d)$ memory overhead at scale.
More importantly, graph embedding aims to preserve structural and attribute relationships with respect to the original graph.
By evaluating consistency with neighbors (Eq.~\ref{quality_identify}), {\sf FeeST} aligns the feedback signal with this objective.
In contrast, self-stability alone can yield false convergence: an embedding may change little across rounds yet remain inconsistent with its neighborhood, providing misleading guidance to the sampler.
Finally, training-parameter signals such as gradient magnitudes are not reliable indicators of structural consistency, as they mainly capture within-iteration optimization dynamics and minibatch stochasticity \cite{ziyin2022strength,spidercache_2025}, and lack a cross-round structural view for updating the sampling frontier (i.e., the active set).

If \( \Psi(v_i) < \mu \), the embedding of \( v_i \) is considered undertrained, and the node will be included in the active node set for the next round of sampling \( V_a^{(r+1)} \). 
Otherwise, it is excluded from further sampling. 
The active node set for round \( r+1 \) is formally defined as:
\vspace{-1mm}
\begin{equation}
\small
V_a^{(r+1)} = \left\{ v_i \in V_a^{(r)} \,\middle|\, \Psi(v_i) < \mu \right\}
\end{equation}
% \vspace{-0.7mm}
Empirically, we set the default \(\mu\) to 0.65. With this feedback mechanism, {\sf FeeST} treats active nodes as a round-level operational state rather than a static graph property or homophily-based label category: after each training round, a node remains active if \(\Psi(v_i)<\mu\), indicating that its representation may still benefit from further sampling and training. 
This helps {\sf FeeST} identify insufficiently stable embeddings, avoiding redundant sampling and computation without sacrificing downstream accuracy (\S \ref{sec:experiments}).
While Eq.~\ref{quality_identify} is most directly aligned with graphs where local similarity is informative, its scale becomes more dataset-dependent in heterophilic graphs because neighbors may not be semantically similar. 
Therefore, in heterophilic settings, we treat \(\mu\) as a validation-calibrated decision boundary rather than a universally fixed threshold, optionally guided by structural homophily metrics~\cite{platonov2023characterizing, gong2024survey}. 
Specifically, we calibrate \(\mu\) over a small candidate set using only the held-out validation split under the standard \(50\%/25\%/25\%\) split, select the best validation quality-efficiency trade-off, and use the test set only for final reporting.

As shown in Figure~\ref{cpu_gpu_util_active_nodes}(b), the number of active nodes decreases rapidly during the early rounds on the {\em YouTube} graph.
This observation aligns with the power-law distribution commonly found in real-world graphs, where most nodes have a low degree. These low-degree nodes become sufficiently trained quickly due to sparse connectivity and limited information content, while high-degree nodes require more rounds of sampling to capture their neighborhood context.
This validates that {\sf FeeST} eliminates redundant exploration from the early rounds, unlike decoupled paradigms where all nodes continue to be sampled regardless of representation quality. Importantly, the active frontier is recomputed every round, enabling nodes to be reactivated when they become inconsistent with their neighborhoods. 
Inactive nodes may still be revisited through random walks or neighbor aggregation, allowing their embeddings to keep evolving.
On the {\em Com-Orkut} graph, we observe that 6.4\% of nodes are reactivated per round on average. Moreover, Figure~\ref{cpu_gpu_util_active_nodes}(c) shows that although {\sf FeeST} starts pruning from the first round and rapidly shrinks the active frontier, its gradient variance remains comparable to {\sf w/o-FeeST} in the first few rounds and becomes consistently lower afterward, suggesting that the induced non-stationary sampling does not introduce harmful optimization instability in practice.

\spara{Implementation.}  
Algorithm~\ref{feedback_sampling} presents the workflow of {\sf FeeST}, which iteratively performs sampling, training, and feedback until the active node set becomes empty. 
To ensure generalizability, {\sf FeeST} abstracts both sampling and training through model-agnostic interfaces. 
The $\texttt{Sample}(v_i, \theta) \rightarrow W_{v_i}^L$ function generates either random-walk sequences (e.g., {\sf DeepWalk}, {\sf node2vec}, {\sf HuGE}) or GNN-based neighborhood subgraphs with configurable fan-out, where \(v_i\) is the source node, \(\theta\) denotes configurable parameters such as walk bias or neighbor budget, and \(W_{v_i}^L\) is the resulting walk or subgraph. 
The $\texttt{Train}(W_{v_i}^L, model) \rightarrow \vec{V}$ function then consumes these samples to update embeddings using the corresponding model, such as Skip-Gram with negative sampling for random walks, or message passing and aggregation for GNNs. 
This unified abstraction allows {\sf FeeST} to plug feedback into different embedding pipelines (details in \S\ref{sec:integration}).

\spara{Complexity Analysis.}
The computational complexity can be interpreted directly from the following phases:  
\textbf{Sampling (line 3).} 
Let $V_a^r$ denote the active node set in round $r$, for each active node $v_i \in V_a^r$, the \texttt{Sample} function generates a walk or subgraph of average length $L$. The cost is $O(|V_a^r|\cdot L)$.  
\textbf{Training (line 4).} Embeddings are updated on the collected samples. With walk length $L$, window size $w$, and embedding dimension $d$, the training cost is $O(|V_a^r|\cdot L \cdot w \cdot d)$.  
\textbf{Feedback (lines 5-6).}
For each active node, {\sf FeeST} computes the neighborhood-consistency score \(\Psi(v_i)\) using a subsampled neighborhood \(\tilde{N}(v_i)\) of size \(\bar{k}_s\), costing \(O(|V_a^r|\cdot \bar{k}_s \cdot d)\).
Since \(|V_a^r|\) shrinks across rounds as sufficiently trained nodes are removed, the total runtime is
$O\!\left(\sum_{r=1}^{R} |V_a^r|\cdot(L \cdot w \cdot d + \bar{k}_s \cdot d)\right)$,
which can be substantially smaller than the decoupled baseline that repeatedly samples and trains over all nodes.
The space complexity is dominated by embeddings $O(|V|\cdot d)$ at the current round, with temporary memory $O(\max_r |V_a^r|\cdot(L+\bar{k}_s))$ for sampled walks and neighbor subsets.  
% The space complexity is dominated by embeddings \(O(|V|\cdot d)\), plus temporary storage for the current round’s samples and neighbor subsets, i.e., \(O(\max_r |V_a^r|\cdot(rL+\bar{k}_s))\).

\begin{algorithm}[t!]
\caption{Feedback-driven Sampling-Training Coupling}
\label{feedback_sampling}
\small
\begin{algorithmic}[1]
\Require Graph $G=(V,E)$, embedding dimension $d$, parameter $\mu$
\Ensure Node embeddings $\vec{V}$
\State Initialize active node set $V_a \gets V$
\While{$V_a \neq \emptyset$}
    \State \textbf{Sampling:} For $v_i \in V_a$, generate samples $W_{v_i}^L \gets \texttt{Sample}(v_i, \theta)$
    \State \textbf{Training:} Update embeddings $\vec{V} \gets \texttt{Train}(W_{v_i}^L, Model)$
    \State \textbf{Feedback:} For $v_i \in V_a$, evaluate embedding quality $\Psi(v_i)$
    \State Form next active set $V_a^{\text{next}} = \{v_i \in V_a \mid \Psi(v_i) < \mu \}$
    \State $V_a \gets V_a^{\text{next}}$
\EndWhile
\State \Return $\vec{V}$
\end{algorithmic}
\end{algorithm}
% \vspace{-2mm}

\vspace{-2mm}
\subsection{Activity-aware Communication Mechanism}
\label{sec:acm}
In large-scale graph embedding systems, efficient communication is essential to overcome the bottlenecks introduced by both intra-machine sampling-training transmission and inter-machine embedding synchronization. As observed in \S~\ref{sec:DistGPU}, communication overhead, particularly between the CPU and GPU, as well as across distributed machines, comprises a significant portion of the overall execution time (up to 78\%), leading to low resource utilization and scalability issues. To address these challenges, we propose an {\em activity-aware communication mechanism} ({\sf ACM}), which encompasses a {\em frequency-aware bitmap compression strategy} ({\sf FaBC}) to alleviate PCIe bandwidth pressure and a {\em hotspot-aware synchronization strategy} ({\sf HaSyn}) to reduce network traffic.

\subsubsection{Analysis of Data Flow in FeeST}
\label{dataflow_analysis_FeedST}

\begin{figure}
 \centering
 \includegraphics[width=3in]{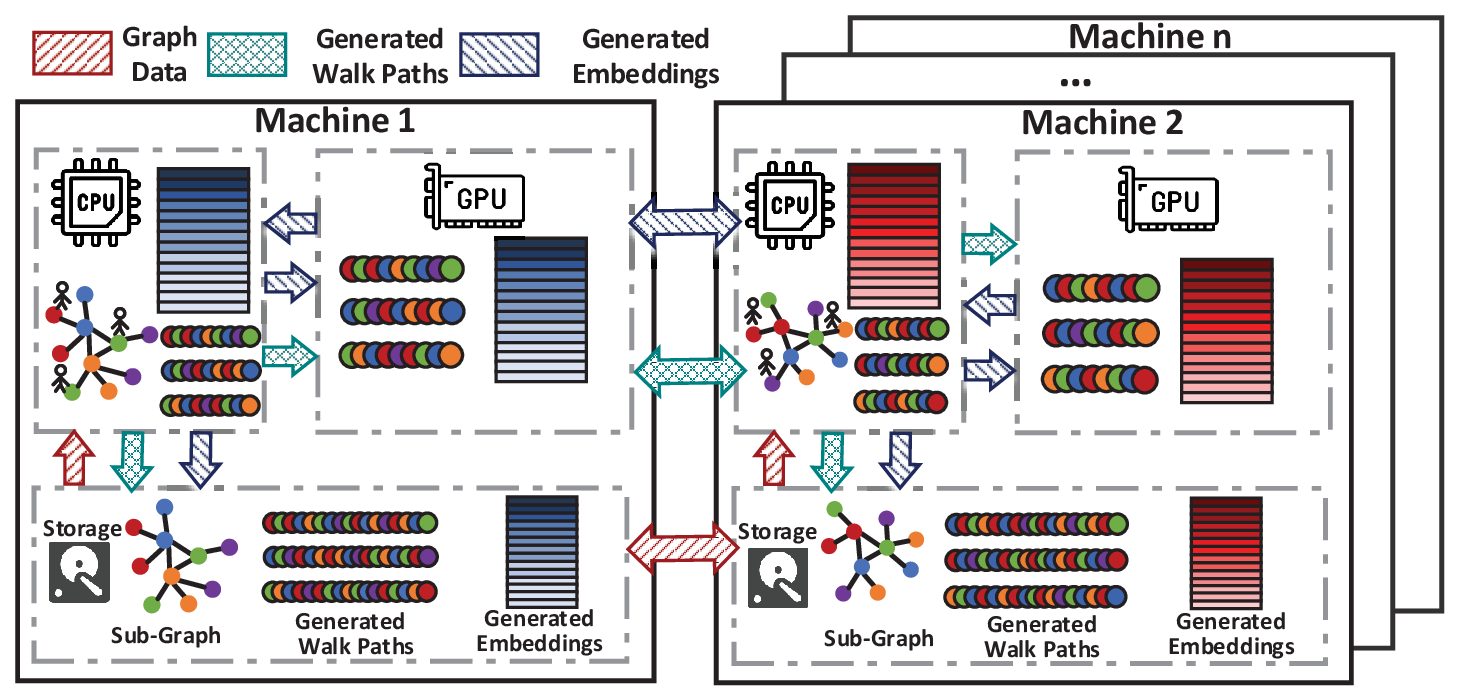}
 \vspace{-2mm}
 \caption{Data flow in the feedback-coupled sampling-training.}
 \vspace{-4mm}
 \label{figure_dataflow_FeedST}
\end{figure}

As shown in Figure~\ref{figure_dataflow_FeedST}, the communication process can be summarized in the following stages:  
(1) \textbf{\textit{Graph data loading and distribution:}} Raw graph data is read from storage into CPU memory and partitioned across worker nodes (red arrows).  
(2) \textbf{\textit{Sampling and path generation:}} Each worker performs random walks on its local subgraph, and path fragments are exchanged if walks cross partitions. The generated walk paths are then aggregated for training (green arrows).  
(3) \textbf{\textit{CPU-GPU transfer for training:}} Aggregated sequences are transmitted from CPU to GPU memory as training input. This PCIe transfer grows with the volume of samples and often emerges as a major bottleneck (green arrows).  
(4) \textbf{\textit{Distributed synchronization:}} During training, updated gradients or embeddings are copied from GPU to CPU, aggregated across machines, and written back to GPUs for parameter updates. This synchronization dominates communication costs in distributed environments (blue arrows).  
(5) \textbf{\textit{Persistence:}} After convergence, final embeddings are transferred back to CPU memory and persisted to storage (blue arrows). In addition, generated walk paths are periodically checkpointed for reproducibility and recovery purposes (green arrows).  
Among these stages, (3) CPU--GPU transfer and (4) distributed synchronization are the two dominant bottlenecks during communication, which motivate the design of {\sf FaBC} and {\sf HaSyn}.

\subsubsection{Frequency-aware Bitmap Compression Strategy}
\label{sec:FaBC}

% In graph embedding systems, random walks generate node sequences or subgraphs as training input. 
% Due to the skewed degree distribution in real-world graphs, high-degree nodes tend to be revisited more frequently, and walks starting from such nodes are likely to acquire richer structural information by traversing their local neighborhoods \cite{HuGE+_2022, common_neighbor_aware_icde_2019}.
% Existing strategies (e.g., biased transitions in {\sf node2vec} or backtracking in {\sf HuGE}) exacerbate this issue by favoring repeated exploration of local neighborhoods. 
% As a result, sampled paths often contain many duplicate node occurrences, which must be transferred from CPU to GPU for training and thus impose significant PCIe bandwidth overhead. 

% In current systems, walk paths are stored as arrays of node IDs, a naive representation that ignores redundancy.
% This not only inflates memory consumption but also aggravates PCIe transfer bottlenecks, since repeated nodes are transmitted multiple times without compression. 
% Given that redundancy primarily arises from frequent nodes, encoding their positions with compact bitmap structures \cite{chan1998bitmap} is far more efficient than repeatedly storing node identifiers. 

In graph embedding systems, sampling generates node sequences or subgraphs as training input.
Due to the skewed degree distribution in real-world graphs, high-degree nodes tend to be revisited more frequently, and random walks are more likely to repeatedly traverse their local neighborhoods \cite{HuGE+_2022, common_neighbor_aware_icde_2019}.
Existing strategies (e.g., biased transitions in {\sf node2vec} or backtracking in {\sf HuGE}) exacerbate this issue by favoring repeated exploration of local neighborhoods. 
As a result, sampled paths or subgraphs contain many duplicate node occurrences that must be transferred from CPU to GPU for training, imposing substantial PCIe bandwidth overhead.
However, existing systems typically materialize samples as plain arrays of node IDs, which fails to exploit this repetition.
Since redundancy is dominated by a small set of frequent nodes, encoding their occurrence patterns with compact bitmap representations \cite{chan1998bitmap} is often far more efficient than repeatedly sending the same identifiers.

% We propose the {\em Frequency-aware Bitmap Compression} ({\sf FaBC}) strategy, which identifies high-frequency nodes and records their positions as bitmaps, thereby transforming repeated identities into lightweight positional markers. 
% While finer-grained frequency ranges may yield higher compression ratios, they also introduce extra bitmaps and computation, offsetting the benefits.
% Hence, {\sf FaBC} strikes a balance by selecting an appropriate frequency range that maximizes compression efficiency without incurring prohibitive overhead.  

We propose the {\em Frequency-aware Bitmap Compression} ({\sf FaBC}) strategy, which identifies high-frequency nodes and records their positions as compact bitmaps, turning repeated identities into lightweight markers for CPU-GPU transfer.
Unlike prior bitmap techniques that mainly target static bitvectors or graph adjacencies, {\sf FaBC} is workload-driven and tailored to the highly repetitive access patterns in sampling-generated walk paths, and is co-designed with {\sf FeLoG}'s workflow to reduce the dominant transfer bottleneck.
While finer-grained frequency buckets can improve compression, they also introduce extra bitmap construction and decoding overhead. {\sf FaBC} therefore selects a small number of buckets (or ranges) based on the observed node-frequency distribution to balance compression effectiveness and runtime cost.

\begin{figure}
  \centering
  \begin{subfigure}{.48\linewidth}
    \centering
    \includegraphics[width=1.12\linewidth]{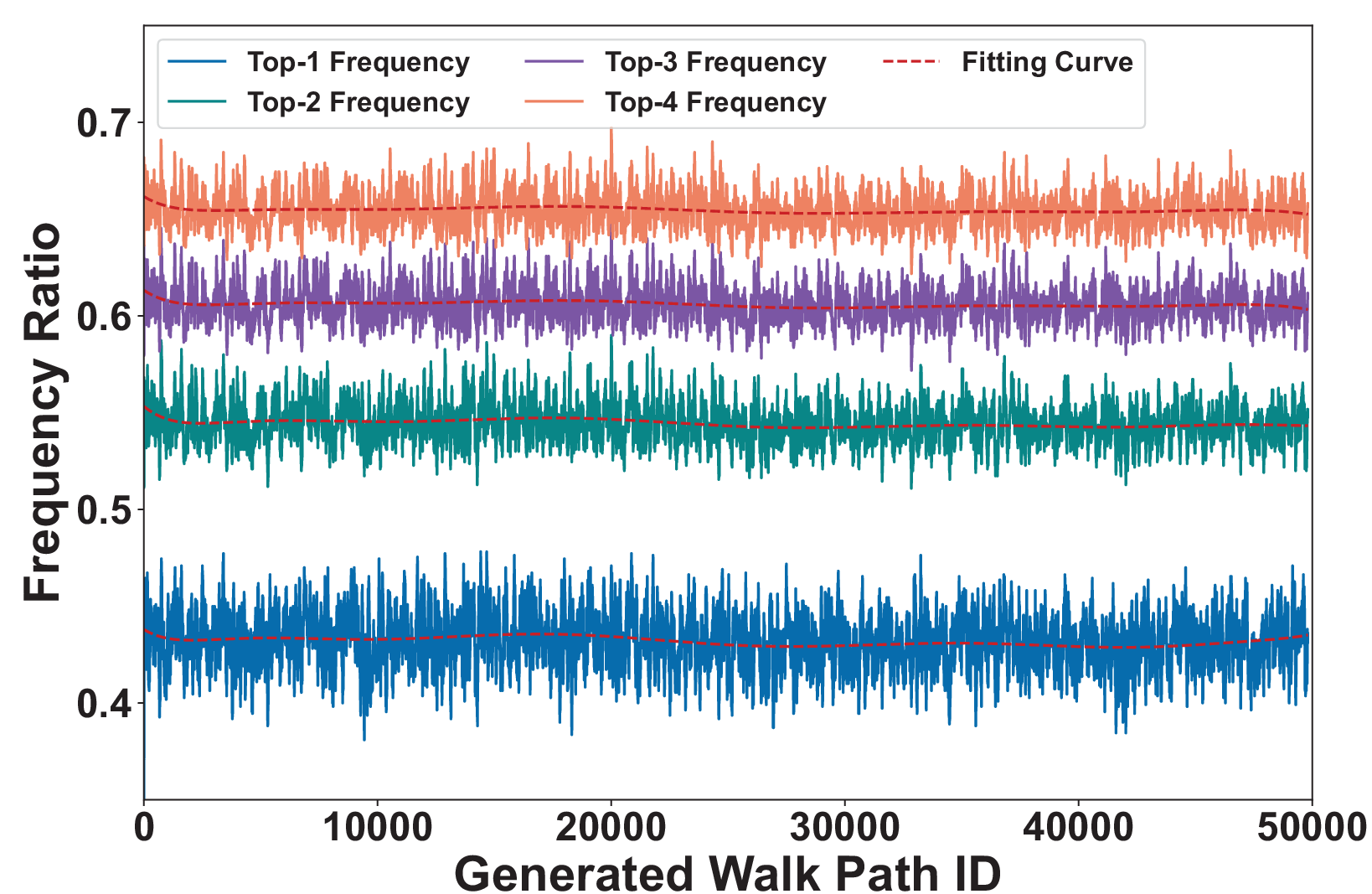}
    \vspace{-4mm}
    \caption{\small Node Occurrences}
    % \label{performance_bottleneck}
  \end{subfigure} 
  \begin{subfigure}{.48\linewidth}
    \centering
    \includegraphics[width=1.12\linewidth]{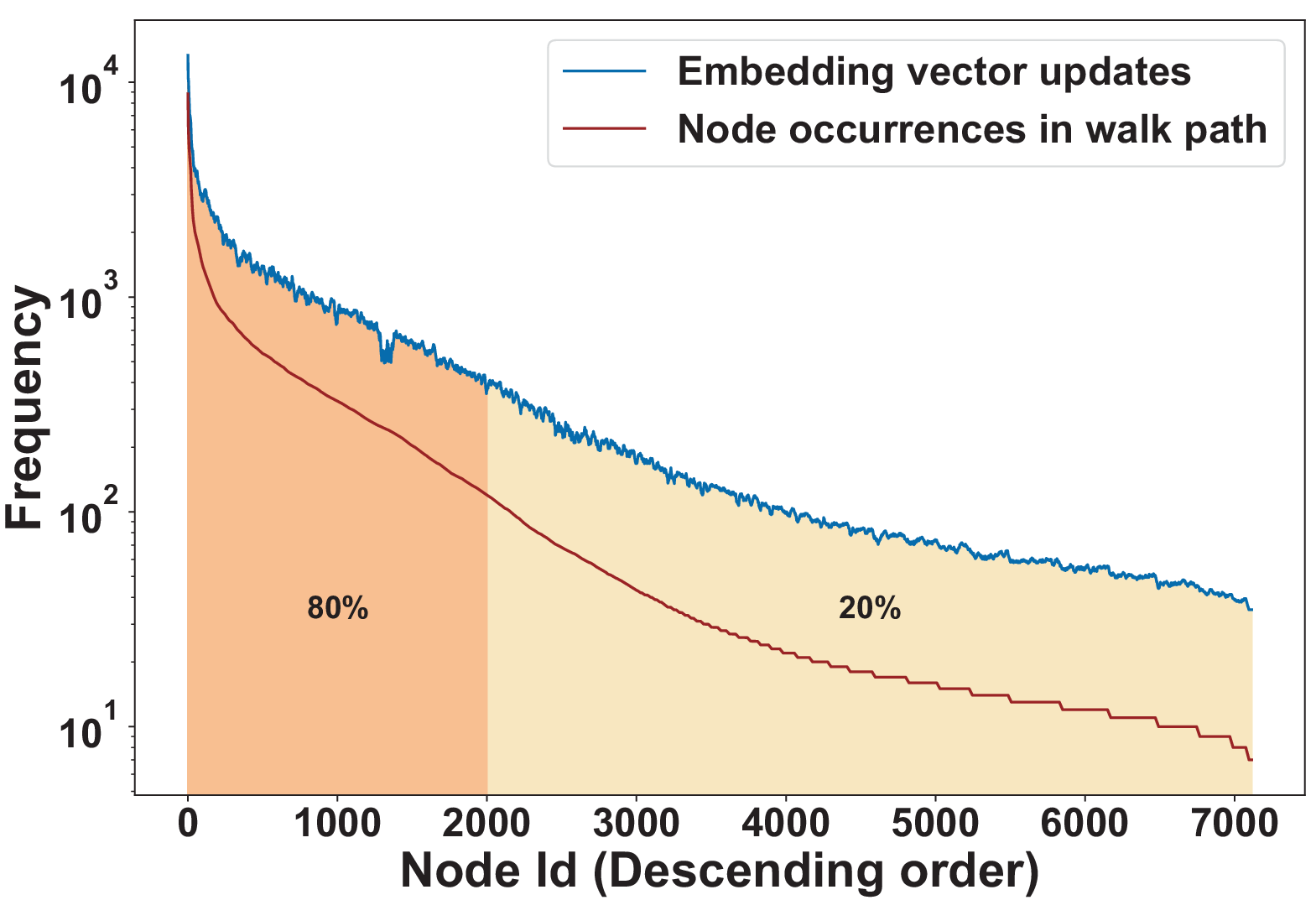}
    \vspace{-4mm}
    \caption{{\small Embedding Updates}}
    % \label{fig:loss-bit-alpha}
  \end{subfigure} 
  \vspace{-3mm}
  \caption{{Distribution characteristics of node occurrences and embedding updates.}}
  \label{node_freq_emb_distri}
  \vspace{-6mm}
\end{figure}

% To this end, {\sf FaBC} adaptively determines compression ranges based on the node frequency distribution observed in the first round of sampling-training. 
% Formally, after the initial sampling, we collect a representative set of random walk sequences and perform frequency analysis on all visited nodes. 
% Let the nodes be sorted in descending order of their frequencies as \( v_{(1)}, v_{(2)}, \dots \). 
% We define the cumulative frequency of the top-\(k\) nodes as $F_k = \sum_{i=1}^{k} v_{(i)}$.
% The accumulation is terminated when the marginal gain of the \((k+1)\)-th node satisfies
% $\frac{F_{k+1} - F_k}{F_k} \leq 0.1$.
% The resulting top-\(k\) frequent nodes are then selected as compression targets and encoded with bitmap structures in all subsequent training data. 
% Since random walks across different rounds share the same parameters, their node frequency distributions remain stable, allowing the compression configuration to be reused without repeated analysis.

To this end, {\sf FaBC} determines the compression range by profiling the node-frequency distribution in the first sampling--training round.
After the initial sampling, we compute normalized node frequencies on the sampled sequences and sort them as \(p(1)\ge p(2)\ge \cdots\).
We define the cumulative frequency of the top-\(k\) nodes as \(F_k=\sum_{i=1}^{k} p(i)\).
Under {\sf FaBC}, selecting the top-\(k\) nodes compresses their repeated occurrences into per-sequence bitmaps and stores the remaining (infrequent) nodes in the original ID form.
For \(N\) sequences with average length \(L\), the baseline storage is \(S_0=4NL\) bytes (4 bytes per node ID).
After compressing the top-\(k\) nodes, the uncompressed tail accounts for a \((1-F_k)\) fraction of positions and costs \(4NL(1-F_k)\).
We store the \(k\) selected node IDs once per sequence, costing \(4Nk\), and store one length-\(L\) bitmap per selected node per sequence, costing \(Nk\cdot (L/8)\) bytes.
Thus, the compressed size is
\begin{equation}
S(k)=4NL(1-F_k)+4Nk+Nk\frac{L}{8} 
\end{equation}
and the compression ratio is
\begin{equation}
g(k)=\frac{S(k)}{S_0}=1-F_k+\frac{k}{L}+\frac{k}{32}
\label{eq:fabC_ratio_full}
\end{equation}
Here, \(1-F_k\) is the fraction of positions that remain uncompressed, \(\frac{k}{L}\) corresponds to storing the \(k\) selected IDs once per sequence, and \(\frac{k}{32}\) is the bitmap overhead (1 bit per position).

{\sf FaBC} expands the compression target set only when the \((k{+}1)\)-th node provides sufficient marginal benefit.
Let \(c=\frac{1}{L}+\frac{1}{32}\) denote the amortized overhead of adding one more compressed node, so \(g(k)=1-F_k+kc\).
Since \(F_{k+1}=F_k+p(k{+}1)\), we have
$g(k{+}1)=g(k)-p(k{+}1)+c$.
To avoid diminishing returns, {\sf FaBC} selects the largest \(k\) satisfying
\begin{equation}
\frac{g(k)-g(k{+}1)}{g(k)} \ge 0.1
\quad \Longleftrightarrow \quad
p(k{+}1)\ge c+0.1g(k)
\label{eq:fabC_stop_full}
\end{equation}
which requires the marginal reduction to be at least \(10\%\) of the current ratio.
The selected top-\(k\) nodes are then used as compression targets, and their occurrences are encoded with bitmaps in subsequent samples.
In practice, the frequency profile is largely stable under fixed sampling parameters, so the configuration can be reused without repeated analysis; if needed, {\sf FaBC} can refresh the profile periodically with low overhead.

\textit{\textbf{Example.}}  
Figure~\ref{node_freq_emb_distri}(a) illustrates the node frequency curve of walks generated by {\sf FeeST} on the {\em YouTube} graph. 
The top-1 and top-2 nodes together contribute over 50\% of the total frequency. 
Including the third most frequent node increases the cumulative frequency by less than 10\%, and adding the fourth brings only marginal improvement. 
Hence, according to the above criterion, {\sf FaBC} selects the top-1 and top-2 nodes as compression targets. 
This demonstrates how the adaptive rule captures dominant redundancy without extra bitmaps. 
The impact of different frequency ranges on compression ratio is further evaluated in Table \ref{compression_ratio}. 

\begin{figure}[!t]
  \centering
  \includegraphics[width=3.35in]{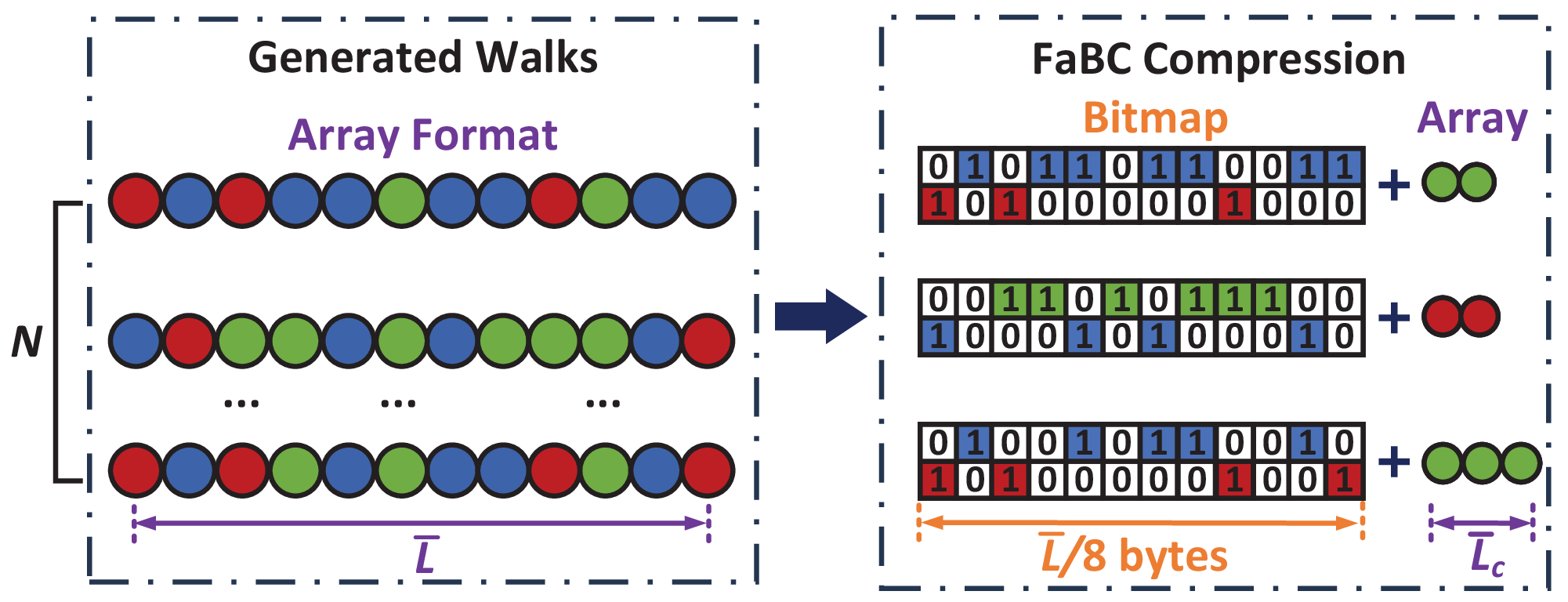}
  \vspace{-4mm}
  \caption{Frequency-aware compression process.}
  \vspace{-2mm}
  \label{compression_diagram}
\end{figure}

As illustrated in Figure~\ref{compression_diagram}, {\sf FaBC} achieves notable compression. 
In a set of \( N \) random walks with average length \( \overline L \), storing nodes as 4-byte unsigned integers results in a raw size of \( 4N\overline L \) bytes. 
After compression, the positions of the top-\(k\) frequent nodes are stored using \(k\) bitmaps of length \(N\), while the remaining nodes are stored as an ID stream.
Let \( \overline{L_c} \) denote the average length of the remaining sequence. 
From Figure~\ref{node_freq_emb_distri}(a), when the top-2 nodes cover approximately 60\% of the walk content, we have \( \overline{L_c} \approx 0.4\overline L \). 
Therefore, the compressed data size is approximately
$\frac{k\cdot N\overline{L}}{8} + 4N\overline{L_c} \approx 0.25N\overline L \ + 1.6N\overline L \ \text{bytes}$,
which significantly reduces the training data size while incurring minimal computational overhead (details in \ref{sec:individual}).

\vspace{-1mm}
\subsubsection{Hotspot-aware Synchronization Strategy}
\label{sec:HaSyn}

Synchronization of embedding vectors is a critical factor influencing both accuracy and system performance in distributed training.
Existing methods such as parameter-server based designs (e.g., {\sf PBG} \cite{PBG_2019} and {\sf NuPS} \cite{NuPS_2022}) and optimized communication topologies (e.g., {\sf TidalMesh} \cite{lim2025tidalmesh}) improve efficiency, but they typically treat all embeddings equally, ignoring the skewed update activity across nodes.
In real-world graphs, the sampling workload is highly skewed, and a small set of high-frequency nodes appears far more often in sampled sequences, leading to disproportionately more embedding updates.
As shown in Figure~\ref{node_freq_emb_distri}(b), more than 80\% of updates on the {\em Wiki-Vote} graph \cite{leskovec2010predicting} are concentrated on a small set of high-frequency nodes, while the majority of embeddings are rarely updated.
This makes uniform full-model synchronization inefficient, as most bandwidth is wasted transmitting nearly static vectors, motivating hotness-aware selective synchronization that prioritizes frequently updated embeddings.

% To address this issue, we propose the \textit{Hotspot-aware Synchronization Strategy} ({\sf HaSyn}), which dynamically adjusts the synchronization frequency of embedding vectors based on their update hotness.
% As illustrated in Figure~\ref{fig:hasyn_strategy}, during the initialization phase, the embedding matrix is constructed by sorting nodes in descending order of their frequencies observed in the training data.
% During synchronization, we partition \(\vec{V}\) into hotspot blocks \(B_f=\{v \mid f(v)=f\}\) indexed by distinct frequency values \(f\in\mathcal{F}\), yielding \(|\mathcal{F}|\) blocks in total.
% One node is randomly selected from each block \(B_f\), forming a synchronization set of size \(|\mathcal{F}|\), and the corresponding embeddings are synchronized across all machines in the current synchronization cycle.
% Due to the skewed frequency distribution, a node in block \(B_f\) is selected with probability \(1/|B_f|\), making hot embeddings (typically in smaller blocks) more likely to be synchronized, while cold embeddings are synchronized less frequently.
% Compared to full-model synchronization, the communication cost of {\sf HaSyn} is reduced to \(\mathcal{O}(|\mathcal{F}|\cdot d \cdot M)\), where \(d\) is the embedding dimension and \(M\) is the number of machines, with \(|\mathcal{F}| \ll |V|\) in practice.

To address this issue, we propose the \textit{Hotspot-aware Synchronization Strategy} ({\sf HaSyn}), which dynamically adjusts the synchronization frequency of embedding vectors based on their update hotness. 
As illustrated in Figure~\ref{fig:hasyn_strategy}, during the initialization phase, the embedding matrix is constructed by sorting nodes in descending order of their frequencies observed in the training data, where the frequency statistics are collected asynchronously during sampling by a background thread. 
During synchronization, we partition \(\vec{V}\) into hotspot blocks \(B_f=\{v \mid f(v)=f\}\) indexed by distinct frequency values \(f\in\mathcal{F}\), yielding \(|\mathcal{F}|\) blocks in total, and randomly select one node from each block \(B_f\) to form a synchronization set of size \(|\mathcal{F}|\).
Due to the skewed frequency distribution, a node in block \(B_f\) is selected with probability \(1/|B_f|\), making hot embeddings (typically in smaller blocks) more likely to be synchronized, while cold embeddings are synchronized less frequently.
Thus, {\sf HaSyn} avoids repeated global graph analysis or full re-ranking in each synchronization period, and maintaining the synchronization set only requires lightweight block-level lookup/update operations.
Compared to full-model synchronization, the communication cost of {\sf HaSyn} is reduced to \(\mathcal{O}(|\mathcal{F}|\cdot d \cdot M)\), where \(d\) is the embedding dimension and \(M\) is the number of machines, with \(|\mathcal{F}| \ll |V|\) in practice (details in \S\ref{sec:individual}).

% Furthermore, under {\sf FeeST} (\S\ref{sec:FeedST}), the active frontier shrinks across rounds, so update activity concentrates on a progressively smaller subset of embeddings and shifts over time.
% Accordingly, hotspots in {\sf FeLoG} are \emph{embedding update hotspots} driven by real-time sampling/training dynamics rather than static graph properties such as degree, making degree-based top-$k$ synchronization wasteful.
% {\sf HaSyn} therefore prioritizes embeddings with high update frequency and adapts to cross-round workload shifts, reducing \emph{inter-machine} synchronization volume while keeping staleness bounded via periodic synchronization of cold embeddings.

Furthermore, under {\sf FeeST} (\S\ref{sec:FeedST}), the active frontier shrinks across rounds, so update activity concentrates on a progressively smaller subset of embeddings and shifts over time.
{\sf HaSyn} therefore prioritizes embeddings with high update frequency and adapts to cross-round workload shifts, reducing \emph{inter-machine} synchronization volume while avoiding indefinite staleness through periodic synchronization of cold embeddings.
As a lightweight hotspot-based strategy, {\sf HaSyn} is most effective when update activity is skewed; under nearly uniform updates or stricter consistency requirements, more sophisticated graph partitioning and adaptive synchronization techniques can be complementary to {\sf FeLoG}.
We analyze the resulting quality-efficiency trade-off of selective synchronization in \S\ref{sec:individual}.

\begin{figure}
 \centering
 \includegraphics[width=3.3in]{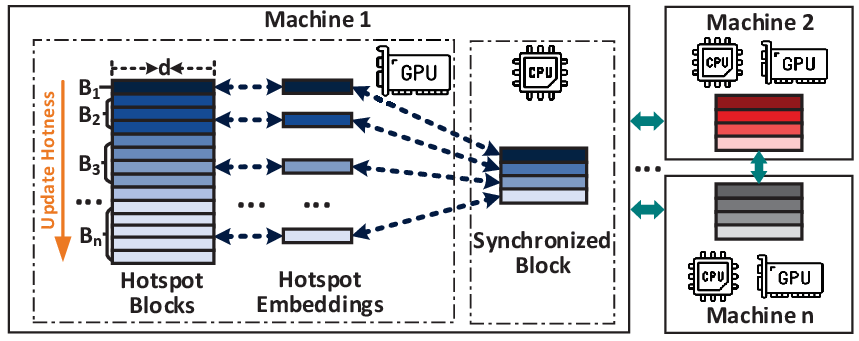}
 \vspace{-2mm}
 \caption{Hotspot-aware Synchronization Strategy across distributed machines.}
 \label{fig:hasyn_strategy}
 \vspace{-2mm}
\end{figure}

\vspace{-2mm}
\subsection{Round-interleaved Pipeline Parallelism}
\label{sec:rip}

While {\sf FeLoG} improves end-to-end efficiency via the {\sf FeeST} model (\S\ref{sec:FeedST}) and the {\sf ACM} mechanism (\S\ref{sec:acm}), its execution still remains \emph{round-synchronous} and \emph{decoupled} between sampling and training.
As discussed in \S\ref{sec:DistGPU}, training starts only after sampling for the current round is fully completed and materialized, which limits CPU-GPU concurrency and leading to a low average utilization ratio (less than 26\%). 

\spara{Execution Analysis of {\sf FeLoG} Components. } Figure~\ref{fig:serial_vs_pipeline}(a) depicts the decoupled execution mode of {\sf FeLoG}, where in each round the training phase is launched only after the sampling phase finishes completely. Furthermore, there exists an intermediate phase, \textit{corpus management}, between sampling and training. 
This includes: (1) the compression and decompression of walk paths using {\sf FaBC}, as introduced in \S\ref{sec:FaBC}, and (2) checkpointing the sampled data to persistent storage for fault tolerance.
Specifically, after each round of sampling, the system performs compression on the CPU. This step significantly reduces the size of the walk paths by identifying and encoding high-frequency nodes via compact bitmap indices, thereby lowering the PCIe transmission load from CPU to GPU. The compressed walk sequences are then serialized and checkpointed to persistent storage to ensure fault tolerance.
Later, during training, the decompression is executed on the GPU side. The GPU first transfers the compressed sequences into its memory, then reconstructs the original node sequences by decoding the bitmap and reassembling the full walk paths. 
Afterwards, the updated embeddings are synchronized across machines by the hotspot-aware strategy (\S\ref{sec:HaSyn}).
% Under the decoupled execution model, both the sampler (CPU) and the trainer (GPU) remain idle during these intermediate corpus management steps, leading to an overall low resource utilization.
Because sampling and training are separated by round-level barriers and corpus materialization, both CPU and GPU can be idle around these intermediate steps, resulting in poor overall resource utilization.

\spara{Pipeline Execution Design.} 
To overcome the limitations of decoupled execution, {\sf FeLoG} introduces a \textit{round-interleaved pipeline parallelism} ({\sf RiPP}) to fully exploit the parallelism of heterogeneous systems and reduce idle time. 
The core idea of {\sf RiPP} is to analyze the data dependencies across the sampling and training phases, decouple their operators into fine-grained execution units, and schedule them in an interleaved, round-wise pipeline to maximize concurrency.
As illustrated in Figure~\ref{fig:serial_vs_pipeline}(b), each sampling-training round in {\sf FeLoG} is decomposed into six operators: {\em random walk generation} (\(R\)), {\em data compression} (\(C\)), {\em checkpointing} (\(P\)), {\em data decompression} (\(D\)), {\em model training} (\(T\)), and {\em synchronization} (\(S\)). 
For instance, \(R_i\) denotes the random walk generation operator in round \(i\), and \(T_i\) corresponds to the associated training operator that consumes the walk corpus.
Unlike {\sf DistGER-Pipe}~\cite{fang2025information}, which overlaps sampling and training in a decoupled CPU-only pipeline driven by pre-defined graph-structural signals, {\sf RiPP} targets feedback-coupled CPU-GPU execution in {\sf FeLoG}.
Specifically, the sampling frontier of round \(i+1\) depends on the embedding-quality feedback produced after training round \(i\), so \(R_{i+1}\) must preserve this one-round feedback dependency rather than being launched independently of \(T_i\).
This feedback dependency also differentiates {\sf RiPP} from general GNN pipeline schedulers such as {\sf NeutronOrch}~\cite{ai2024neutronorch}, {\sf DAHA}~\cite{li2024daha}, and {\sf LeapGNN}~\cite{chen2025leapgnn}, which mainly optimize execution scheduling without modeling {\sf FeeST}'s round-level embedding-quality feedback.

% \begin{figure}
%   \centering
%   \includegraphics[width=3.2in]{Figures/Rip_2.eps}
%   \caption{Execution flow comparison: (a) Serial execution model; (b) Round-interleaved pipeline execution in {\sf FeLoG}.}
%   \label{fig:serial_vs_pipeline}
% \end{figure}

\begin{figure}
  \centering
  \begin{subfigure}{\linewidth}
    \centering
    \includegraphics[width=0.98\linewidth]{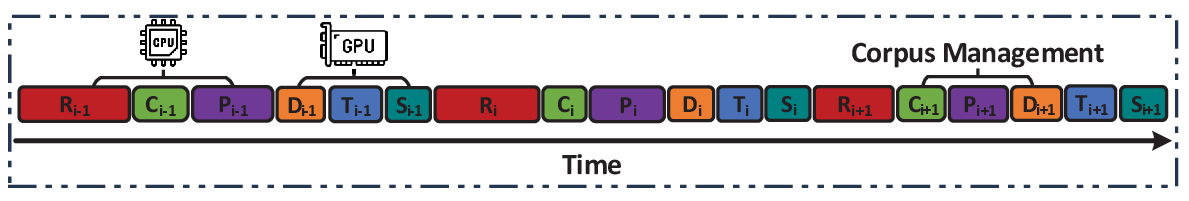}
    \vspace{-2mm}
    \caption{Decoupled Execution}
  \end{subfigure} 
  \begin{subfigure}{\linewidth}
    \centering
    \includegraphics[width=0.98\linewidth]{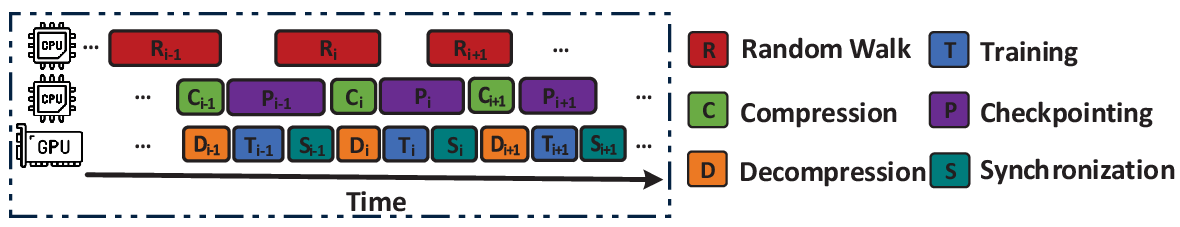}
    \vspace{-2mm}
    \caption{Round-interleaved Pipeline Execution}
  \end{subfigure} 
  \vspace{-6mm}
  \caption{Execution flow comparison in \felog.}
  \label{fig:serial_vs_pipeline}
  \vspace{-4mm}
\end{figure}

To maximize concurrency under this feedback dependency, {\sf RiPP} adopts a three-group operator scheduling strategy.
It groups and dispatches: \ding{202}\normalsize~CPU-side random walk generation (\(R\)), \ding{203}\normalsize~CPU-side data compression (\(C\)) and checkpointing (\(P\)), and \ding{204}\normalsize~GPU-side data decompression (\(D\)), model training (\(T\)), and synchronization (\(S\)) into three asynchronous operator pipelines, each mapped to the most appropriate hardware resources.
These operators form a hybrid parallelism scheme: \emph{intra-round} serial execution for dependent operators on the same data chunk, and \emph{inter-round} parallel execution across different threads and devices whenever their dependencies are satisfied.
For example, once walk chunks of round \(i\) are generated on the CPU, \(C_i\) and \(P_i\) proceed on CPU-side threads, while ready chunks from adjacent rounds are decompressed, trained, and synchronized on the GPU.
Under {\sf RiPP}, active-frontier construction is separated from random-walk generation.
After \(T_{i-1}\), each machine computes \(\Psi(v)\) for its local candidate nodes using the locally visible embedding snapshot, which contains local training updates and previously synchronized remote embeddings.
Each machine then materializes its local portion of \(V_a^{(i)}\), allowing \(R_i\) to start without waiting for the current synchronization \(S_{i-1}\) to complete.
This use of locally visible or slightly stale embeddings follows common distributed graph learning practice~\cite{DistDGL_2020,pipegcn2022,bai2025staleness}.
Once \(V_a^{(i)}\) is fixed, \(R_i\) no longer computes \(\Psi(v)\) or accesses embeddings, but only performs graph-structure-based walk generation from active nodes.
Thus, \(R_i\) can overlap with \(S_{i-1}\), which asynchronously propagates previous-round embedding updates for subsequent training and later active-frontier construction.
This design keeps CPU and GPU resources continuously utilized while preserving the feedback dependency required by {\sf FeeST}, thereby improving end-to-end training efficiency.

\vspace{-2mm}
\subsection{Integration of {\sf FeLoG} with Existing Systems}
\label{sec:integration}

Modern graph embedding frameworks can be broadly classified into two categories: 
random walk-based systems (e.g., {\sf DeepWalk}~\cite{DeepWalk_2014}, {\sf node2vec}~\cite{node2vec_2016}, {\sf HuGE}~\cite{HuGE_2021}, {\sf DistGER}~\cite{fang2023distributed}) 
and GNN-based systems (e.g., {\sf GraphSAGE}~\cite{GraphSAGE_2017}, {\sf DistDGL} \cite{DistDGL_2020}, {\sf NeutronTP} \cite{ai2024neutrontp}). 
{\sf FeLoG} is designed as a lightweight optimization layer that can be seamlessly incorporated into both paradigms 
through two unified interfaces, \texttt{Sample} and \texttt{Train}. 
This abstraction decouples state-of-the-art embedding models from our proposed feedback loop, enabling system-level integration without modifying the core model design.

\spara{Sample Interface.}  
The sampling process is abstracted as $\texttt{Sample}(v_i,$ $ \theta)$ $ \rightarrow W_{v_i}^L$,
% \[
% \small
% \texttt{Sample}(v_i, \theta) \rightarrow W_{v_i}^L,
% \]
where $v_i$ is the source node, $\theta$ denotes model-specific parameters (e.g., bias factor, neighbor budget, entropy criterion), 
and $W_{v_i}^L$ is either a node sequence or a neighborhood subgraph of length $L$.  
This unified form supports:  
(1) \emph{Random-walk systems}, where \texttt{Sample} generates walks such as uniform walks in {\sf DeepWalk}, second-order walks in {\sf node2vec}, or biased walks toward informative neighbors in {\sf HuGE}/{\sf DistGER};  
(2) \emph{GNN systems}, where \texttt{Sample} extracts subgraphs with configurable fan-out for message passing.  
Within {\sf FeLoG}, the \texttt{Sample} stage is enhanced by two mechanisms:  
\emph{FeeST} dynamically updates the active node sets to avoid redundant walks, 
and \emph{FaBC} compresses generated sequences to reduce CPU-GPU transmission.

\spara{Train Interface.}  
The training process is abstracted as $\texttt{Train}(W_{v_i}^L,$ ${model}, {feedback=True}) \rightarrow \vec{V}$,
% \[
% \small
% \texttt{Train}(W_{v_i}^L, \text{model}, \text{feedback=True}) \rightarrow \vec{V},
% \]
where $W_{v_i}^L$ denotes the sampled corpus, {\em model} is the embedding architecture, and $\vec{V}$ is the resulting embedding matrix.  
For random walk systems, this corresponds to Skip-Gram with negative sampling (Eq.~\ref{skim_gram_eq}, Eq.~\ref{negative_sampling_Eq}), 
while for GNN systems it corresponds to iterative message passing and aggregation (Eq.~\ref{GNN_train}).  
{\sf FeLoG} integrates feedback directly into this step:  
after each update, \emph{FeeST} evaluates embedding quality and prunes converged nodes, 
while \emph{HaSyn} adaptively synchronizes high-frequency embeddings across machines to reduce network costs. 
In GNNs, \emph{FaBC} can be reused to compact redundant neighbor lists, and \emph{HaSyn} applies to hotspot embedding vectors arising from skewed node access.  

\spara{Unified Deployment View.}  
From a system perspective, the deployment of {\sf FeLoG} can be summarized as:  
\[
\small
\begin{aligned}
& \texttt{S = Sample}(G, \theta) \quad (\textsf{FeeST + FaBC}) \\
& \texttt{$\vec{V}$ = Train}(S, {model}, {feedback=True}) \quad (\textsf{FeeST + HaSyn}) \\
& \texttt{Persist}(S, \vec{V}) \\
& \texttt{RiPP}(\{{Sample}, {Train}, {Persist}\}) \quad (\textsf{Pipeline execution})
\end{aligned}
\]

This API-style abstraction emphasizes that {\sf FeLoG} is not a standalone framework, 
but a modular optimization layer that can be embedded into both CPU-GPU and multi-machine distributed environments. 
In random-walk frameworks such as {\sf DistGER}, all three mechanisms (\textsf{FeeST, FaBC, HaSyn}) are integrated together with pipeline execution (\textsf{RiPP}).  
In GNN frameworks, {\sf FeLoG} mainly leverages \textsf{FeeST} and can extend \textsf{FaBC} and \textsf{HaSyn} to reduce neighbor redundancy and hotspot synchronization. 
Our experiments in \S\ref{sec:experiments} demonstrate the end-to-end benefits of this integration when applied to {\sf DistGER}, 
while also validating its extensibility to GNN-based systems such as {\sf DistDGL} and {\sf NeutronTP}.

\vspace{-2mm}
\section{Experimental Results}
\label{sec:experiments}
We evaluate the efficiency (\S \ref{sec:overall}) and scalability (\S \ref{sec:scalability}) of our proposed method, {\sf FeLoG},
by comparing with {\sf PyTorch-BigGraph} ({\sf PBG}) \cite{PBG_2019}, {\sf Distributed DGL}
({\sf DistDGL}) \cite{DistDGL_2020}, {\sf DistGER} \cite{fang2023distributed}, {\sf DistGER-Pipe} \cite{fang2025information}, {\sf LeapGNN} \cite{chen2025leapgnn},
% \textcolor{red}{[why not DistGER-Pipe?]} \textcolor{blue}{(DistGER-GPU is implemented by DistGER-Pipe, has been clarified in competitors.)}
and  {\sf NeutronTP} \cite{ai2024neutrontp}. We also compare the effectiveness (\S \ref{sec:effectiveness}) of generated embeddings
on link prediction, graph-based RAG, and node classification.
Finally, we analyze the generalizability of {\sf FeLoG} for the GNN-based embeddings (\S \ref{sec:generality}) and efficiency due to individual
parts of {\sf FeLoG} (\S \ref{sec:individual}).
\textbf{Our codes and datasets are available at \cite{code}}.

\subsection{Experimental Setup}
\label{sec:setup}
\spara{{Environment.}} 
We conduct experiments on a cluster of 8 machines, each equipped with a 2.60GHz Intel$^\circledR$ Xeon$^\circledR$ Gold 6240 CPU (36 cores with 72 threads), a 32GB NVIDIA V100 GPU, 192GB DDR4 memory, and interconnected via full-duplex 10 Gbps networking.
The machines run Ubuntu 20.04, with GCC 9.4.0 used for compiling {\sf DistGER}, {\sf DistGER-G}, and {\sf FeLoG}, and Python 3.10 with torch 1.11.0 used as the backend deep learning framework for {\sf PyTorch-BigGraph}, {\sf DistDGL}, {\sf LeapGNN}, {\sf NeutronTP}, and {\sf GraNNDis}.
The multi-GPU comparison with {\sf GraNNDis} is conducted on a server equipped with four NVIDIA A40 GPUs.

\spara{Datasets. }
We employ seven widely-used, real-world graphs
(Table~\ref{graph_datasets_memory_usage}): {\em Flickr} (FL) \cite{Flickr_Youtube_Graph},
{\em Youtube} (YT) \cite{Flickr_Youtube_Graph},
{\em LiveJournal} (LJ) \cite{BlogCatalog_Twitter_LiveJournal_Graph},
{\em Com-Orkut} (OR) \cite{com-orkut_2012}, {\em Twitter} (TW) \cite{twitter_2010}, {\em U.K.-2007} (UK) \cite{BSVLTAG} and {\em OGB-papers100M} (OGB) \cite{hu2020open}.
The {\em FL} and {\em YT} datasets are labeled, with {\em FL} having 195 labels for user-content interactions and {\em YT} 47 labels for video engagement.
The {\em OGB} dataset has 128-dimensional node features, representing citation networks of scientific papers. For heterophilic settings, we use two labeled benchmarks, {\em Amazon-ratings} (AR: \(|V|=24.49\)K, \(|E|=93.05\)K, 5 classes) and {\em Roman-empire} (RE: \(|V|=22.66\)K, \(|E|=32.93\)K, 18 classes)~\cite{platonov2023critical}, to evaluate the sensitivity of {\sf FeeST} to \(\mu\) via downstream node classification.
We also use synthetic graphs \cite{RMAT_2004} (up to 1 billion nodes, 10 billion edges) to assess the scalability of {\sf FeLoG}.
Considering the default settings of popular graph embedding methods, we use their undirected version. 

\begin{table}[tb!]
\centering
\caption{Dataset statistics ($K{=}10^3, M{=}10^6, B{=}10^9$) and Avg. memory footprint (GB) per machine. ``\xmark'' indicates failure due to out of memory or $>1$ day runtime.}
\label{graph_datasets_memory_usage}
\vspace{-2mm}
\footnotesize
\setlength{\tabcolsep}{0.9pt}
\renewcommand{\arraystretch}{0.92}
\begin{adjustbox}{max width=\columnwidth}
\begin{tabular}{lcc@{\hspace{2pt}}ccccccc}
\toprule
\multirow{2}{*}{\textbf{Graph}} &
\multicolumn{2}{c}{\textbf{Dataset statistics}} &
\multicolumn{7}{c}{\textbf{Avg. memory footprint per machine (GB, CPU/GPU)}} \\
\cmidrule{2-3}\cmidrule{4-10}
& \textbf{\#Nodes} & \textbf{\#Edges}
& {\sf PBG} & {\sf DistDGL} & {\sf DistGER} & {\sf DistGER-G} & {\sf NeutronTP} & {\sf LeapGNN} & {\sf FeLoG} \\
\midrule
{\em FL}  & 80.51 K  & 5.90 M   & 8.9/-   & 5.4/1.4 & 0.9/-   & 0.5/0.9  & 8.6/1.2   & 6.8/1.2    & 0.5/0.5 \\
{\em YT}  & 1.14 M   & 2.99 M   & 9.7/-   & 5.5/1.5 & 4.3/-   & 2.3/5.3  & 9.0/1.6   & 17.0/3.4   & 1.9/0.6 \\
{\em LJ}  & 2.24 M   & 14.61 M  & 10.3/-  & 5.8/1.7 & 5.5/-   & 2.6/6.6  & 9.3/2.2   & 30.7/7.7   & 2.1/0.9 \\
{\em OR}  & 3.07 M   & 117.19 M & 11.4/-  & 6.3/2.1 & 6.9/-   & 6.0/8.0  & 9.7/2.6   & 39.4/20.1  & 6.1/1.8 \\
{\em TW}  & 41.65 M  & 1.47 B   & 31.7/-  & \xmark  & 67.2/-  & \xmark   & 19.6/19.4 & \xmark      & 59.7/7.2 \\
{\em OGB} & 111.06 M & 1.62 B   & \xmark  & \xmark  & 72.3/-  & \xmark   & 30.7/31.2 & \xmark      & 71.5/9.4 \\
{\em UK}  & 105.15 M & 3.72 B   & \xmark  & \xmark  & 125.8/- & \xmark   & \xmark     & \xmark      & 107.3/9.0 \\
\bottomrule
\end{tabular}
\end{adjustbox}
\vspace{-2mm}
\end{table}

\spara{Baselines.} We select six state-of-the-art distributed graph embedding frameworks: the multi-relation embedding system {\sf PyTorch-BigGraph} ({\sf PBG}) developed by Facebook {\scriptsize\url{https://github.com/facebookresearch/PyTorch-BigGraph}}~\cite{PBG_2019}; the GNN-based system {\sf DistDGL} from Amazon {\scriptsize\url{https://github.com/dmlc/dgl}}~\cite{DistDGL_2020}; the recent GNN-based framework {\sf NeutronTP} {\scriptsize\url{https://github.com/iDC-NEU/NeutronTP}}~\cite{ai2024neutrontp} and {\sf LeapGNN} {}{\scriptsize\url{https://github.com/ISCS-ZJU/LeapGNN-AE}}~\cite{chen2025leapgnn}; and the random walk-based system {\sf DistGER} {\scriptsize\url{https://github.com/RocmFang/DistGER}}~\cite{fang2023distributed}.
In addition, we include {\sf DistGER-Pipe} \cite{fang2025information}, a CPU-only pipeline-optimized extension of {\sf DistGER}, and further implement its hybrid CPU-GPU variant, {\sf DistGER-G}, which performs sampling on CPUs and offloads training to GPUs.
We additionally compare with {\sf GraNNDis} {\scriptsize\url{https://github.com/AIS-SNU/GraNNDis_Artifact}}~\cite{song2024granndis}, a recent multi-GPU GNN training framework.
We do not compare against single-machine execution planners such as {\sf NeutronOrch} \cite{ai2024neutronorch}, {\sf DAHA} \cite{li2024daha}, and {\sf GIDS} \cite{park2024accelerating}, as they target intra-node scheduling and are not designed for {\sf FeLoG}'s multi-machine setting. Empirically, in our setup on {\em OR}, SSD-based {\sf GIDS} takes 3095\,s while {\sf FeLoG} completes in 61.8\,s, and the in-memory {\sf NeutronOrch}/{\sf DAHA} run out of memory on the billion-scale {\em UK} graph.

% \spara{Parameters.} 
% For {\sf FeLoG}, in its default mode, we use random-walk based graph embedding. However, we also demonstrate its generality over GNN-based embeddings in \S \ref{sec:generality}. We 
% set $\mu$=0.65 in \feest, and compress the top-2 most frequent nodes in \fabc (sensitivity analysis w.r.t. these parameters is provided in the Appendix B).
% For {\sf DistGER} and {\sf DistGER-G},
% we set the sliding window size $w$=10, number of negative samples $K$=5, and synchronization period=0.1 sec \cite{Pword2vec_2019}.
% For {\sf PBG}, {\sf DistDGL}, and {\sf NeutronTP}, we follow the default settings reported in their respective papers \cite{PBG_2019, DistDGL_2020, ai2024neutrontp}.
% For fair comparison across all systems,
% we set the embedding dimension $d$=128 that is commonly used \cite{HuGE_2021,node2vec_2016,DeepWalk_2014,Verse_2018,ProNE_2019},
% and report the average end-to-end running time for each epoch. For task effectiveness evaluations,
% we find the best results from a grid search over learning rates from 0.001-0.1, \# epochs from 1-30,
% and \# dimensions from 32-512.

\spara{Parameters.}
By default, {\sf FeLoG} uses random-walk based graph embedding, while we also evaluate its generality over GNN-based embeddings in \S \ref{sec:generality}.
We set $\mu{=}0.65$ in \feest, and compress the top-2 most frequent nodes in \fabc (the sensitivity analysis is reported in \S~\ref{sec:exp_sensitivity_analysis}).
For {\sf DistGER} and {\sf DistGER-G}, we set the sliding window size $w{=}10$, the number of negative samples $K{=}5$, and the synchronization period to 0.1 sec~\cite{Pword2vec_2019}.
For {\sf PBG}, {\sf DistDGL}, {\sf LeapGNN} and {\sf NeutronTP}, we follow the default settings in their papers~\cite{PBG_2019, DistDGL_2020, ai2024neutrontp, chen2025leapgnn}.
For fair system comparison, we use embedding dimension $d{=}128$ that is commonly adopted~\cite{HuGE_2021,node2vec_2016,DeepWalk_2014,Verse_2018,ProNE_2019}, and report the average end-to-end epoch time.
For task effectiveness, we select the best results via grid search over learning rates in $[0.001, 0.1]$, training epochs in $[1, 30]$, embedding dimensions in $\{32, 64, 128, 256, 512\}$, and, for neighbor-sampling based GNN settings, sampling hyperparameters including the number of layers $\in \{2,3,4\}$ and per-layer fanout $\in \{5,10,15,25\}$. 
% (applied consistently to {\sf FeLoG} in its GNN mode and the GNN-based baselines).

\begin{figure}
    \centering
    \includegraphics[width= 3 in]{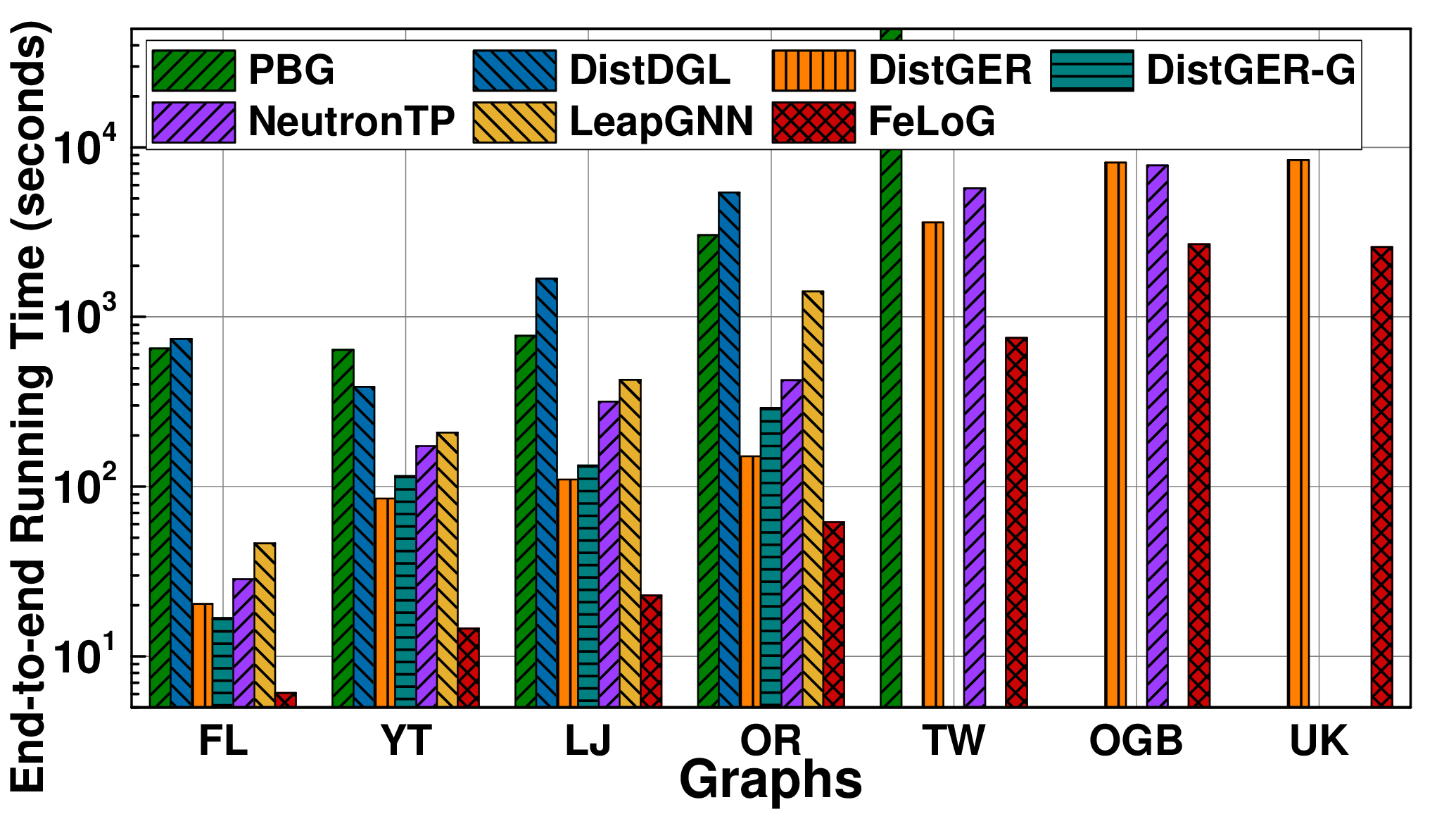}
    \vspace{-2mm}
    \captionof{figure}
      {Efficiency of {\sf PBG} \cite{PBG_2019}, {\sf DistDGL} \cite{DistDGL_2020}, {\sf DistGER} \cite{fang2023distributed}, {\sf DistGER-G} \cite{fang2025information}, {\sf NeutronTP} \cite{ai2024neutrontp}, {\sf LeapGNN} \cite{chen2025leapgnn} and {\sf FeLoG} (ours).
        \label{overall_performance}
      }
      \vspace{-6mm}
\end{figure}

\subsection{Efficiency and Memory Requirement}
%w.r.t. Competitors}
\label{sec:overall}

We report the end-to-end running times of {\sf PBG}, {\sf DistDGL}, {\sf DistGER}, {\sf DistGER-G}, {\sf NeutronTP}, {\sf LeapGNN}, and {\sf FeLoG}
on seven real-world graphs with the cluster of 8 machines in Figure~\ref{overall_performance}.
The reported average epoch end-to-end time includes sampling, training, and communication overheads.
{\sf FeLoG} significantly outperforms the competitors
on all these graphs, achieving an average acceleration of $27.9 \times$.
{\sf PBG} leverages a parameter server to synchronize embeddings between clients,
resulting in more load on the communication network. As a result, {\sf PBG} is on average
68.4$\times$ slower than {\sf FeLoG}.
As a system of the same type, {\sf FeLoG} outperforms the CPU-based {\sf DistGER} and the CPU-GPU hybrid {\sf DistGER-G} by 3.9$\times$ and 5.3$\times$ on average, respectively.
{\sf DistGER} uses information-theoretic walks to adapt sampling, but without training feedback it applies a uniform configuration to all nodes, causing coarse-grained control and redundancy.
{\sf DistGER-G}, though accelerating training via GPU, incurs heavy CPU-GPU data transfer overhead, which reduces GPU utilization and degrades overall performance.
Notice that {\sf DistGER-G} fails on larger datasets such as {\em TW}, {\em OGB}, and {\em UK} due to its redundant corpus generation during sampling and suboptimal system support for training, which lead to memory consumption exceeding available capacity.
For GNN-based system {\sf DistDGL}, due to the long running time of graph sampling (e.g., taking 80\% of the overhead),
it is much less efficient than the other systems. Similarly to {\sf DistGER-G}, it cannot run on billion-edge graphs, as it fails to terminate within one day.
Two recent GNN systems, {\sf NeutronTP} and {\sf LeapGNN}, optimize communication and feature operations.
However, their lack of real-time embedding quality awareness and uniform treatment of node features across iterations lead to redundant computation and hinder convergence speed. Hence, they are on average 7.9$\times$ and $15.8\times$ slower than {\sf FeLoG}, and cannot run on the largest dataset {\em UK} due to out-of-memory issues.
We further discuss accuracy-time convergence in \S\ref{sec:effectiveness} to quantify the time-to-quality tradeoff.

Considering the resource consumption that affects scalability, Table~\ref{graph_datasets_memory_usage} reports the average per-machine memory footprint of {\sf FeLoG} and baseline systems on each graph.
{\sf FeLoG} consistently achieves the lowest or one of the lowest memory usage across both CPU and GPU components, with particularly low usage on GPU.
For the billion-scale graphs like {\em UK}, {\sf FeLoG} is one of the few systems able to run without exceeding memory limits, highlighting its efficiency and scalability in distributed settings.

% \begin{figure}
%  \centering
%  \includegraphics[width= 3.2 in]{Figures/FeLoG_scalability_1.eps}
%  \vspace{-3mm}
%  \caption{Scalability analysis of \felog.}
% \label{Dist_scalability}
% \end{figure}

\begin{figure}
  \centering
  \begin{subfigure}{.52\linewidth}
    \centering
    \includegraphics[width=1.1\linewidth]{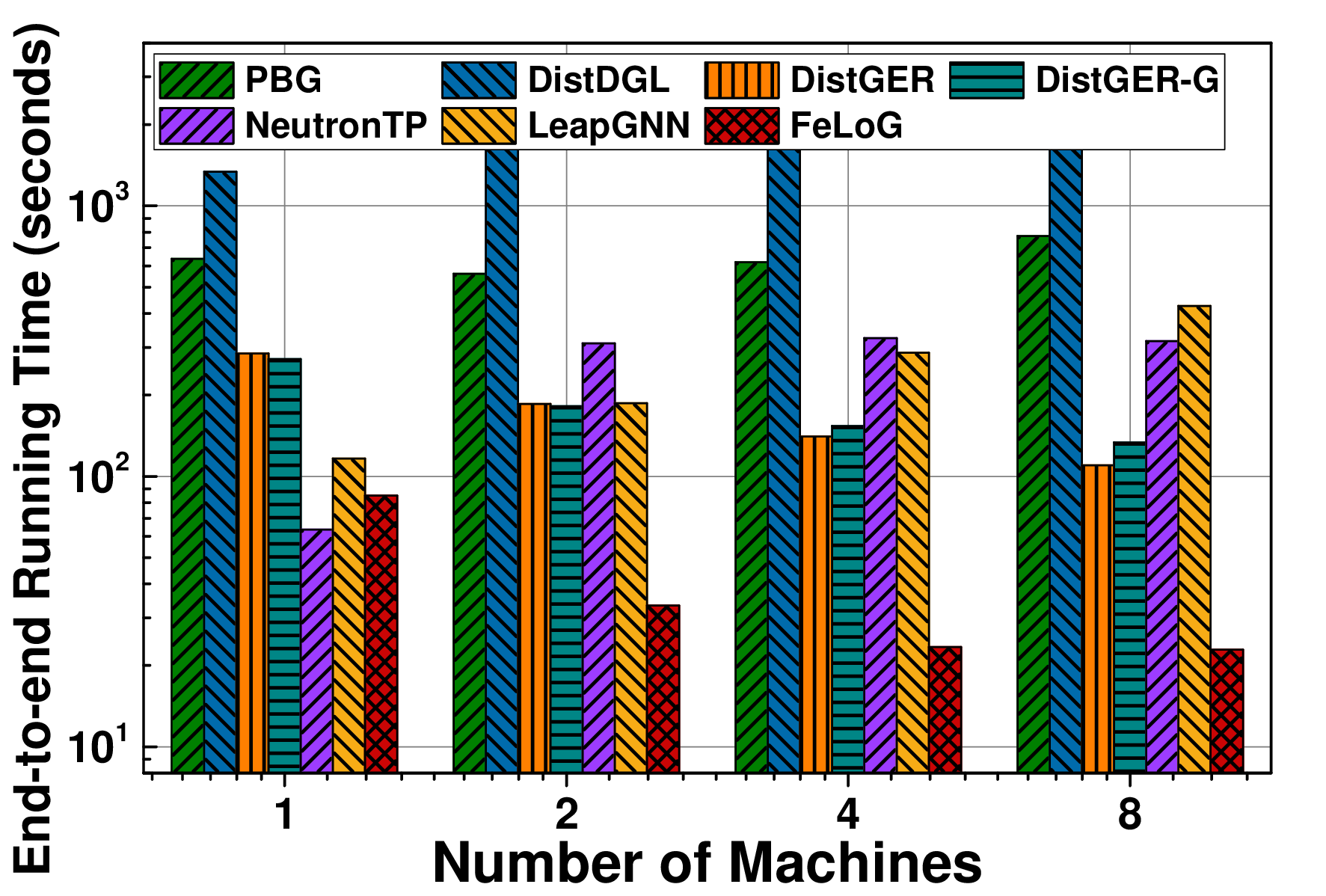}
    \vspace{-5mm}
    \caption{Scalability across Machine Scales}
    % \label{performance_bottleneck}
  \end{subfigure} 
  \begin{subfigure}{.47\linewidth}
    \centering
    \includegraphics[width=0.89\linewidth]{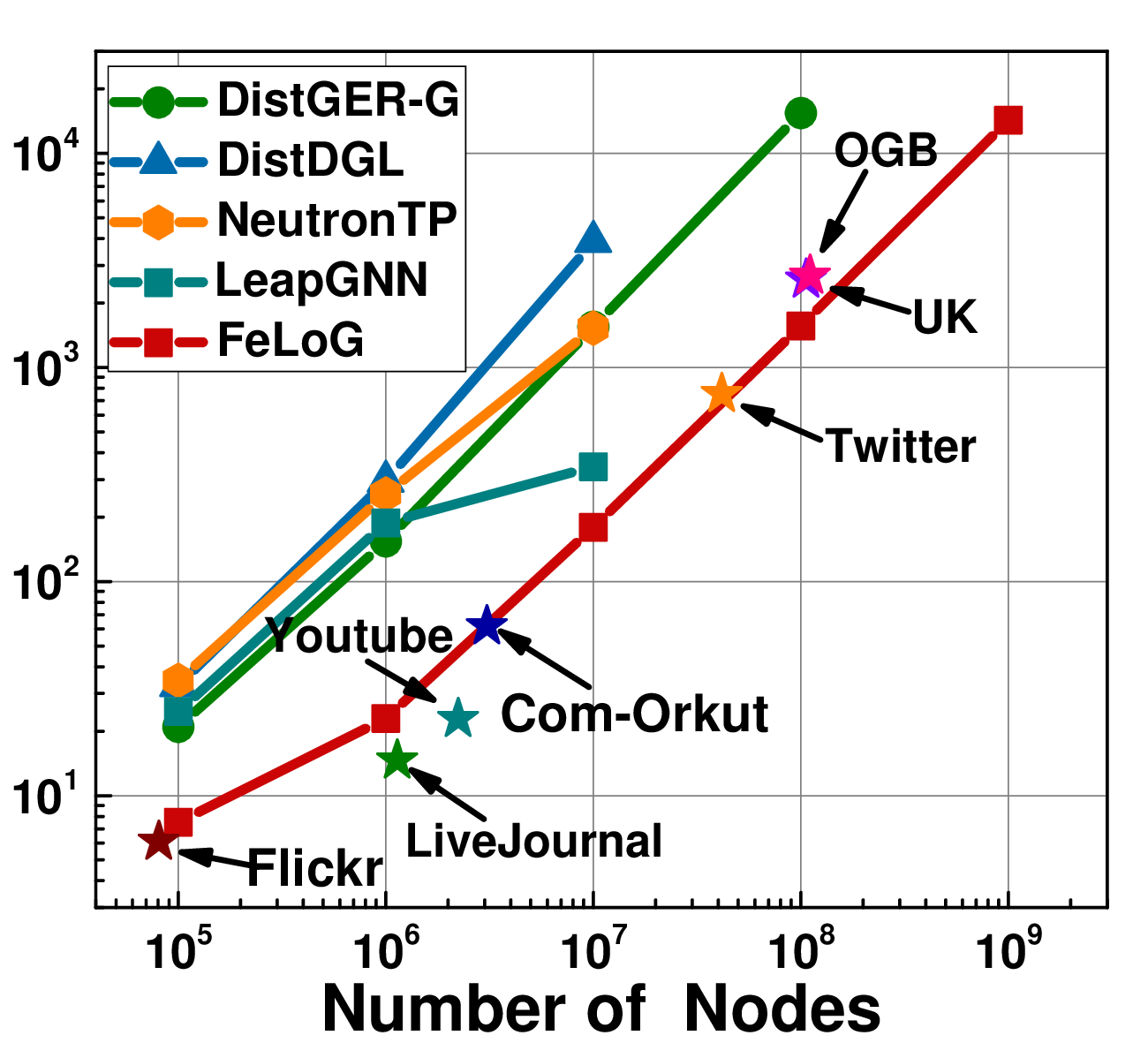}
    \vspace{-1mm}
    \caption{Scalability across Graph Sizes}
    % \label{fig:loss-bit-alpha}
  \end{subfigure}
  \vspace{-6mm}
  \caption{Scalability analysis of \felog.}
  \label{Dist_scalability}
  \vspace{-2mm}
\end{figure}

\subsection{Scalability}
% w.r.t. Competitors}
\label{sec:scalability}

Figure~\ref{Dist_scalability}(a) shows end-to-end running times of all competing
systems on the {\em LiveJournal} graph, as we increase \# machines
from 1 to 8 to evaluate scalability. {\sf FeLoG} achieves better scalability than the other
six distributed systems. Due to space limitations, we omit results on other graph datasets, 
which either exhibit similar trends or are too large for the competing systems to scale.
{\sf PBG} leverages a parameter server and a shared network filesystem
to synchronize the parameters in the distributed model. When the number of machines increases, {\sf PBG} puts more load
on the communications network, resulting in poor scalability. 
{\sf DistDGL} suffers from limited scalability due to its full-graph parameter synchronization and frequent gradient communication across machines.
The CPU-based {\sf DistGER} exhibits better scalability than the CPU-GPU hybrid {\sf DistGER-G}, as the latter suffers from substantial CPU-GPU data transfer overhead and a sequential execution pattern that undermines the performance benefits of GPU acceleration.
{\sf NeutronTP}'s end-to-end running time remains nearly stable as the number of machines increases, indicating limited scalability. This is mainly due to its uniform processing of all node features, leading to redundant computation and underutilization of distributed resources.
{\sf LeapGNN} scales poorly due to frequent step-wise global synchronization that amplifies stragglers and idle GPUs as machines increase.
In comparison, {\sf FeLoG} incorporates a feedback-driven sampling-training coupling model and an activity-aware communication mechanism to reduce both computation and communication costs, it also employs  a round-interleaved pipeline to maximize CPU-GPU utilization. Hence, {\sf FeLoG} achieves better scalability than other systems.
% Due to space limitations, we omit {\sf FeLoG}'s scalability results on other graphs, which exhibit similar trends. On {\em Twitter}, the end-to-end running times {\sf FeLoG} on 1, 2, 4, and 8 machines are 3090s, 1739s, 1197s, and 746s, respectively,
% while on {\em Com-Orkut}, the results are 549s, 439s, 301s, and 174s, respectively. 
% The results show a good linear relationship.

To further assess the scalability of {\sf FeLoG}, we generate synthetic graphs \cite{RMAT_2004} with a fixed node degree of 10 and the number of nodes from $10^5$ to $10^9$. Figure~\ref{Dist_scalability}(b) presents the running times on these synthetic graphs using a cluster of 8 machines, suggesting that the running time increases linearly with the size of a graph, and {\sf FeLoG} has the capability to handle even billion-node graphs. Moreover, the running times for seven real-world graphs (including the billion-scale graphs, for which the competing systems do not terminate in 1 day or crash due to memory limitation) are inserted into the plot, which is consistent with the trend on synthetic data.

\begin{table}[t]
  \caption{$AUC$ of {\sf FeLoG} and competitors for link prediction.}
  \vspace{-2mm}
  \label{AUC_results}
  \centering
  \footnotesize
  \setlength{\tabcolsep}{4pt}
  \begin{tabular*}{\columnwidth}{@{\extracolsep{\fill}}lcccc@{}}
    \toprule
    Method & Youtube & LiveJournal & Com-Orkut & Twitter \\
    \midrule
    {\sf PBG}     & 0.753 & 0.882 & 0.955 & 0.912 \\
    {\sf DistDGL} & 0.894 & 0.718 & 0.815 & running time $>$ 1 day \\
    {\sf DistGER} & \textbf{0.966} & 0.976 & 0.921 & 0.919 \\
    {\sf FeLoG}   & 0.949 & \textbf{0.979} & \textbf{0.965} & \textbf{0.942} \\
    \bottomrule
  \end{tabular*}
  \vspace{-2mm}
\end{table}

\begin{table}[t]
\centering
\vspace{2mm}
\caption{RAG retrieval performance of {\sf FeLoG} and {\sf NeutronTP} 
on {\em Flickr} and {\em Youtube} datasets ($K{=}20$, 2-hop retrieval).}
% \vspace{-2mm}
\footnotesize
\begin{tabular}{lcccc}
\toprule
\multirow{2}{*}{Metric} & 
\multicolumn{2}{c}{Flickr} & 
\multicolumn{2}{c}{Youtube} \\
\cmidrule(lr){2-3} \cmidrule(lr){4-5}
 & {\sf NeutronTP} & {\sf FeLoG} & {\sf NeutronTP} & {\sf FeLoG} \\
\midrule
Hit@K   & 0.817 & \textbf{0.955} (+16.9\%) & 0.790 & \textbf{0.860} (+8.9\%) \\
nDCG@K  & 0.202 & \textbf{0.499} (+147.0\%) & 0.267 & \textbf{0.677} (+153.6\%) \\
MRR@K   & 0.343 & \textbf{0.714} (+108.2\%) & 0.430 & \textbf{0.924} (+114.9\%) \\
\bottomrule
\end{tabular}
\label{tab:rag_comparison_compact}
\vspace{-2mm}
\end{table}

\subsection{Effectiveness}
%%% ooibc: effectiveness? meaning accuracy?
%w.r.t. Competitors}
%%% Fang: Yes, here effectiveness refers to the quality of the generated embeddings, measured by the accuracy on downstream tasks.
\label{sec:effectiveness}
\spara{Application in Link Prediction.}
To perform link prediction on a given graph $G$, following \cite{Verse_2018, NRP_2020,fang2023distributed},
we first uniformly at random remove 50\% edges as positive test edges, and the rest are used as positive training edges.
We also provide negative training and test edges by considering those node pairs between which no edge exists in $G$.
We ensure that the positive and negative set sizes are similar. For a pair of nodes $(u, v)$, let $\varphi(u)$ and
$\varphi(v)$ be the vectors learned by embedding methods.
The link prediction is conducted as a classification task
based on the similarity of $u$ and $v$, i.e., $\varphi(u)\cdot\varphi(v)$.
The effectiveness of link prediction is measured via the $AUC$ (Area Under Curve) score \cite{yang2022auc}, the higher the better.
We repeat this procedure 50 times to offset the randomness of edge removal and report the average $AUC$ in
Table~\ref{AUC_results}.
%shows $AUC$ for all the methods on five real-world graphs.
%, respectively, where a ``$-$'' indicates that the method fails due to the limitation of computing resources or because its running time exceeds 1 day.
{\sf FeLoG} outperforms all competitors on these graphs, except for {\sf DistGER} on {\em Youtube}, where {\sf FeLoG} ranks second.
On average, {\sf FeLoG} achieves a 9.8\% higher $AUC$ score compared with the other three systems. %, thanks to our sampling-training coupling model. 
Figure~\ref{Dist_time_accuracy}(a) exhibits accuracy-efficiency tradeoffs of {\sf FeLoG} and competitors, i.e., their $AUC$ score convergence curves w.r.t. increasing running times, over {\em LiveJournal}, indicating that {\sf FeLoG} tends to reach its peak accuracy in fewer epochs (around 4).
In addition, Figure~\ref{Dist_time_accuracy}(b) shows consistently fast accuracy-efficiency convergence of {\sf FeLoG} across graphs, reaching high $AUC$ within a short running time on {\em Youtube} (3 epochs) and {\em Com-Orkut} (6 epochs).

\begin{figure}
  \centering
  \begin{subfigure}{.325\linewidth}
    \centering
    \includegraphics[width=0.95\linewidth]{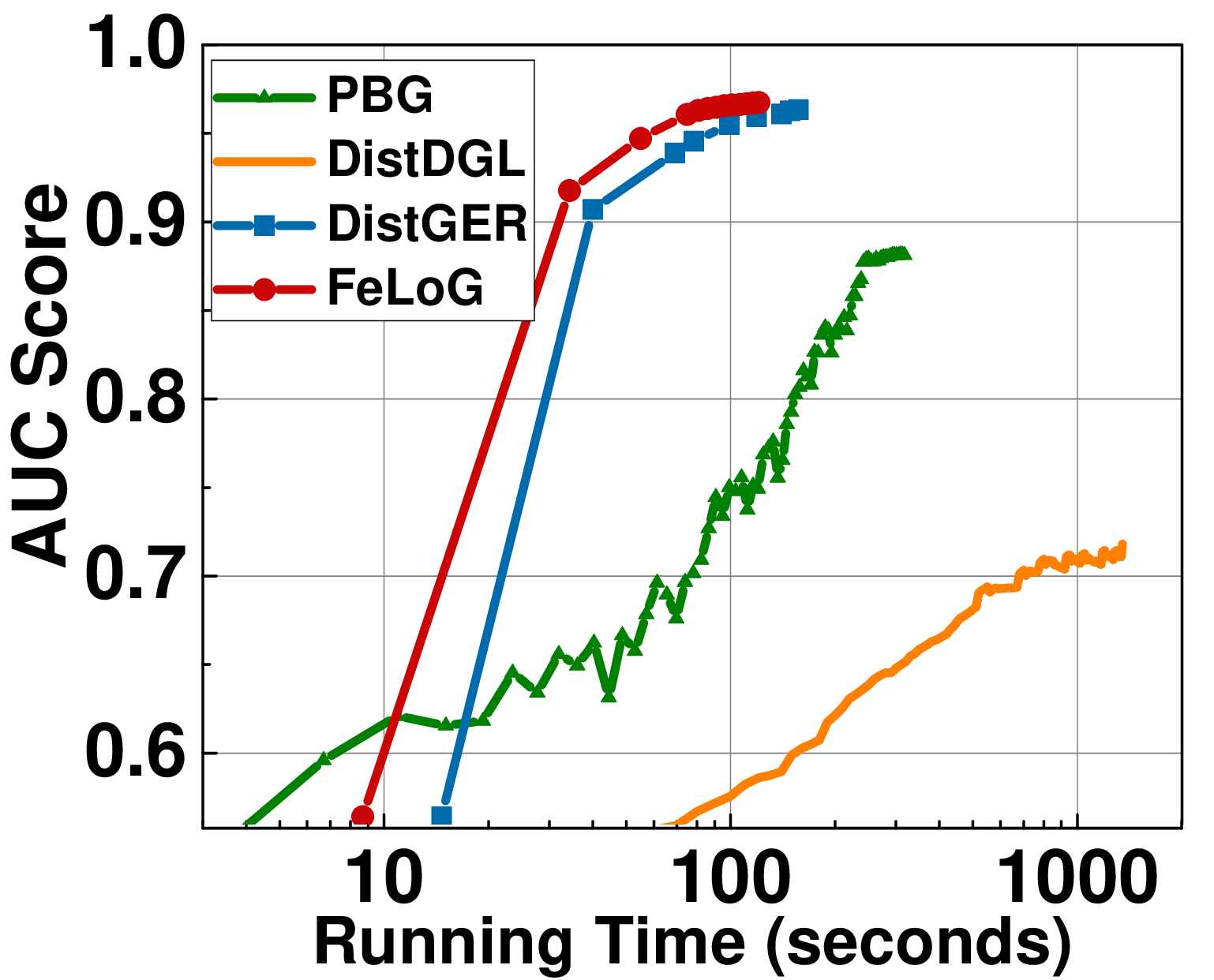}
    % \vspace{-5mm}
    \caption{\footnotesize Time-AUC}
    % \label{fig:loss-bit-alpha}
  \end{subfigure}
  \begin{subfigure}{.325\linewidth}
    \centering
    \includegraphics[width=0.95\linewidth]{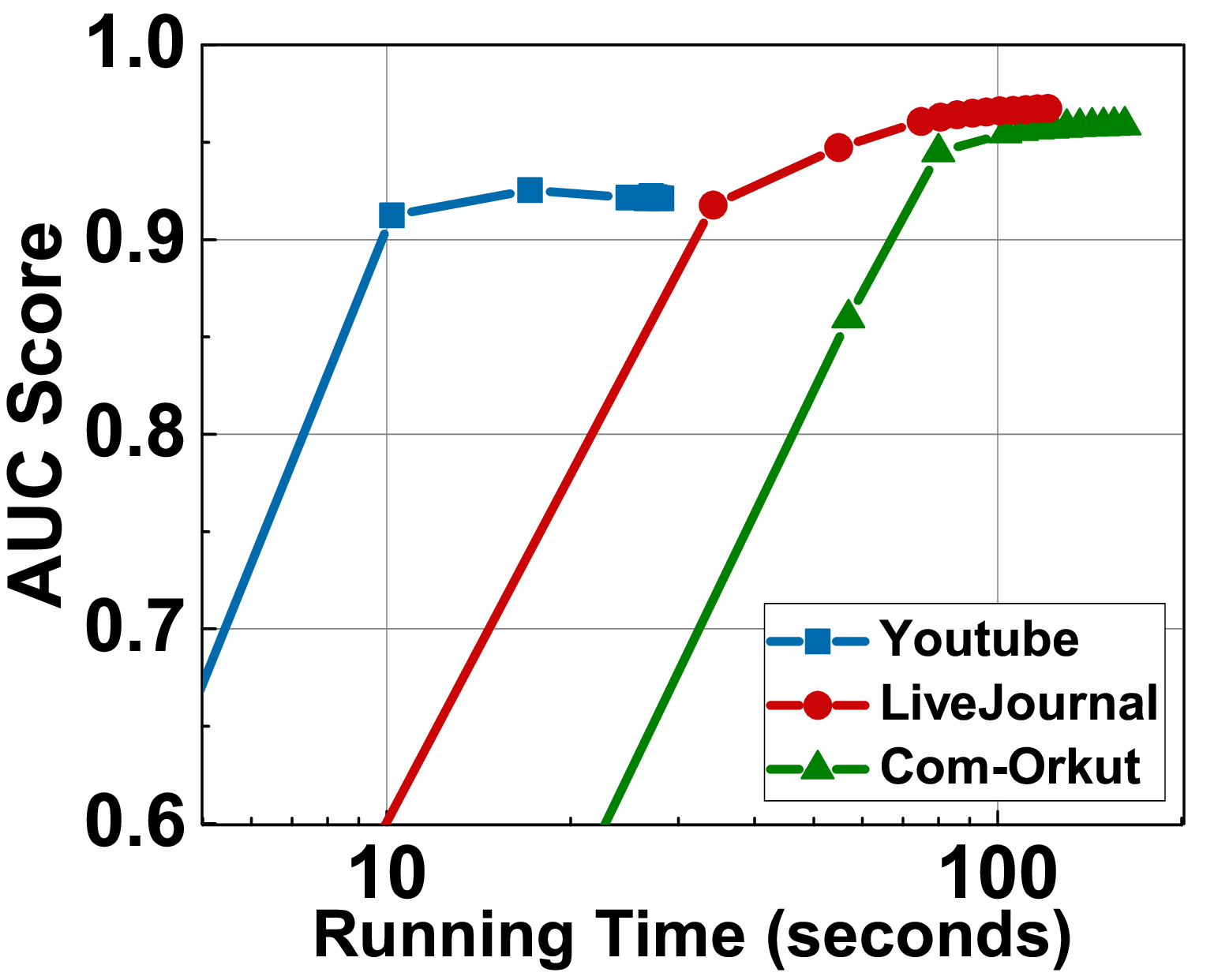}
    % \vspace{-5mm}
    \caption{\footnotesize \felog Across Graphs}
    % \label{fig:loss-bit-alpha}
  \end{subfigure}
   \begin{subfigure}{.325\linewidth}
  % \vspace{-2mm}
    \centering
    \includegraphics[width=0.95\linewidth]{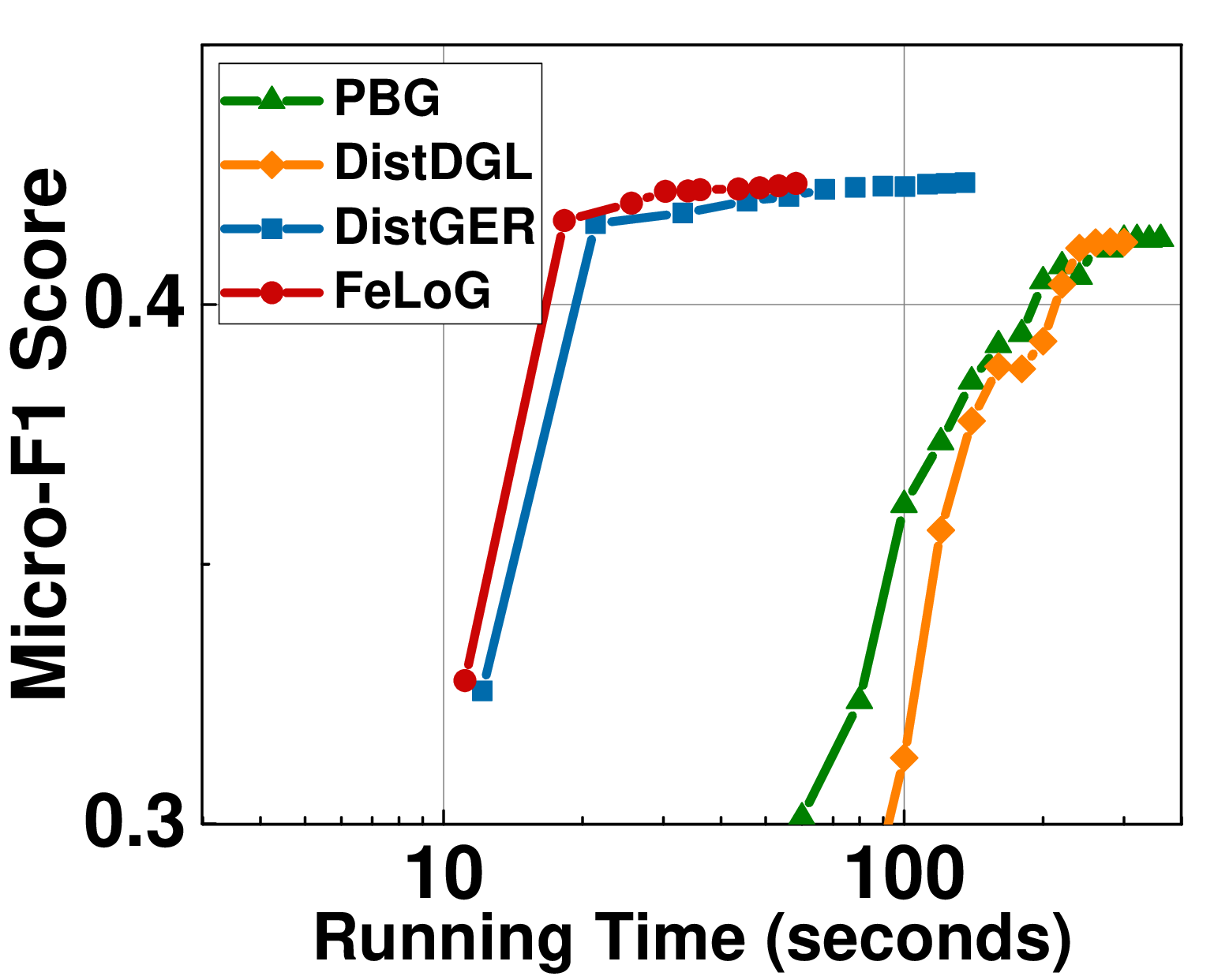}
    % \vspace{-2mm}
    \caption{\footnotesize{{\footnotesize{Time-Micro-F1}}}}
  \end{subfigure} 
  \vspace{-4mm}
  \caption{{The influence of running time on embedding quality for {\sf FeLoG} and competitors. Each point corresponds to one training epoch.}}
  \vspace{-2mm}
  \label{Dist_time_accuracy}
\end{figure}

\spara{Application in Multi-label Classification.}
This task predicts one or more labels for each graph node and has applications in %modern applications ranging from
text categorization \cite{zhang2006multilabel} and bioinformatics \cite{zhang2018ontological}.
We use embedding vectors and a one-vs-rest logistic regression classifier
with L2 regularization \cite{MLC_LIBLINEAR_2008}, %(using the LIBLINEAR library),
then evaluate the effectiveness by micro-averaged F1 ($Micro-F1$) and macro-averaged F1 ($Macro-F1$) \cite{WangC016}
scores, where $Micro-F1$ gives equal weight to each test instance and $Macro-F1$ assigns equal weight to each label category \cite{keikha2018community}.
%To train a classifier, nodes are uniformly at random split into training and test sets.
Following \cite{HuGE_2021,node2vec_2016,DeepWalk_2014,Verse_2018},
we select 10\% to 90\% training data ratio on {\em Flickr}, and 1\% to 9\% training ratio on {\em Youtube}.
%and the remaining nodes for testing.
We report the averaged $Macro-F1$ and $Micro-F1$ scores in Figure~\ref{MLC_results}.
% shows the $Macro-F1$ and $Micro-F1$ scores achieved by each system as a
%function of the training ratio variation, respectively.
We find that \felog has better $Macro-F1$ and $Micro-F1$ scores
than existing frameworks, %on these graphs, %. In particular, compared with the KnightKing,
%DistGER consistently outperforms the other random walk-based systems on all graphs in $Macro-F1$ and $Micro-F1$ scores,
gaining 9.8\% and 2.8\% average improvements, respectively, due to its more effective information-centric random walks.
%Definition of $Macro-F1$ and $Micro-F1$ are as the following:
Figure~\ref{Dist_time_accuracy} (c) shows the time-accuracy trade-offs of {\sf FeLoG} and competitors on node classification over {\em YT}. The results also show that {\sf FeLoG} reaches the same or higher accuracy in less time.

\begin{figure}
  \centering
  \includegraphics[width= 3.4 in]{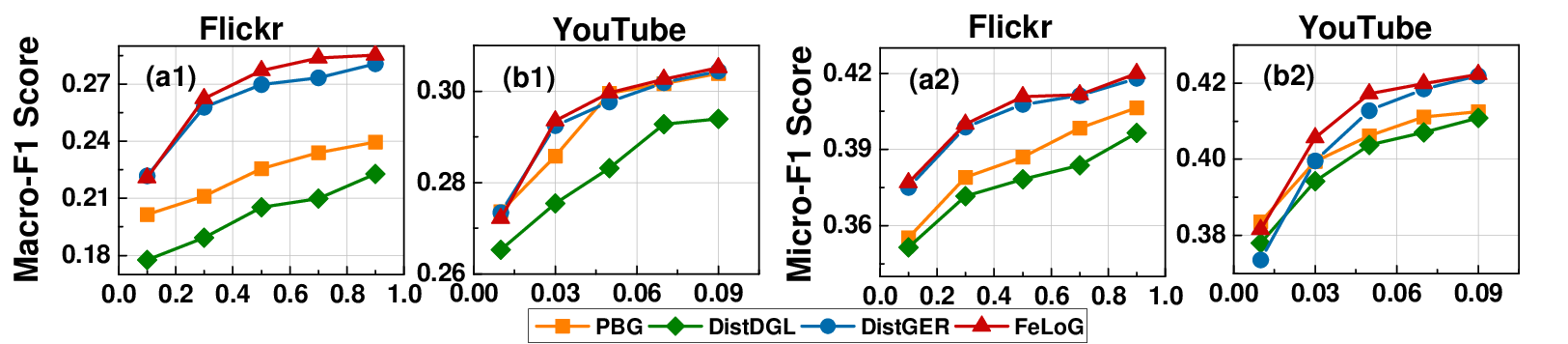}
  \vspace{-6mm}
  \caption{$Macro-F1$ (a1, b1) and $Micro-F1$ (a2, b2) scores for multi-label node classification.}
  \label{MLC_results}
\end{figure}

\spara{Application in Graph-based RAG.}
To evaluate the generalization of {\sf FeLoG} to graph-based retrieval-augmented generation (GraphRAG), where graph embeddings serve as the retrieval backbone for LLMs, we integrate the learned embeddings into a FAISS-based vector retrieval pipeline on {\em FK} and {\em YT}.
{\em FK} captures user-content interactions, while {\em YT} models large-scale user-item engagement, enabling us to construct ground-truth relevance based on shared semantic attributes reflected by these interactions.
Given a query node, we retrieve top-$K$ candidate nodes by embedding similarity and evaluate retrieval quality using {\em Hit@K}, {\em nDCG@K}, and {\em MRR@K} (Table~\ref{tab:rag_comparison_compact}).
Across both datasets, {\sf FeLoG} consistently outperforms {\sf NeutronTP}, achieving an average improvement of 150.3\% in {\em nDCG@20}, 12.9\% in {\em Hit@20}, and 111.6\% in {\em MRR@20}.
These results indicate that {\sf FeLoG} yields higher-quality retrieval embeddings for GraphRAG.

\begin{figure}
  \centering
  \begin{subfigure}{.38\linewidth}
    \centering
    \vspace{-1mm}
    \includegraphics[width=1.05\linewidth, height= 0.86 in]{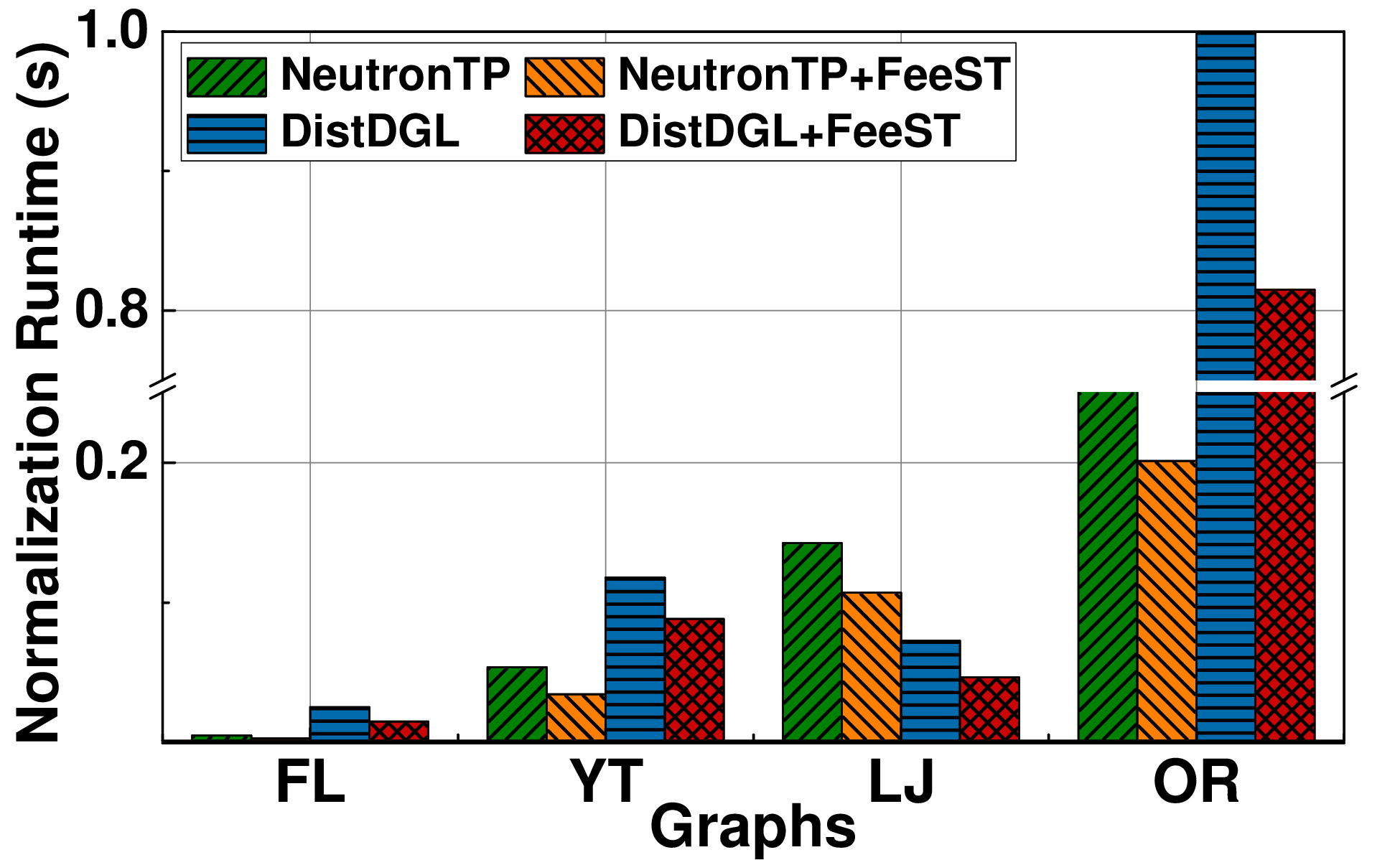}
    \vspace{-5mm}
    \caption{\small Generalizability}
    \vspace{-3mm}
    % \label{fig:felog_breakdown_train}
  \end{subfigure} 
  \begin{subfigure}{.3\linewidth}
    \centering
    \vspace{-1mm}
    \includegraphics[width=1.05\linewidth, height= 0.9 in]{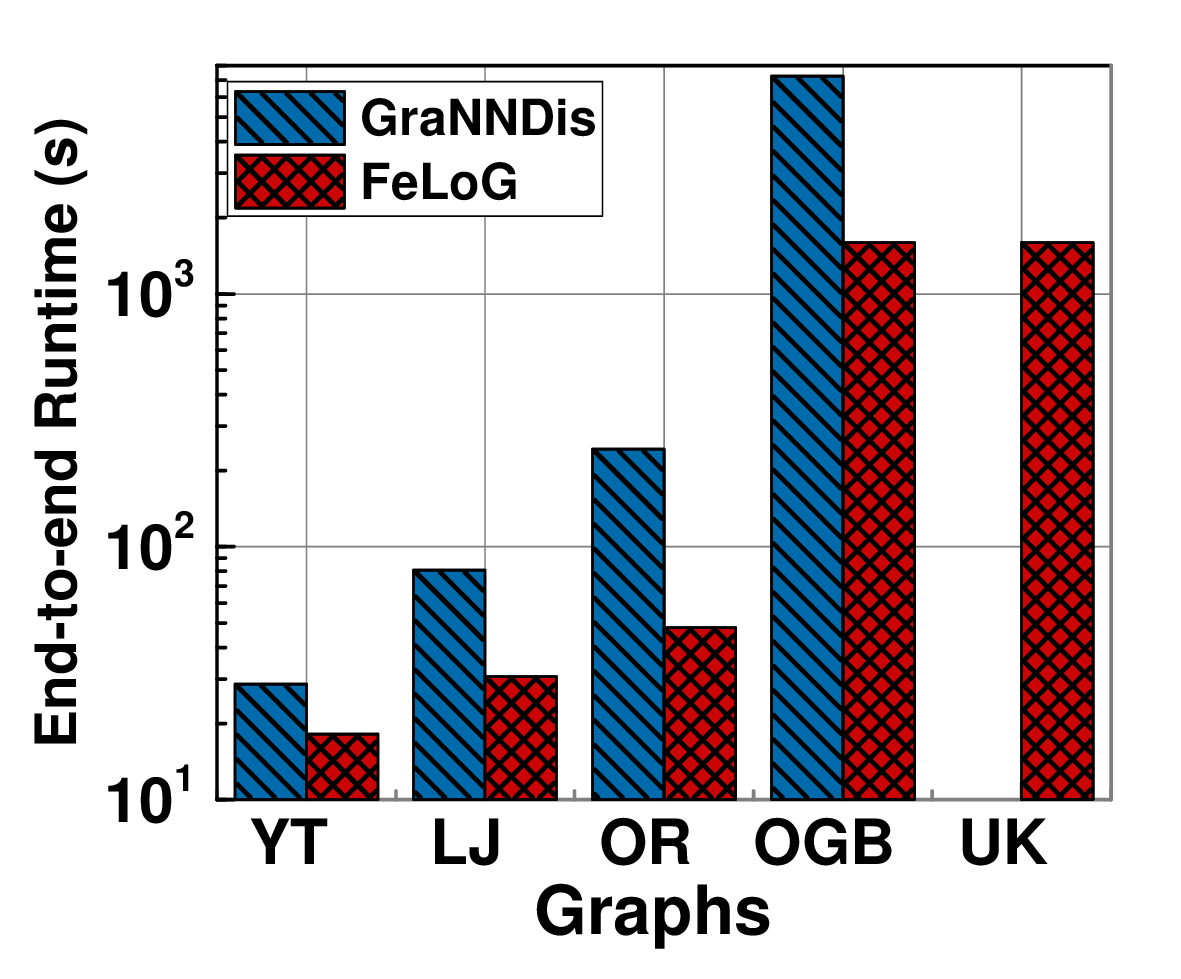}
    \vspace{-5mm}
    \caption{{{\small Multi-GPU}}}
    \vspace{-3mm}
    % \label{Dist_efficiency_FeedST}
  \end{subfigure} 
  \begin{subfigure}{.3\linewidth}
    \centering
    \vspace{-1mm}
    \includegraphics[width=1.05\linewidth, height= 0.91 in]{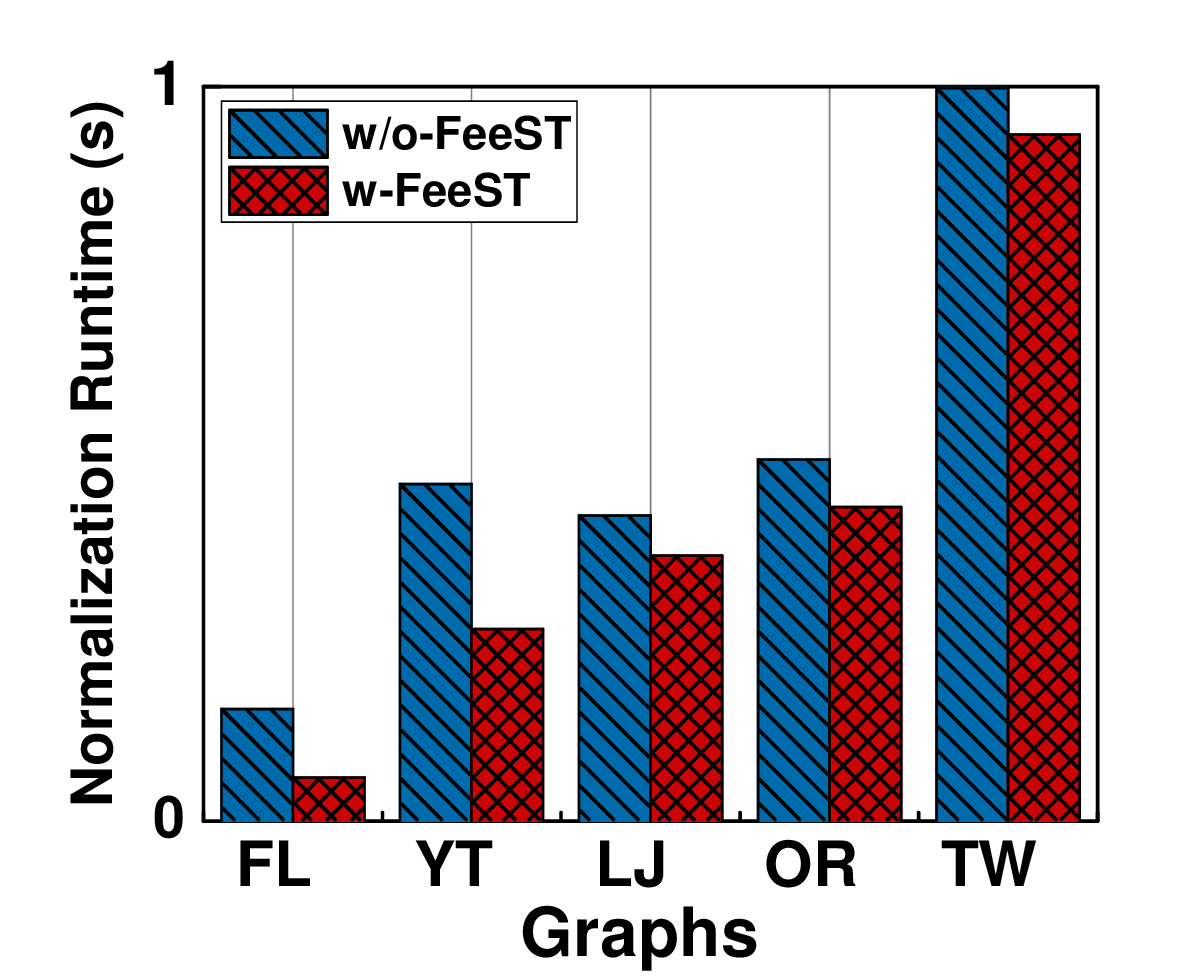}
    \vspace{-5mm}
    \caption{{\small FeeST Efficiency}}
    \vspace{-3mm}
    % \label{Dist_efficiency_FeedST}
  \end{subfigure}
  % \vspace{-2mm}
  \caption{{\felog generalizability and \feest impact on efficiency.}}
  \vspace{-2mm}
  \label{fig:felog_generlizability_feeST}
\end{figure}

\subsection{Generalizability of FeLoG}
%%% ooibc: Generalisability?
%%% Fang: Has been modified.
\label{sec:generality}

% \begin{figure}
%  \centering
%  \includegraphics[width= 2.8 in]{Figures/FeLoG_API.eps}
%  \caption{Generalizability of {\sf FeLoG}: Performance improvements on {\sf NeutronTP} and {\sf DistDGL} using the unified API.}
%  \label{FeLoG_generality}
% \end{figure}

% \begin{table}[tb!]
% \begin{minipage}{0.36\linewidth}

% \end{minipage}%
%  \quad
% \begin{minipage}{.57\linewidth}
%     \centering
%     \includegraphics[width= 1.8 in]{Figures/FeLoG_API.eps}%
%     \vspace{-4mm}
%     \captionof{figure}
%       {\felog Generalizability. %Performance gains on {\sf NeutronTP} and {\sf DistDGL} with the unified API.
%         \label{FeLoG_generality}
%       }
% \end{minipage}%\hfill
% \vspace{-5mm}
% \end{table}

A key feature of {\sf FeLoG} is its general API design, which supports seamless integration with both random walk-based and GNN-based systems (\S\ref{sec:FeedST}). 
To validate this generalizability, we extend two representative distributed GNN systems, {\sf NeutronTP} and {\sf DistDGL}, by incorporating the feedback-coupled sampling-training model ({\sf FeeST}) via the unified API of {\sf FeLoG}, resulting in {\sf NeutronTP+FeeST} and {\sf DistDGL+FeeST}. 
As shown in Figure~\ref{fig:felog_generlizability_feeST}(a), {\sf FeLoG} consistently reduces normalization runtime across datasets. 
For {\sf NeutronTP}, it achieves speedups between $1.33\times$ and $1.69\times$, with an average of $1.50\times$. 
For {\sf DistDGL}, the speedups range from $1.23\times$ to $1.67\times$, averaging $1.45\times$. 
These results demonstrate that the proposed feedback loop mechanism generalizes beyond random walk-based solutions and can be effectively applied to GNN-based pipelines, yielding significant efficiency gains without altering their underlying training logic.
We further compare {\sf FeLoG} with {\sf GraNNDis} \cite{song2024granndis}, a recent multi-GPU GNN training framework, on a server with four NVIDIA A40 GPUs.
As shown in Figure~\ref{fig:felog_generlizability_feeST}(b),
{\sf FeLoG} achieves an average $3.45\times$ speedup over {\sf GraNNDis} across datasets, further confirming its ability to support and accelerate multi-GPU training.

\subsection{Performance due to Individual Parts of FeLoG}
\label{sec:individual}
%%% ooibc: for this, an histogram of overall cost with breakdown would have served the purpose
%%% else it could be repetitive of the earlier subsections?
%%% Fang: Previous section presents the overall performance evaluation of FeLoG, 
%%% while this section analyzes the component-level contributions of its key mechanisms (FeeST, ACM, and RiPP).
%%% Thanks for your suggestion, we can add an overall runtime breakdown figure summarizing the contributions of each part at the beginning of this section, and then discuss each individual parts. Let me try to acchieve this.
% \textcolor{blue}{will add a breakdown analysis for \felog}
% We begin with a breakdown analysis of {\sf FeLoG}'s performance, highlighting the contribution of individual components in sampling and training. As shown in Figure~\ref{fig:felog_breakdown_train_feeST} (a) and (b), we analyze the running time ratios of sampling ({\sf FeLoG-S}), compression ({\sf FaBC-C}), training ({{\sf FeeST-T}), decompression ({\sf FaBC-D}), and synchronization ({\sf HaSyn}) across different graph datasets. 
% These results demonstrate that {\sf FeLoG} efficiently focuses on sampling and training ({\sf FeeST}), with minimal interference from communication operations (\fabc and \hasyn).

\spara{Feedback-driven Sampling-Training Efficiency.}
To evaluate the effectiveness of {\sf FeLoG}'s design (\S\ref{sec:FeLoG}), we first examine the impact of the feedback-driven sampling-training coupling mechanism ({\sf FeeST}, \S \ref{sec:FeedST}). We compare the execution time of {\sf FeLoG} with and without {\sf FeeST} (denoted as {\sf w-FeeST} and {\sf w/o-FeeST}) across all graphs. In {\sf w/o-FeeST}, the number of sampling iterations is determined using the same strategy as {\sf DistGER}, while all other settings remain consistent.
As shown in Figure~\ref{fig:felog_generlizability_feeST}(c), {\sf w-FeeST} achieves significant acceleration, ranging from 1.6$\times$ to 5.5$\times$. This improvement comes from the dynamic feedback loop between sampling and training, where sampling is adaptively guided by real-time embedding quality indicators, effectively reducing redundant sampling and unnecessary computation.
{We further assess the sensitivity of \feest to the similarity parameter $\mu$ in \S \ref{sec:exp_sensitivity_analysis}.}

\eat{
To further assess the sensitivity of {\sf FeeST} to the similarity convergence parameter $\mu$, we report the $AUC$ scores and execution times of {\sf FeLoG} on the {\em Flickr} graph with varying $\mu$ values. While a smaller $\mu$ often requires more sampling to reach convergence, increasing $\mu$ leads to longer training time without consistently improving accuracy. 
As shown in Figure~\ref{Dist_efficiency_FeedST}(b) and (c), the $AUC$ score peaks around $\mu = 0.6$-$0.65$, beyond which accuracy slightly drops, while execution time continues to grow. 
This observation suggests that $\mu$ values in this range provide a favorable trade-off between efficiency and accuracy. 
Similar trends are observed on other datasets, so we use $\mu=0.65$ as the default configuration in our experiments.
}
% In addition, we validate the generalizability of the {\sf FeeST} by applying its API to GNN-based systems (\S \ref{sec:generality}), where it also demonstrates consistent efficiency improvements.

% \begin{figure}
%   \centering
%   \includegraphics[width= 3.4 in]{Figures/ACM_1.eps}
%   \vspace{-6mm}
%   \caption{{Impact of \acm on communication performance.}}
%   \label{Dist_efficiency_communicaition}
% \end{figure}

\begin{figure}
  \centering
  \begin{subfigure}{.325\linewidth}
    \centering
    \includegraphics[width=\linewidth, height=1 in]{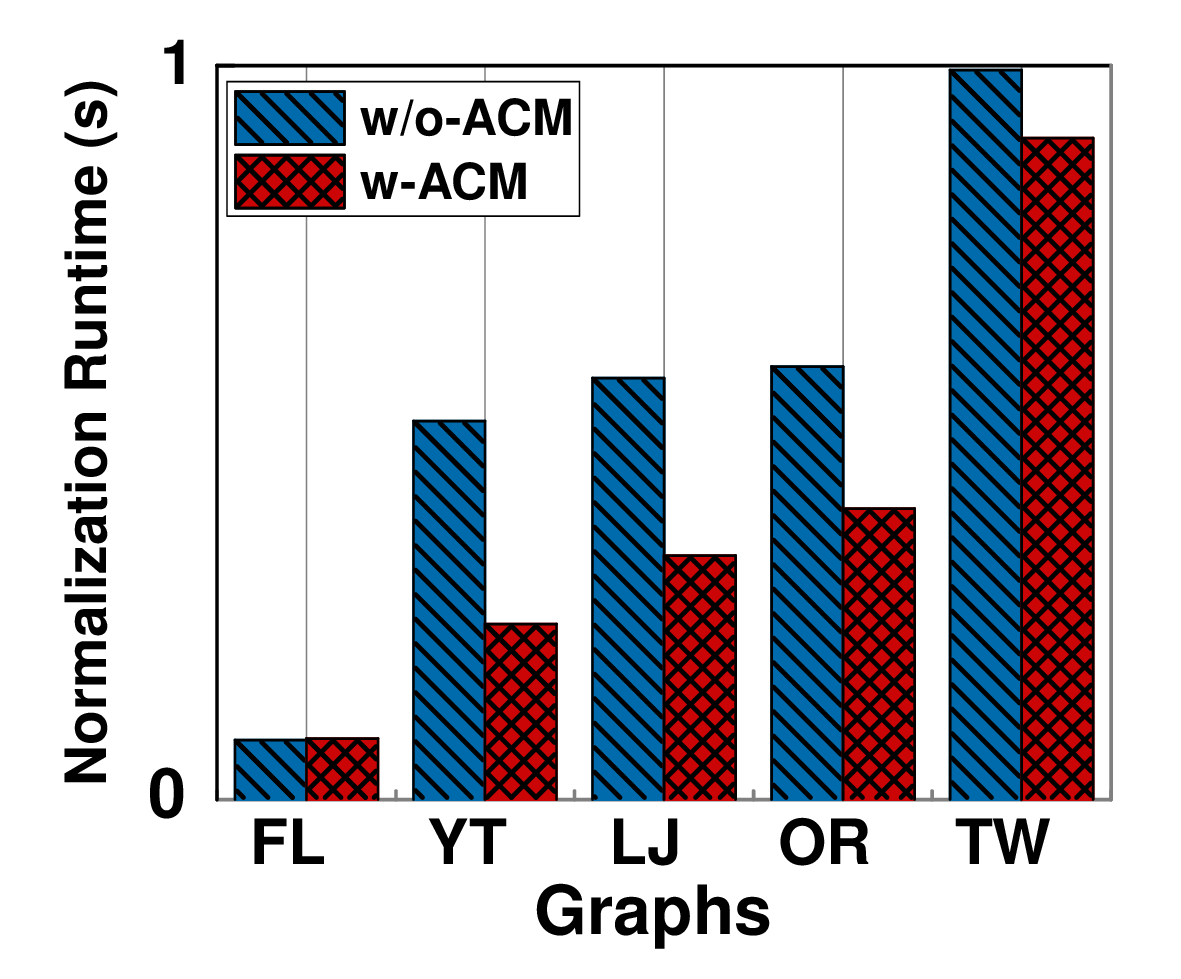}
    \vspace{-6mm}
    \caption{ACM Efficiency}
    % \label{performance_bottleneck}
  \end{subfigure} 
  \begin{subfigure}{.325\linewidth}
    \centering
    \includegraphics[width=\linewidth, height=0.98 in]{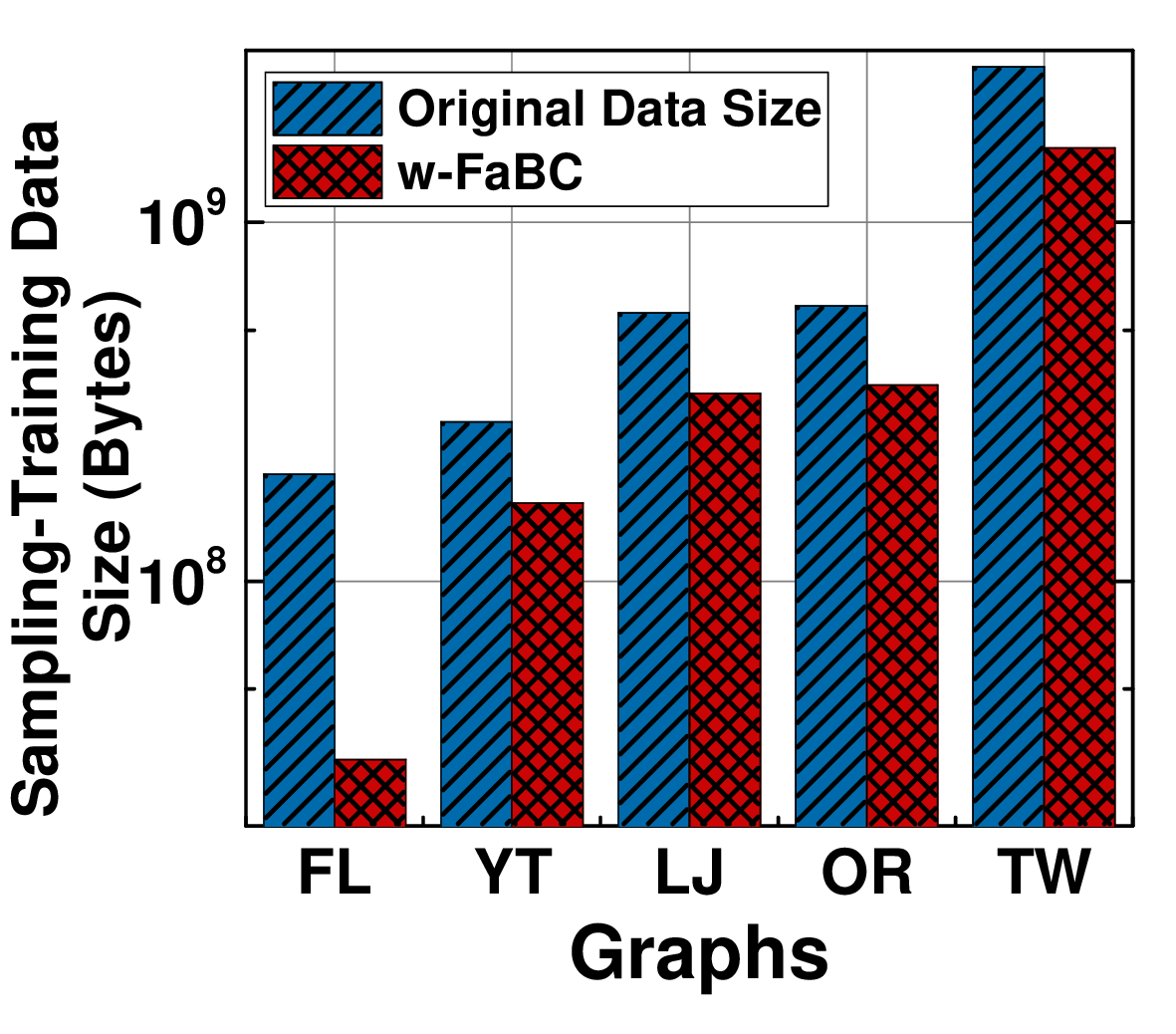}
    \vspace{-6mm}
    \caption{Training Reduction}
    % \label{fig:loss-bit-alpha}
  \end{subfigure}
  \begin{subfigure}{.325\linewidth}
    \centering
    \includegraphics[width=\linewidth, height=1 in]{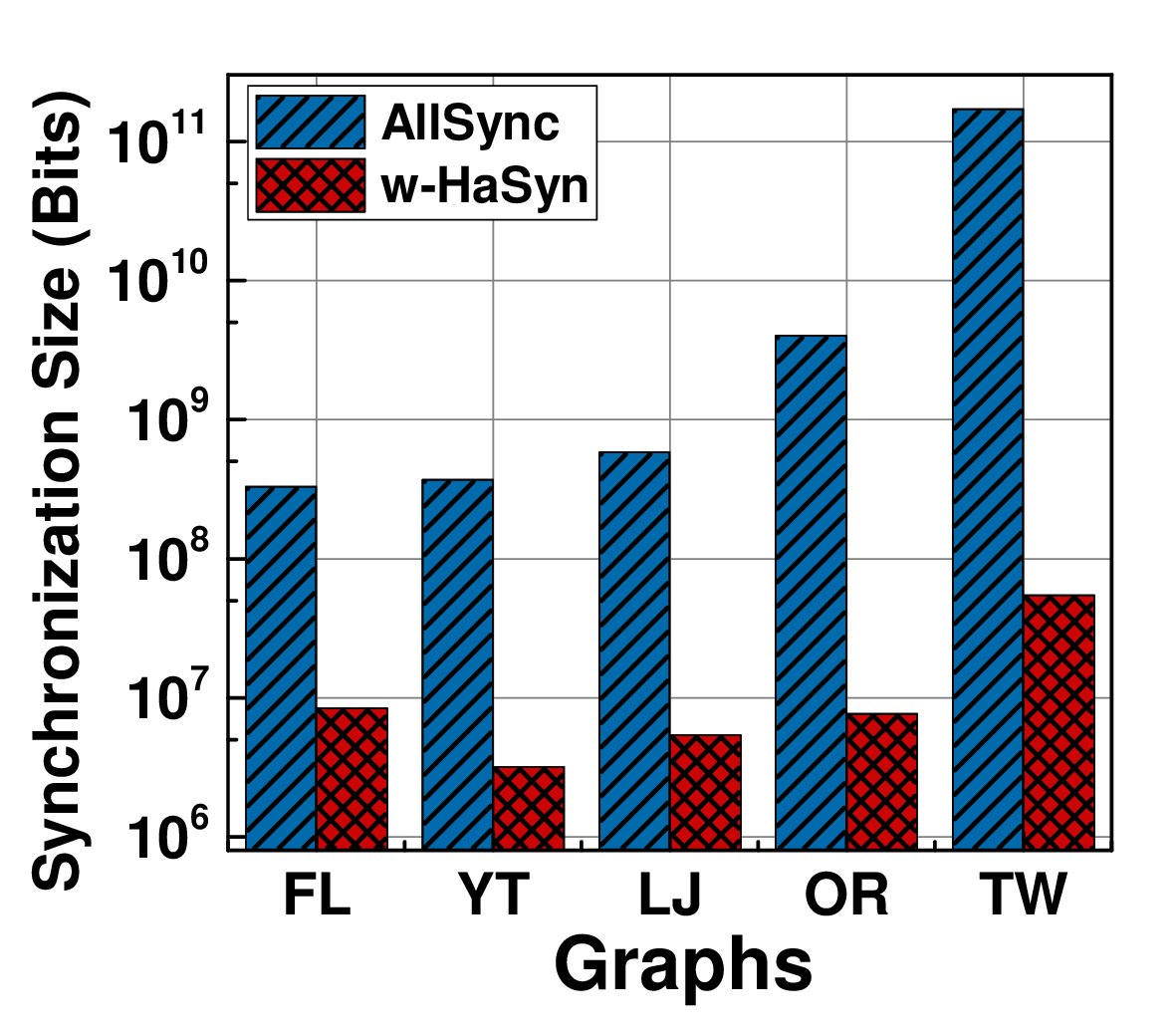}
    \vspace{-6mm}
    \caption{Sync Reduction}
    % \label{fig:loss-bit-alpha}
  \end{subfigure} 
  \vspace{-6mm}
  \caption{Impact of \acm on communication performance.}
  \label{Dist_efficiency_communicaition}
  \vspace{-2mm}
\end{figure}

\spara{Communication Efficiency.}
%Building upon {\sf FeLoG} equipped with {\sf FeeST}, 
We next evaluate the impact of the activity-aware communication mechanism ({\sf ACM}, \S\ref{sec:acm}) on {\sf FeLoG}'s communication performance in distributed settings.
Specifically, we compare the end-to-end runtime of {\sf FeLoG} with and without {\sf ACM}, denoted as {\sf w-ACM} and {\sf w/o-ACM}, respectively.
As shown in Figure~\ref{Dist_efficiency_communicaition}(a), enabling {\sf ACM} yields an average speedup of 6.19$\times$ across all datasets.
This substantial gain comes from two complementary strategies in {\sf ACM}: {\sf FaBC} (\S\ref{sec:FaBC}) alleviates PCIe bottlenecks within each machine, while {\sf HaSyn} (\S\ref{sec:HaSyn}) reduces network traffic across machines.
To better understand the contributions of these two components, we further break down their individual impact.
Figure~\ref{Dist_efficiency_communicaition}(b) shows the size of training data transferred between CPU and GPU, before and after applying {\sf FaBC}, denoted as {\sf Original Data Size} and {\sf w-FaBC}, respectively. On average, {\sf FaBC} reduces intra-machine data transfer volume by 53.1\%, demonstrating its effectiveness in relieving PCIe overhead.
Finally, Figure~\ref{Dist_efficiency_communicaition}(c) compares {\sf HaSyn} against the conventional full synchronization method {\sf AllSync} adopted by systems like {\sf DistDGL} and {\sf PBG}. The results show that {\sf HaSyn} reduces communication cost by over 90\% during each synchronization period.
We compare {\sf HaSyn} with {\sf AllSync} and {\sf RandomSync} on {\em LJ}.
Compared with {\sf AllSync}, {\sf HaSyn} reduces runtime from 509.39s to 111.26s, achieving a $4.58\times$ speedup with only a slight AUC drop.
It also achieves higher AUC than {\sf RandomSync} under the same synchronization size, indicating that update hotness is effective for selecting embeddings to synchronize.
In addition, we assess how compression frequency range affects the compression ratio of \fabc in \S~\ref{sec:exp_sensitivity_analysis}.

\begin{figure}
  \begin{subfigure}{.323\linewidth}
    \centering
    \includegraphics[width=\linewidth]{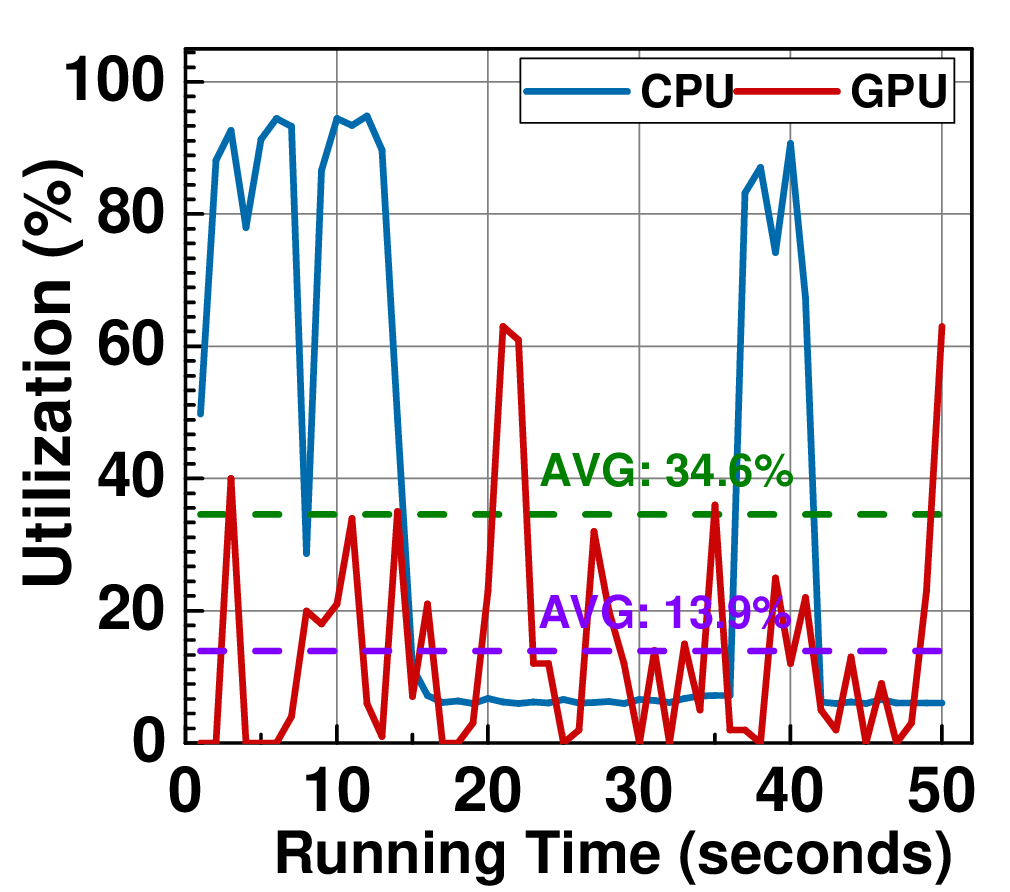}
    \vspace{-5mm}
    \caption{\small Decoupled}
    % \label{fig:loss-bit-alpha}
  \end{subfigure}
  \begin{subfigure}{.323\linewidth}
    \centering
    \includegraphics[width=\linewidth]{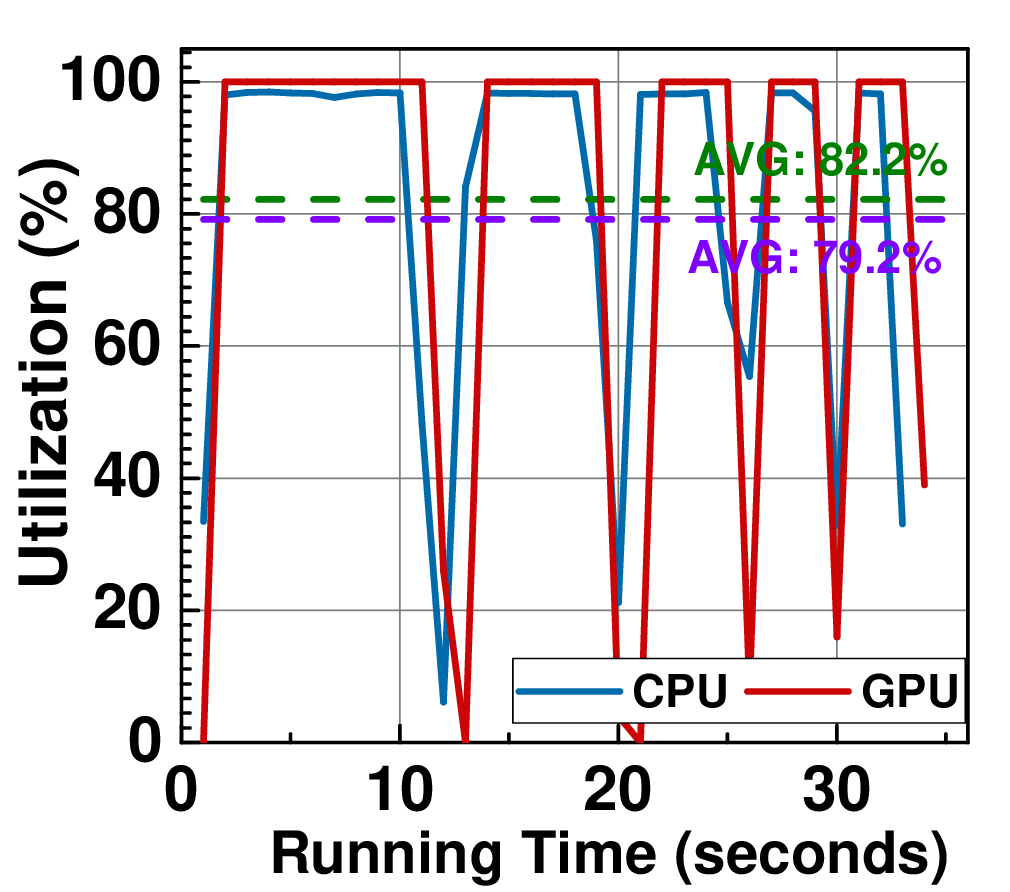}
    \vspace{-5mm}
    \caption{\small Pipeline}
    % \label{fig:loss-bit-alpha}
  \end{subfigure}
  \begin{subfigure}{.335\linewidth}
    \centering
    \includegraphics[width=\linewidth]{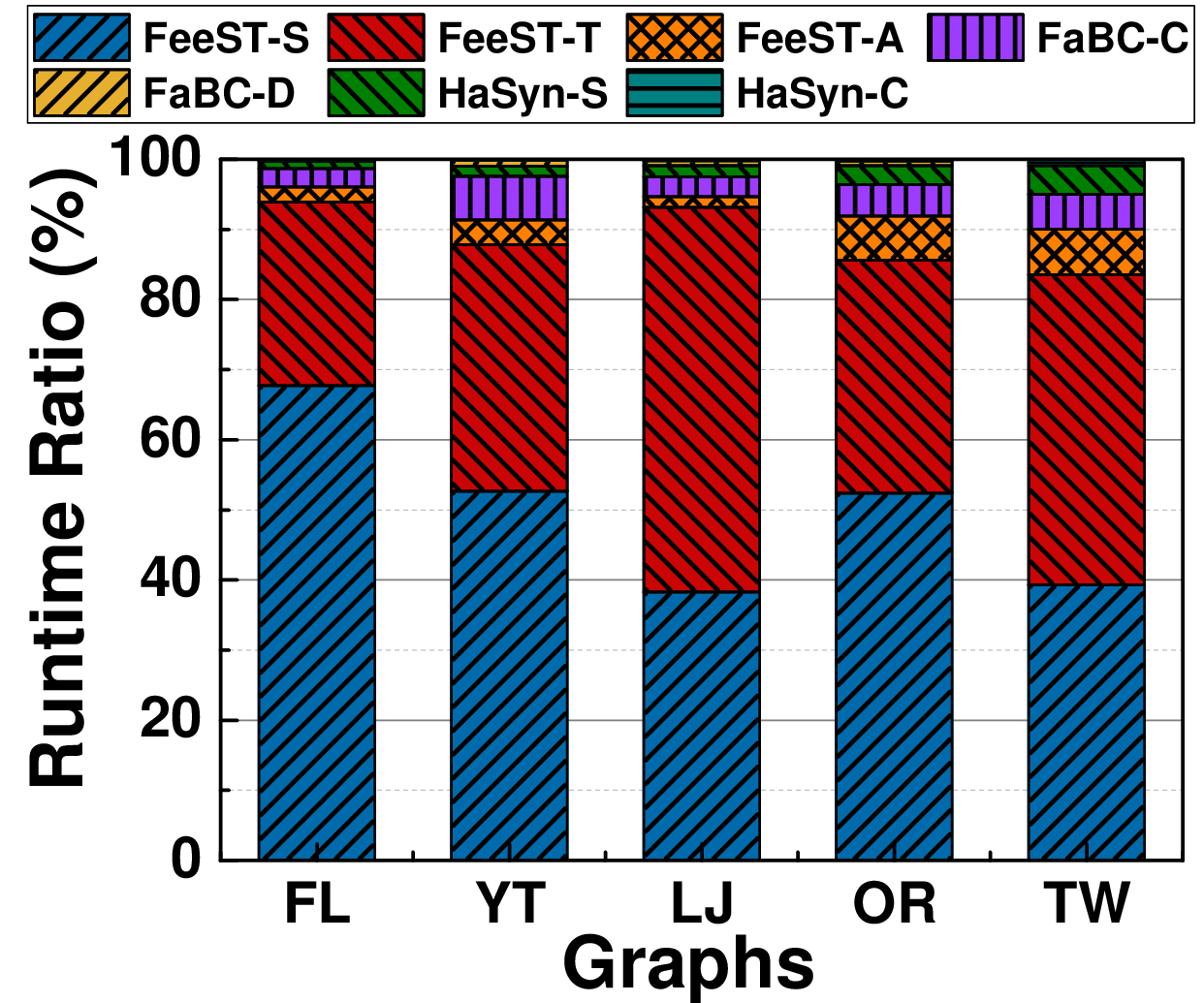}
    \vspace{-5mm}
    \caption{{\small {Breakdown Analysis}}}
    % \label{fig:loss-bit-alpha}
  \end{subfigure} 
  \vspace{-6mm}
  \caption{{Resource utilization and runtime breakdown analysis.}}
  \label{fig:Dist_efficiency_rip_breakdown}
  \vspace{-6mm}
\end{figure}

\spara{Computing Resources Utilization.}
To improve computational efficiency, {\sf FeLoG} introduces a round-interleaved pipeline mechanism ({\sf RiPP}, \S \ref{sec:rip}) by decoupling the sampling and training operators. To evaluate its effectiveness, we measure CPU and GPU utilization over time on the {\em LJ} dataset under both decoupled and pipelined execution.
As shown in Figure~\ref{fig:Dist_efficiency_rip_breakdown}(a), under decoupled execution, CPU utilization drops significantly after the initial sampling phase, leaving resources idle. Similarly, GPU utilization remains low, as each training round must wait for sampling to complete.
In contrast, {\sf RiPP} enables concurrent execution between rounds; the sampling stage overlaps with the training stage through different threads and devices. This design leads to much higher resource utilization: Figure~\ref{fig:Dist_efficiency_rip_breakdown}(b) shows average CPU and GPU utilizations of 82.2\% and 79.2\%, respectively. 
Overall, {\sf RiPP} effectively boosts system throughput and scalability by maximizing hardware utilization.

% \subsection{Breakdown Analysis of \felog}
% \label{sec:exp_breakdown_analysis}

\spara{{Breakdown Analysis.}}
We further provide a fine-grained runtime breakdown of {\sf FeLoG}. As shown in Figure~\ref{fig:Dist_efficiency_rip_breakdown}(c), we separate the execution into sampling ({\sf FeeST-S}), training ({\sf FeeST-T}), active-set computation ({\sf FeeST-A}), compression ({\sf FaBC-C}), decompression ({\sf FaBC-D}), embedding synchronization ({\sf HaSyn-S}), and synchronization-set selection ({\sf HaSyn-C}) across different graphs. The normalized breakdown shows that sampling and training remain the dominant components. In contrast, {\sf FeeST-A}, {\sf FaBC}, and {\sf HaSyn} introduce only limited runtime overhead, accounting for 4.0\%, 4.7\%, and 2.4\% on average, respectively.
In addition, the dominant extra memory of {\sf HaSyn} comes from the synchronization-set buffer, whose size is $|\mathcal{F}| \times d \times 4$ bytes for $d$-dimensional float embeddings. For example, on {\em LJ}, $|\mathcal{F}|=1316$, so it requires only $0.64$ MB of memory.

\begin{figure}
  \centering
    \begin{subfigure}{.48\linewidth}
    \centering
    \includegraphics[width=\linewidth]{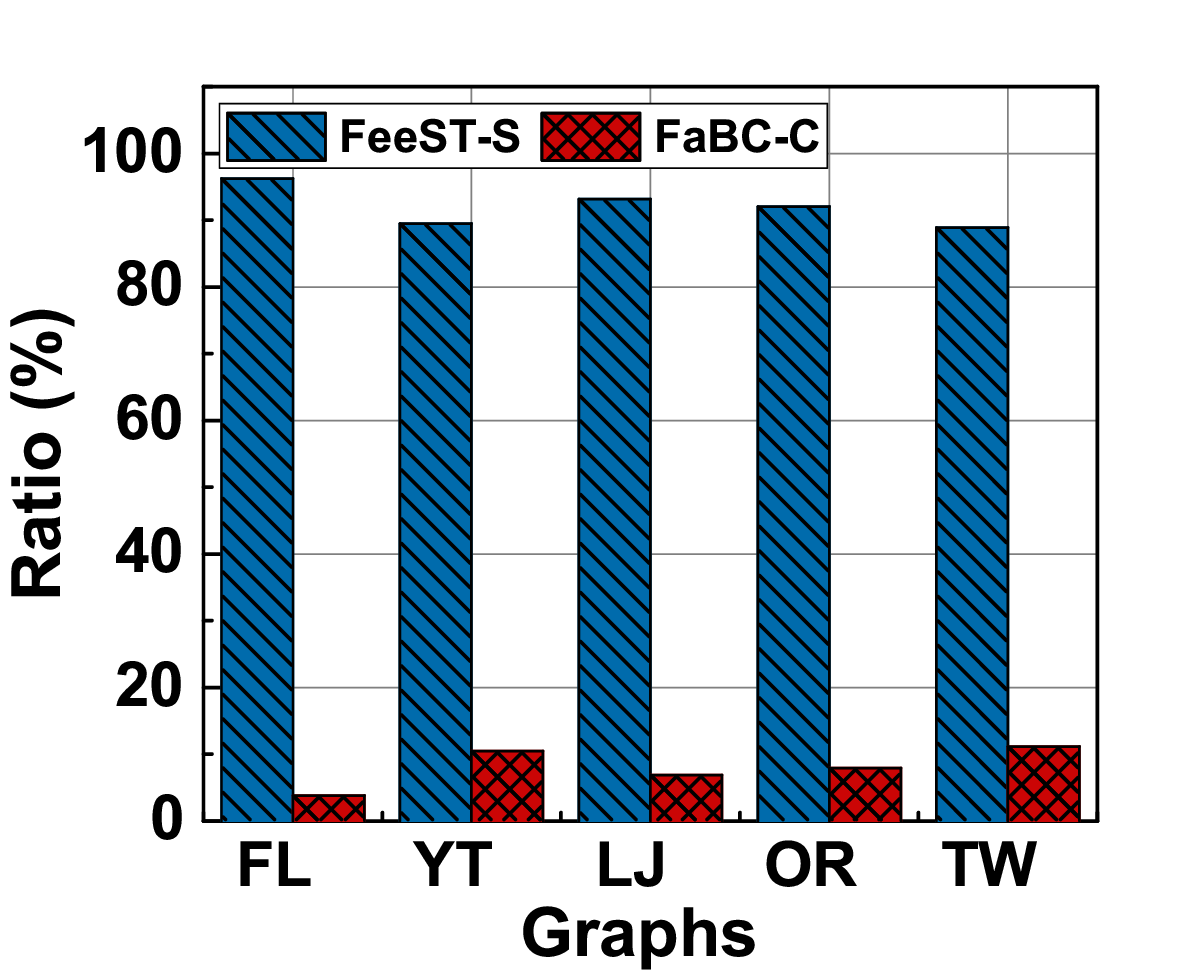}
    \vspace{-5mm}
    \caption{Sampling}
    % \label{fig:felog_breakdown_sample}
  \end{subfigure}
  \begin{subfigure}{.48\linewidth}
    \centering
    \includegraphics[width=\linewidth]{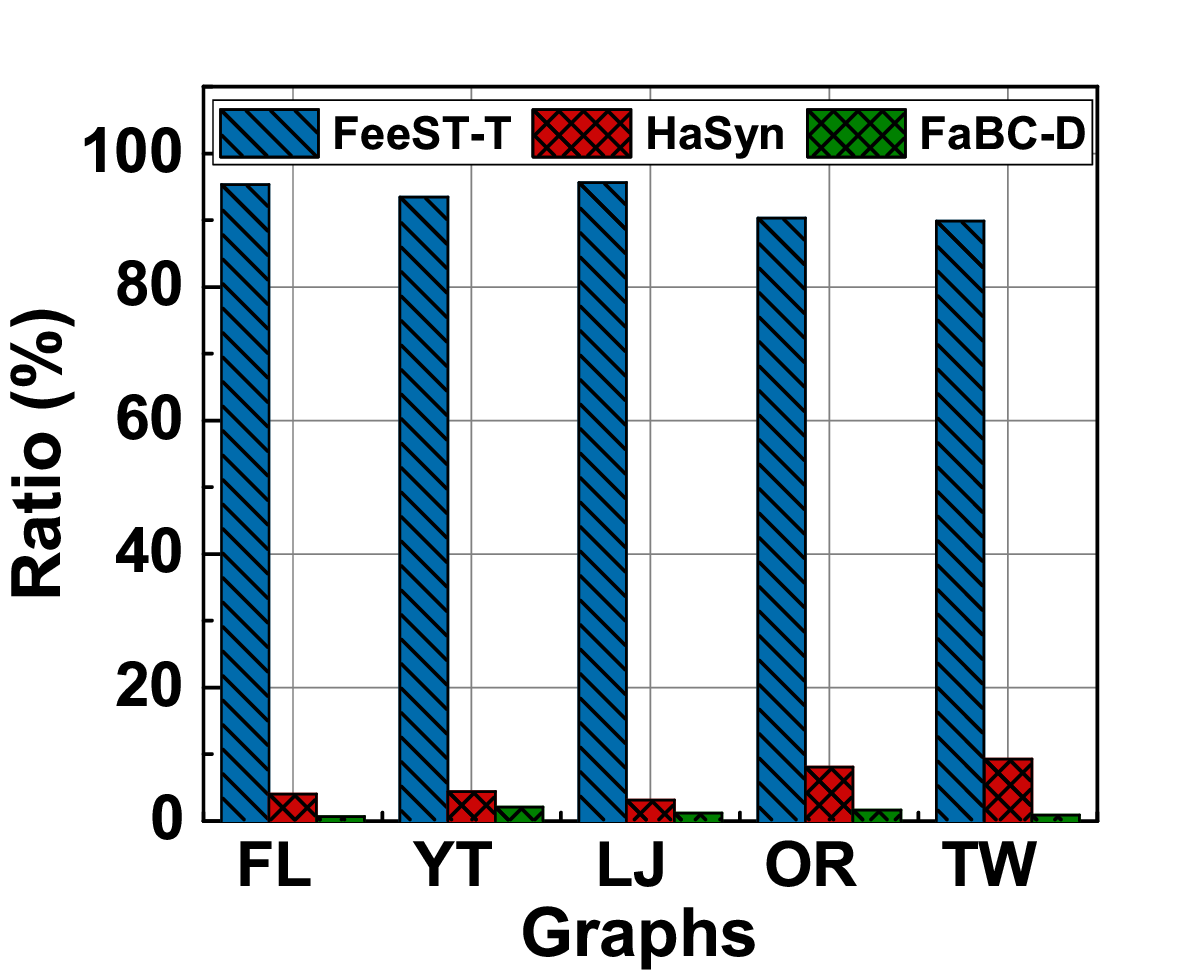}
    \vspace{-5mm}
    \caption{Training}
    % \label{fig:felog_breakdown_train}
  \end{subfigure} 
  \vspace{-2mm}
  \caption{Breakdown analysis of \felog for sampling and training, and impact of \feest on efficiency.}
  \vspace{-4mm}
  \label{fig:felog_breakdown_train_feeST}
\end{figure}

\subsection{Parameter Sensitivity Analysis}
\label{sec:exp_sensitivity_analysis}
To further assess the sensitivity of {\sf FeeST} to the similarity convergence parameter $\mu$, we report the $AUC$ scores and execution times of {\sf FeLoG} on the {\em Flickr} graph with varying $\mu$ values. A smaller \(\mu\) prunes the active frontier more aggressively, whereas increasing \(\mu\) retains more nodes for subsequent sampling and leads to longer training time.
As shown in Figure~\ref{fig:Dist_efficiency_FeedST_mu_heter}(a) and (b), the $AUC$ score peaks around $\mu = 0.6$-$0.65$, beyond which accuracy slightly drops, while execution time continues to grow. 
This observation suggests that $\mu$ values in this range provide a favorable trade-off between efficiency and accuracy for this dataset. 
Similar trends are observed on other datasets used in our experiments, so we use $\mu=0.65$ as the default configuration for our datasets in the experiments.
For heterophilic settings, we further evaluate {\sf FeeST} on {\em Amazon-ratings} (AR) and {\em Roman-empire} (RE). 
As shown in Figure~\ref{fig:Dist_efficiency_FeedST_mu_heter}(c) and (d),
{\sf FeeST} provides a meaningful quality-efficiency trade-off under varying \(\mu\): {\em AR} is broadly stable, while {\em RE} favors a lower threshold. Runtime increases with \(\mu\), and \(\mu=1\) corresponds to the {\sf w/o-FeeST} setting.
These results support validation-calibrated \(\mu\) in heterophilic settings, with only 1.85s and 1.68s per candidate on {\em AR} and {\em RE}, respectively.

To assess how compression frequency range affects the compression ratio, Table~\ref{compression_ratio} shows the results when compressing the top-1, top-2, and top-3 most frequent nodes in the training data. The results indicate that expanding the compression range does not significantly improve the compression ratio. Thus, a balance must be struck between compression efficiency and computation overhead. {\sf FeLoG} addresses this by analyzing the first-round sampling to dynamically configure compression parameters, effectively balancing compression gain and processing cost.

\begin{figure}
% \vspace{-2mm}
  \centering
  \begin{subfigure}{.45\linewidth}
  % \vspace{-2mm}
    \centering
    \includegraphics[width=1\linewidth]{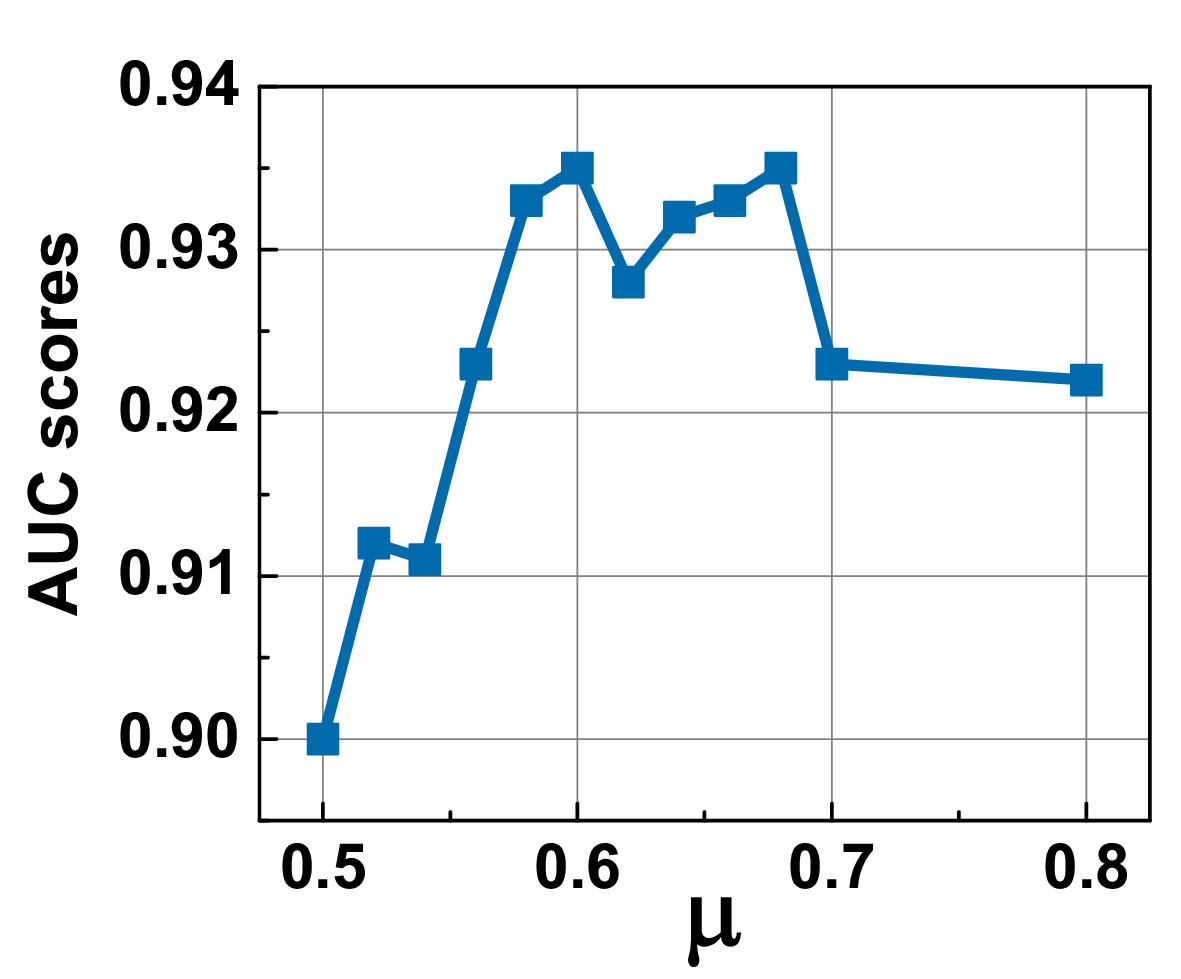}
    % \vspace{-2mm}
    \caption{{\footnotesize{{Accuracy on link prediction}}}}
    % \label{performance_bottleneck}
  \end{subfigure} 
  \begin{subfigure}{.45\linewidth}
  % \vspace{-2mm}
    \centering
    \includegraphics[width=1\linewidth]{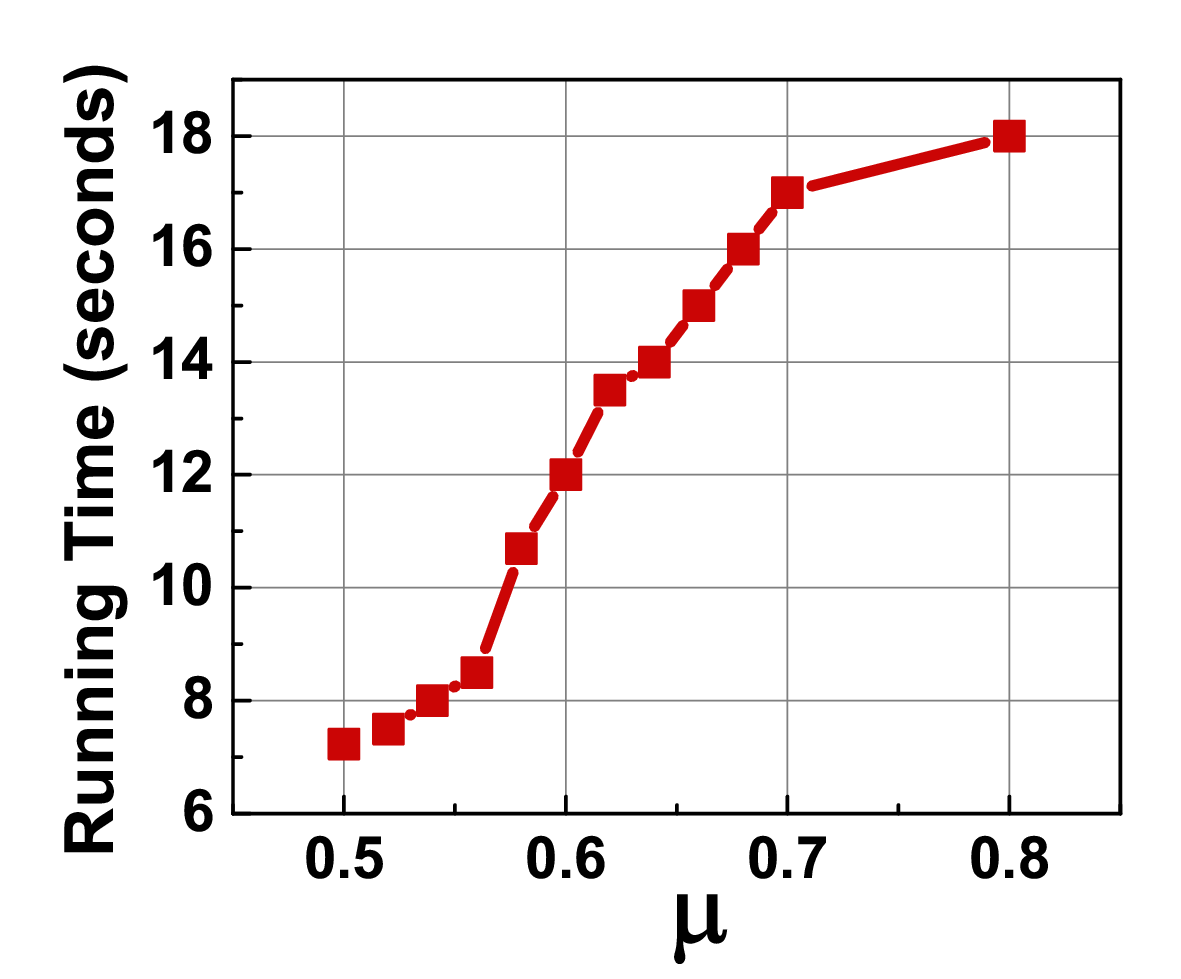}
    % \vspace{-2mm}
    \caption{{\footnotesize{{Efficiency on link prediction}}}}
  \end{subfigure} 
  \begin{subfigure}{.45\linewidth}
  % \vspace{-2mm}
    \centering
    \includegraphics[width=\linewidth]{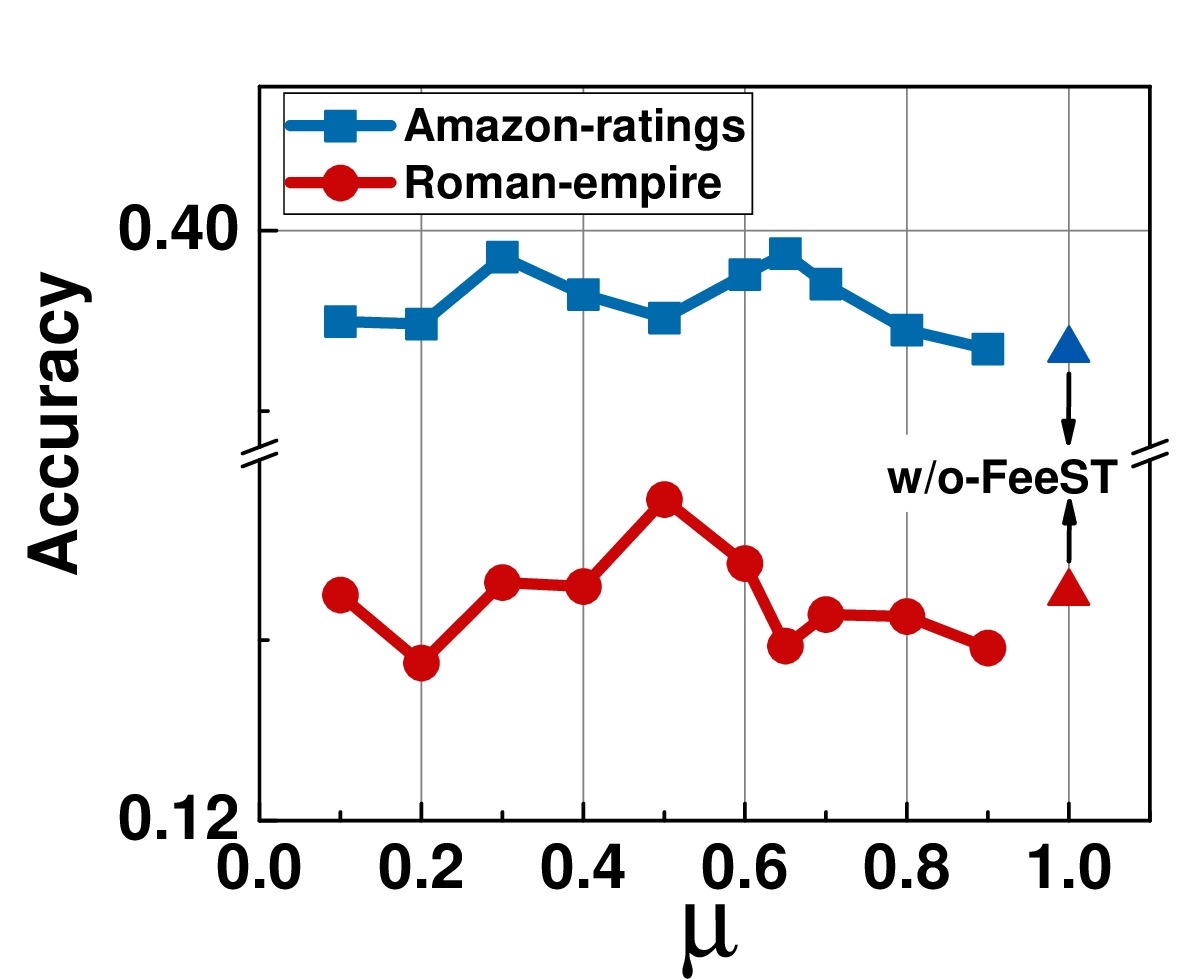}
    % \vspace{-2mm}
    \caption{{\footnotesize{{Accuracy on node classification}}}}
  \end{subfigure}
  \begin{subfigure}{.45\linewidth}
  % \vspace{-2mm}
    \centering
    \includegraphics[width=\linewidth]{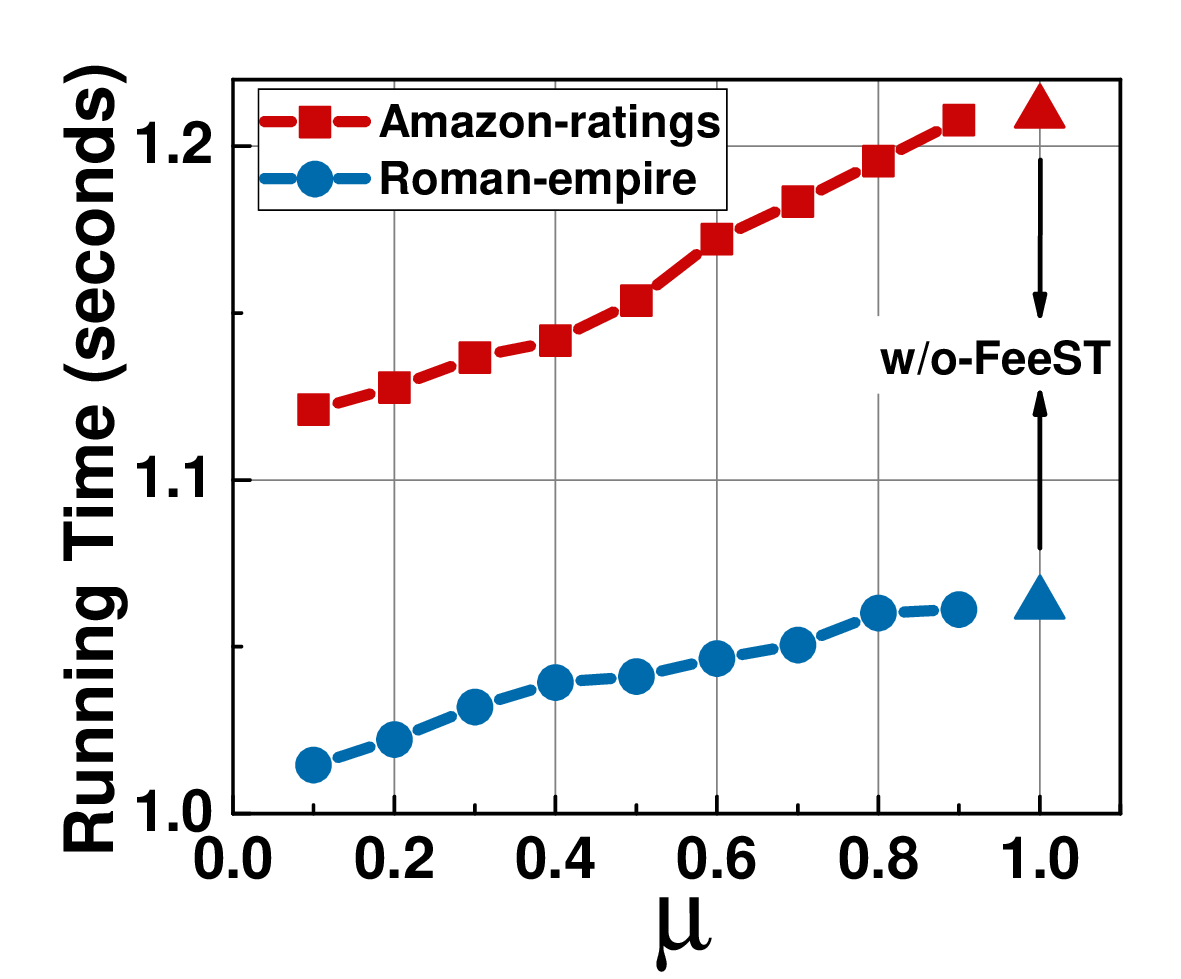}
    % \vspace{-2mm}
    \caption{{\footnotesize{{Efficiency on node classification}}}}
  \end{subfigure} 
   \centering
  \vspace{-2mm}
  \caption{{Sensitivity of \feest w.r.t. parameter $\mu$.}}
  \vspace{-2mm}
  \label{fig:Dist_efficiency_FeedST_mu_heter}
\end{figure}

\begin{table}
\footnotesize
\centering
\caption{Impact of different compression frequency ranges on compression ratio.}
\label{compression_ratio}
\begin{tabular}{lccc}
\hline
\textbf{Dataset} & \textbf{Top-1} & \textbf{Top-2} & \textbf{Top-3} \\
\hline
Flickr       & 0.67 & 0.48 & 0.46 \\
YouTube      & 0.59 & 0.42 & 0.40 \\
LiveJournal  & 0.59 & 0.48 & 0.44 \\
Com-Orkut     & 0.60 & 0.49 & 0.43 \\
Twitter      & 0.59 & 0.47 & 0.41 \\
\hline
\end{tabular}
\end{table}
\section{Conclusions}
\label{sec:conclusions}
% We presented {\sf FeLoG}, a scalable and efficient distributed system for graph embedding with 
% that addresses inefficiency, high communication overhead, and low resource utilization.
We presented {\sf FeLoG}, a scalable and efficient distributed graph embedding system that addresses redundant sampling, excessive communication, and low resource utilization caused by the decoupling between sampling and training.
\felog introduces three innovations: (1) a feedback-coupled sampling-training model that adapts sampling based on real-time embedding quality, (2) an activity-aware communication mechanism that reduces both intra- and inter-machine communication, and (3) a round-interleaved pipeline that overlaps sampling and training to improve CPU-GPU utilization.
% Our experiments on billion-edge graphs demonstrate that {\sf FeLoG} significantly improves system efficiency, scalability, and generalizability over state-of-the-art systems, highlighting the potential of feedback loop design for large-scale machine learning workloads.
Experiments on billion-scale graphs show that {\sf FeLoG} achieves an average speedup of 27.9$\times$, reduces communication cost by over 53.1\%, and sustains over 80\% CPU-GPU utilization, while generalizing to different graph embedding and GNN-based frameworks.
Future work includes incorporating topology-aware optimization for intra-node GPU interconnects and exploring GPU-directed remote data access to further reduce CPU-mediated transfers.

\begin{acks}
This work was supported by the National Natural Science Foundation of China (Nos. 62402187 and U22A2027), the China Postdoctoral Science Foundation (Nos. GZB20240243 and 2024M751009), the Postdoctoral Project of Hubei Province (No. 2024HBBHCXA024), and the Kunpeng\&Ascend Center of Cultivation of HUST.
Peng Fang is especially grateful to Professor Beng Chin Ooi for his invaluable guidance and insightful feedback on this work during his postdoctoral research at the National University of Singapore.
\end{acks}

\bibliographystyle{ACM-Reference-Format}
\bibliography{ref}

\end{document}